\documentstyle[11pt,aaspp4]{article}

\newcommand{\omigm}{\Omega_{\it IGM}}
\newcommand{\subigm}{_{\it IGM}}

\newcommand{\suplf}{^{\it LF}}
\newcommand{\Lya}{Ly$\alpha$}
\newcommand{\Qj}{Q0302-003}
\newcommand{\Qt}{Q1935-69}
\newcommand{\HS}{HS1700+6416}
\newcommand{\HE}{HE2347-4342}
\newcommand{\ME}{Miralda-Escud\'e}
\newcommand{\ea}{et al.~}
\newcommand{\tauhe}{\tau_{\it HeII}}
\newcommand{\nhone}{N_{\it HI}}
\newcommand{\nhetwo}{N_{\it HeII}}
\newcommand{\hethree}{$^3$He}
\newcommand{\hefour}{$^4$He}
\newcommand{\subHe}{_{\it HeII}}
\newcommand{\subH}{_{\it HI}}
\newcommand{\subHLy}{_{\it H Ly}}

\newcommand{\NN}{{\cal N}}
\newcommand{\fnz}{\frac{\partial^2\NN}{\partial \nhone \partial z}}
\newcommand{\fnzt}{\partial^2\NN/\partial \nhone \, \partial z}
\newcommand{\beq}{\begin{equation}}
\newcommand{\eeq}{\end{equation}}
\newcommand{\kms}{km~s$^{-1}$}
\newcommand{\cdu}{cm$^{-2}$}
\newcommand{\fluxu}{erg s$^{-1}$ cm$^{-2}$ Hz$^{-1}$ sr$^{-1}$}
\newcommand{\zem}{z_{\it em}}
\newcommand{\Keck}{{\it Keck}}
\newcommand{\Ghthin}{\Gamma\subH^{\rm thin}}
\newcommand{\Ghethin}{\Gamma\subHe^{\rm thin}}
\newcommand{\substar}{_\ast}
\newcommand{\Wbar}{\overline{W}}
\newcommand{\hetohyd}{He~II/H~I}
\newcommand{\Nobs}{N_{\it obs}}
\newcommand{\Nperp}{N_{\perp}}

\begin{document}

\title{ THE HIGH-REDSHIFT HE~II GUNN-PETERSON EFFECT: \\
IMPLICATIONS AND FUTURE PROSPECTS }

\author{ Mark A. Fardal$^{1}$, Mark L. Giroux$^{1}$, and 
    J. Michael Shull$^{1}$  }
\affil{ Center for Astrophysics and Space Astronomy, \\
Department of Astrophysical and Planetary Sciences \\
University of Colorado, Campus Box 389, Boulder, CO 80309 \\
               \        \\               
$^{1}$ also at JILA, University of Colorado \\
and National Institute of Standards and Technology.  }

\begin{abstract}
  
  Absorption due to He~II \Lya\ has now been detected at low
  resolution in the spectra of four quasars between redshifts
  $z =$ 2.74 -- 3.29.  We assess these observations, giving
  particular attention to the radiative transfer of the ionizing 
  background radiation, cloud diffuse emission and ionization physics,
  and statistical fluctuations.  We use high-resolution observations 
  of H~I absorption towards quasars to derive an improved model for
  the opacity of the intergalactic medium (IGM) from the distribution 
  of absorbing clouds in column density and redshift.  We use these 
  models to calculate the H~I and He~II photoionization history,
  the ratio $\eta =$ He~II/H~I in both optically-thin and 
  self-shielded clouds, and the average line-blanketing contribution 
  of the clouds to He~II absorption. The derived ionization rate, 
  $\Gamma_{HI} = (1-3) \times 10^{-12}$ s$^{-1}$ ($z = 2-4$) 
  is consistent with the ionizing background intensity inferred from 
  the ``proximity effect'', but it remains larger than that inferred by
  N-body hydrodynamical simulations of the Ly$\alpha$ absorber distribution. 
  The He~II observations are consistent with 
  line blanketing from clouds having $N\subH \geq 10^{12}$ cm$^{-2}$, 
  although a contribution from a more diffuse IGM would help 
  to explain the observed opacity.  We compute the expected He~II optical 
  depth, $\tauhe(z)$, and examine the implications of the sizable 
  fluctuations that arise from variations in the cloud numbers and 
  ionizing radiation field.  We assess how He~II absorption 
  constrains the intensity and spectrum of the ionizing radiation
  and the fractional contributions of the dominant sources
  (quasars and starburst galaxies).  Finally, we demonstrate how 
  high-resolution ultraviolet observations can distinguish between 
  absorption from the diffuse IGM and the Ly$\alpha$ forest clouds 
  and determine the source of the ionizing background.

\end{abstract} 


\section{INTRODUCTION}

Absorption in the He~II $\lambda$303.78 \Lya\ line has long been
considered a potentially important tool for studying the high-redshift
universe (\ME\ \& Ostriker 1990; \ME\ 1993; Shapiro, Giroux, \& Babul
1994).  As early estimates showed, He~II is more abundant than H~I in
the highly ionized ``\Lya\ forest clouds'' that appear in quasar
spectra, because He~II has a lower photoionization cross section and a
larger recombination rate.  Comparison of the H~I and He~II absorption
could tell us about the photoionizing background, the history of structure
formation, and internal conditions in the \Lya\ clouds.  Also, He~II
absorption from a smoothly distributed intergalactic medium (IGM)
should be more easily detectable than the corresponding H~I
absorption, which has proven difficult to measure unambiguously
(Giallongo \ea 1994; Williger \ea 1994; Fang \& Crotts 1995).

\begin{figure}
\plotone{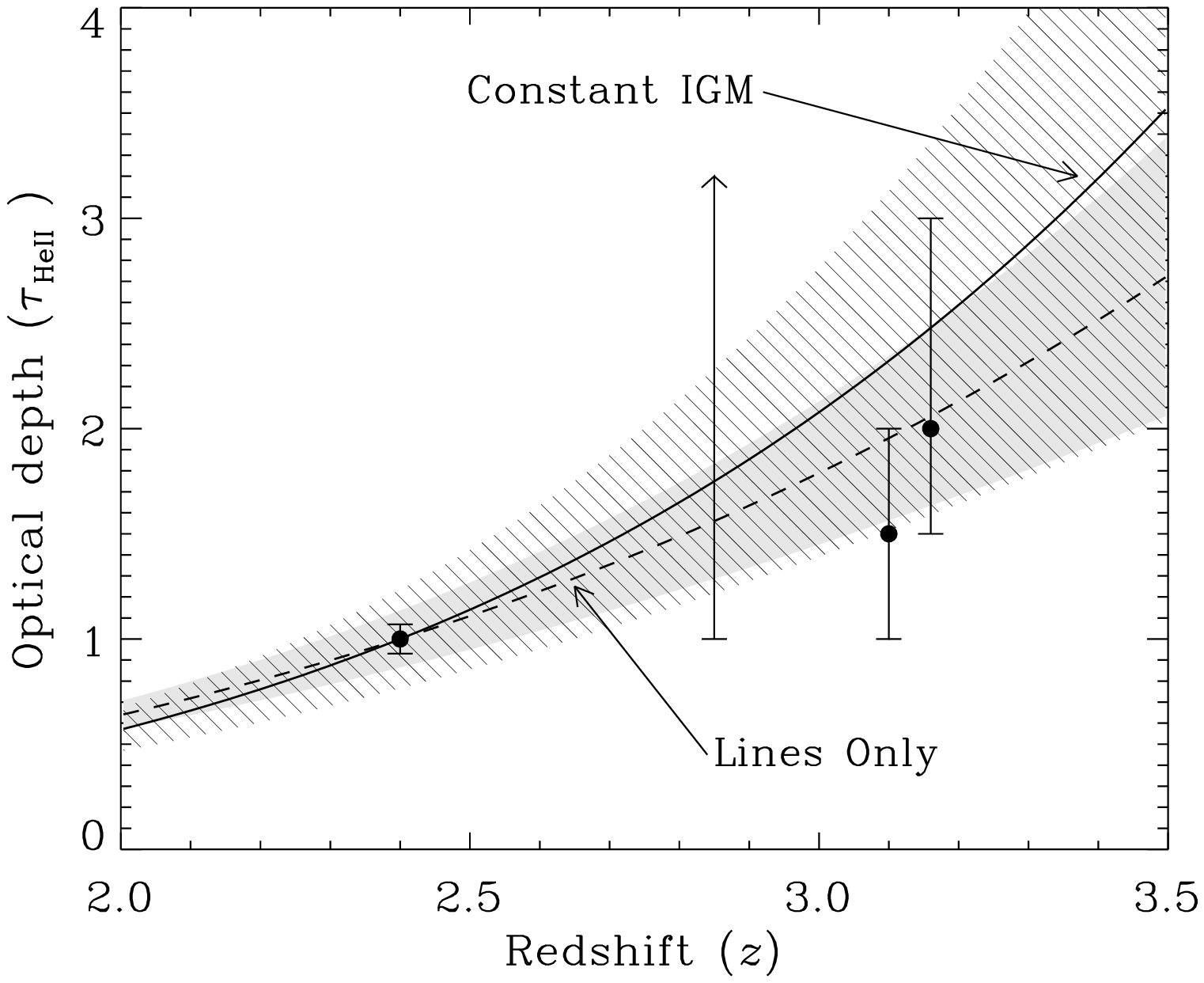}
\caption{
\label{fig:obs-opt-depth}
Mean He~II line opacity $\tau\subHe(z)$.  Points show the
four current measurements:  
\HS\ (Davidsen \ea 1996) at $\zem = 2.74$;
\HE\ (Reimers \ea 1997) at $\zem = 2.90$;
\Qt\ (Tytler \ea 1995) at $\zem = 3.18$; and 
\Qj\ (Jakobsen \ea 1994) at $\zem = 3.29$.  
The points are plotted at the mean redshifts of the He~II
absorption.  The curves show the expected evolution for an IGM of
constant comoving density ({\em solid}) and for pure line-blanketing
({\em dashed}), with constant $\eta$.  We have also indicated the
expected range of fluctuations in optical depth due to ionization 
fluctuations discussed in \S~4, for the IGM-dominated (hatched region) 
and line-dominated (shaded region) cases respectively.  However, since 
the four observed redshift windows are larger than the expected 
correlation length of the fluctuations, the expected fluctuations of 
the observed optical depth $\tau\subHe$ are smaller than the regions 
shown.  }
\end{figure}

These hopes have partially been realized by recent observations of
He~II toward four quasars. \Qj\ was observed with two instruments on
the Hubble Space Telescope (HST): at low resolution with the Faint
Object Camera (FOC) (Jakobsen \ea 1994) and at higher resolution with
the Goddard High Resolution Spectrograph (GHRS) (Hogan, Anderson, \&
Rugers 1997).  \Qt\ was observed with HST using both the Faint Object
Spectrograph (FOS; Tytler \ea 1995) and the FOC (Tytler 1997).  \HS\ 
was observed with the Hopkins Ultraviolet Spectrometer (HUT) during
the Astro-2 mission (Davidsen \ea 1996, hereafter D96).  Most
recently, \HE\ was observed with HST/GHRS (Reimers \ea 1997).  The resolution
of the HUT spectrum is $\sim \!3$~\AA\, while that of the HST/FOS
spectra is $\sim\!15$~\AA\ in the FOC spectra.  The HST/GHRS spectra 
potentially have higher resolution, although the data are usually 
binned to enhance the signal-to-noise ratio.  Thus,
individual absorption lines are
not resolved.  Instead, the results are reported (see also
Figure~\ref{fig:obs-opt-depth}) in terms of an average optical depth:
$1.3 \leq \tauhe \lesssim 4$ towards \Qj\ ($\zem =3.29$), $1.0 \leq
\tauhe \leq 2.0$ ($\zem = 3.18$) towards \Qt, $\tauhe = 1.00 \pm 0.07$
towards \HS\ ($\zem = 2.74$), and $\tauhe$ varying from $\sim\!1$ to
$\geq4.8$ toward \HE. As shown by the error estimates, the HUT result
is the most restrictive of the four, and we normalize our plot to
that point.
We anticipate our discussion in \S 3--4 by placing
these data points in the context of two simple models
for the evolving He~II Ly$\alpha$ opacity.  
As we indicate in Fig. 1 and will discuss in detail below, a single 
high-precision measurement is inadequate to constrain
models for the He~II  opacity since the intrinsic
fluctuations in the opacity at a given epoch are significant.

Early interpretation of these results in the discovery papers, as well
as by Madau \& Meiksin (1994), Zheng \& Davidsen (1994), Giroux,
Fardal, \&~Shull (1995, hereafter Paper I), and Songaila, Hu, \&~Cowie
(1995), has focused on the requirements for reproducing the observed
He~II absorption.  As yet, it has been difficult to place strong
constraints on the required assumptions.  These include the
distribution of absorption-line clouds in redshift, H~I column, and
H~I line width, and the populations of ionizing sources and their
intrinsic spectra.  Assumptions about the physical properties of the
clouds must also be made, both to derive the ionization correction
necessary to convert $\nhone$ to $\nhetwo$ and to estimate the
distribution of He~II line widths.  In addition, an unknown amount of
He~II absorption may arise in diffuse gas in the IGM, i.e. the true
Gunn-Peterson effect (Gunn \& Peterson 1965, hereafter GP).
These uncertainties also affect the analysis of H~I and He~II
absorption by numerical hydrodynamical simulations (cf.\ Zhang \ea 1997).
 
A further complication, discussed by Zheng \& Davidsen (1995) and in
Paper I, is that ionizing radiation from the observed quasar itself
may alter the level of He~II ionization in nearby clouds.  This
``He~II proximity effect'' is expected to be larger than the corresponding
H~I proximity effect if, as is believed, intervening clouds strongly
depress the level of He~II-ionizing radiation for the general
metagalactic background.  This effect is not obvious in the spectrum
of \HS, but it has been claimed in the spectrum of \Qj\ (Hogan \ea
1997; Heap 1997).

Fortunately, a substantial reduction in the volume of model parameter
space is possible.  A high-resolution \Keck\ spectrum of \Qj\
(Songaila, Hu, \& Cowie 1995) has provided the actual set of H~I
absorption lines to which the observed He~II line-blanketing
absorption must correspond.  It is likely that similar line lists will
be generated for the other quasars used to measure He~II absorption.
In addition, high-quality spectra along the lines of sight to many
quasars are providing increasingly good statistics on the average
distribution in column density, Doppler $b$-values, and redshift of
H~I absorbing clouds.  Additional He~II absorption observations are
planned with HST, and significant advances are expected to come from
the Far Ultraviolet Spectroscopic Explorer (FUSE), scheduled for
launch in late 1998, and the Cosmic Origins Spectrograph (COS) scheduled 
for the HST refurbishment mission in late 2002.  The FUSE spectrograph
may be able to resolve individual lines at wavelengths down to 915
\AA\ ($z\subHe \geq 2.01$), providing line widths and directly giving
the ionization correction in the absorbing clouds.  The COS instrument
offers significantly greater throughput, to study faint QSOs at
$z\subHe > 2.9$.

The purpose of this paper is to ask whether any simple model fits all of
the current observations and to identify important questions that
could be answered in the future.  Our approach is phenomenological; we
do not employ a fundamental theory of the \Lya\ forest, as has been
attempted in Gnedin \& Hui (1996) or Bi \& Davidsen (1997), nor do
we use a numerical hydrodynamical approach.  Instead,
we employ the standard, observationally oriented division between
``clouds'' and a diffuse IGM.  
In \S~2, we outline our calculation of the cosmological radiative transfer 
of direct and diffuse photoionizing radiation through the absorbing
clouds.  
We compile a new absorption-line sample and construct from it accurate
models for the distribution of clouds in redshift and H~I column
density, 
constraining the opacity of the universe to H~I- and He~II-ionizing photons. 
We describe our assumptions about the physical properties of 
these absorbers and our models for the evolution and spectral shape
of ionizing sources.
In \S~3, we describe the results of our radiative transfer calculations.
These give the relation between the emitted spectrum of sources
and the spectrum seen by the absorbing matter, as well
as the implied opacity of the universe
to He~II-ionizing photons.
Using these models, we discuss the ionization conditions
necessary to reproduce the mean levels of H~I and He~II
Ly$\alpha$ absorption seen in the current observations.
In \S~4, we discuss the fluctuations in the
He~II absorption expected from variations in the number density of
lines and the metagalactic radiation field.  
In \S~5, we show how the
effects discussed in this paper could be studied by future ultraviolet
observations of He~II absorption.

\section{THE AVERAGE HE~II ABSORPTION}

\subsection{The Observations at Present}
\label{sec:present-obs}

To date, every quasar observation that has been made at appropriate
wavelengths has detected an absorption trough due to redshifted He~II
\Lya.  The observations of \Qj\ by Hogan \ea (1997) and of \HS\ by
Davidsen \ea (1996) are of particular interest because they have
sufficient resolution and signal-to-noise ratio to show some structure
in the absorption.  The new observations of \HE\ (Reimers \ea 1997) may
also prove valuable, because they appear to exhibit a wide variance in
He~II absorption. However, the average transmission and its
corresponding ``effective'' optical depth shown in
Figure~\ref{fig:obs-opt-depth} remain the most useful statistics
derived from these observations.

Hogan \ea (1997) draw some interesting conclusions by comparing their
spectrum with a \Keck\ spectrum of the same quasar (Songaila \ea
1995).  They claim to see two separate He~III ionization
bubbles around the quasar and suggest that the gas far from the quasar
has a significant Gunn-Peterson opacity due to an IGM.  They place a
lower limit of $\eta \equiv N\subHe / N\subH > 100$ far away from the
quasar.  However, these conclusions are based on a potentially
optimistic error estimate for the flux calibration.  While Hogan \ea
quote a 95\% confidence range of $1.5 < \tau < 3.0$ in a region far
from the quasar, a separate analysis of the GHRS data 
(Heap 1997) yields $\tau = 1.5
\pm 0.2$ from the ACCUM mode of exposure and $\tau \approx 4$
for the FLYLIM mode.  We also note that the sudden drop in the
spectrum at $\lambda = 1283 $~\AA\ could be due to a Lyman limit
system at $z=0.41$, although the probability of this is only 
$\sim \! 3\%$. This hypothesis could be checked by looking for the Mg~II
doublet at 3942 and 3952~\AA.

Because of binning, the published spectrum of \HS\ (Davidsen \ea 1996)
does not show the full resolution of the observations on which it was
based.  Nevertheless, some interesting structural features are
apparent.  There is no sign of the He~II proximity effect in this
quasar, contrary to predictions (\ME\ \& Ostriker 1992; Zheng \&
Davidsen 1995; Paper I).  This allows one to set a limit on the
ionizing flux at $h\nu \geq 4$ Ryd. 
Also, there are two bumps in the spectrum which seem statistically
significant.  One matches the He~II Ly$\beta$ emission wavelength, but
by analogy with H~I Ly$\beta$, this line is expected to be weak.
In \S~\ref{sec:abs-fluct} we will consider whether these bumps might
be caused by variations in the ionization level of the clouds.

\subsection{Physical Considerations}
\label{sec:phys}

The problem of modeling the He~II \Lya\ absorption can be broken into
two parts.  First, one estimates the photoionizing background by
making assumptions about the ionizing sources and estimating the
population of quasar absorption line clouds from H~I measurements.
This is an iterative process, as the amounts of He~I and He~II in the
absorbing clouds both depend on and alter the ionizing spectrum.  This
process gives an estimate of the \hetohyd\ ratio $\eta$.  Second, one
derives the He~II line opacity, using $\eta$ and H~I measurements of
the IGM and absorption-line clouds.  The basic principles and
equations involved in this process have been discussed in many papers
(\ME\ \& Ostriker 1990, \ME\ 1993; Jakobsen \ea 1994; Madau \& Meiksin
1994; Paper I; Davidsen \ea 1996).  For reference, we state the main
equations here.

A continuous IGM containing a species $s$ with proper number density
$n_s(z)$ and resonant scattering cross section $\sigma_s \equiv
(\pi e^2 / m_e c^2) f_s \lambda_s$ gives a line optical depth of
\begin{equation}
   \label{igm-opt-depth}
   \tau_s^{\rm GP}(z) =
   \left( \frac{c}{H_0} \right)
   n_s(z) \sigma_s (1 + z)^{-1} (1 + \Omega_0 z)^{-1/2} \; ,
\end{equation}
(Gunn \& Peterson 1965).  We assume $\Omega_0 = 1$ throughout; as
shown below, many of our results turn out to be independent of
$\Omega_0$. The large cross sections for resonant scattering imply
that only a small density of the scattering particles is needed for
sizable GP absorption.  
This density is
sufficiently low that it makes a negligible contribution to the
continuum (bound-free) opacity in H~I and He~II.  
The continuum opacity of a continuous IGM is related to the Gunn-Peterson
optical depth by
$d\tau_s^{\it cont} / dz = (\sigma_s^{\it cont} / \sigma_s^{\it line} )
\, \tau_s^{GP} \, / \, (1+z)$.
For both H~I and He~II,  
$(\sigma_s^{\it cont} / \sigma_{Ly\alpha}^{\it line} ) = 1.42.$
Limits on the smooth GP trough in the H~I
\Lya\ line [$\tau\subH^{GP}(z=3$--$4) \lesssim 0.1$] (Giallongo \ea
1994; Williger \ea 1994; Fang \& Crotts 1995) imply a very low
smoothly distributed density of H~I in the IGM.  
Even if the corresponding
He~II \Lya\ depth were much larger---for example, if it provided almost
all of the optical depth $\tau\subHe(z \approx 2.4) = 1.0$ seen in \HS---the
smoothly distributed He~II density would contribute only a few percent
of the total He~II bound-free opacity, as we shall see 
in \S~\ref{sec:col-dist}.
Indeed, one of the conclusions of
numerical hydrodynamic models (Zhang \ea 1997) is that the He~II
\Lya\ absorption may be dominated by the contribution from low density
regions (``voids'' in the gas distribution), which are far too low in
density to have significant bound-free opacity.  The bound-free
opacity arises almost entirely from the discrete absorbers (\Lya\ 
forest and Lyman limit systems).
  
For the \Lya\ absorption-line forest (LF), the effective line optical 
depth is
\begin{equation}
   \label{line-opt-depth}
   \tau_s^{LF}(z) =
   \frac{(1 + z)}{\lambda_s}
   \int \fnz \: W_\lambda \: d\nhone \; ,
\end{equation}
with $W_\lambda$ the equivalent width of the line in wavelength units.
(Note: the term ``effective'' optical depth denotes the optical depth
corresponding to the average transmission, $\langle T \rangle 
\equiv
\langle \exp(-\tau) \rangle 
= \exp(-\tau_{\rm eff})$.  
This is not the same as the average optical
depth.  We will always refer to the former in this paper.)  The
effective continuum optical depth is
\begin{equation}
   \label{cont-opt-depth}
   { {d\tau_{\it eff}(\nu)} \over {dz} } = 
   \int \fnz
   {\left[ 1 - \exp(-\nhone \sigma_{\it tot}(\nu)) \right] } 
   \; d\nhone \; , 
\end{equation}
where the continuum cross section is summed over species $s$, 
\begin{equation}
   \label{cont-cross-section}
   \sigma_{\it tot}(\nu) = \sum_s \left( \frac{N_s}{\nhone} \right)
   \sigma_s^{\it cont}(\nu) \; .
\end{equation}
Usually only one species will contribute appreciably at any frequency.

As the term $N_s / N\subH$ implies, the relative column densities must
be estimated based upon physical modeling of the clouds.  The usual
approach is to assume that the clouds are highly photoionized, and
that helium is so highly ionized that He~I may be neglected. Thus, both H
and He within the clouds have only two stages to consider.  In
photoionization equilibrium,
\begin{equation}
  n\subH = \frac {n_e n_{\it HII} \alpha_H^{(A)} }{\Gamma\subH},  \; \; \;
  n\subHe = \frac {n_e n_{\it HeIII} \alpha_{\it He}^{(A)} }{\Gamma\subHe} .
\end{equation}
Here, $\alpha_H^{(A)} = (2.51 \times 10^{-13}$ cm$^{3}$~s$^{-1})
T_{4.3}^{-0.76}$ and $\alpha_{\it He}^{(A)} = (1.36 \times 10^{-12}$
cm$^{3}$~s$^{-1}) T_{4.3}^{-0.70}$ are the case-A recombination rate
coefficients, appropriate for the low column density gas in these
absorbers at temperature $T = (10^{4.3}~K)T_{4.3}$.  If we define
$\alpha_1$ and $\alpha_4$ as the spectral indexes of the radiation
field just above the thresholds for ionizing H~I (1~Ryd) and He~II
(4~Ryd), we can write the specific intensities near these thresholds
as $J_{\nu} = J\subH (\nu/\nu\subH)^{-\alpha_1}$ and $J_{\nu} =
J\subHe (\nu/\nu\subHe)^{-\alpha_4}$.  In the approximation that the
photoionization cross sections scale as $\nu^{-3}$ above threshold,
the photoionization rates of the two species are $\Gamma\subH \approx
4\pi J\subH \sigma\subH / h (3 + \alpha_1)$ and $\Gamma\subHe \approx
4\pi J\subHe \sigma\subHe / h (3 + \alpha_4)$.  The ratio of He~II to
H~I is then
\begin{equation}
  \label{eta-general}
  \eta \equiv \frac{N\subHe}{N\subH} = 
         \left( \frac {n_{\it HeIII}} {n_{\it HII}} \right)
         \left( \frac {\alpha\subHe} {\alpha\subH} \right)
         \left( \frac {\Gamma\subH} {\Gamma\subHe} \right)  .
\end{equation}
If the ionization level is high enough that $n_{\it HeIII} \approx
n_{\it He}$ and $n_{\it HII} \approx n_H$, the first ratio simplifies
to the abundance ratio $n_{\it He} / n_H = (Y/4)/(1 - Y)$, where
estimates of the primordial helium mass abundance range from 
$Y = 0.22$--0.25 (Schramm \& Turner 1997).  
Recent analyses of metal-poor
extragalactic H~II regions find $Y = 0.232 \pm 0.003 {\rm (stat)} \pm
0.005 {\rm (sys)}$ (Olive \& Steigman 1995) and $Y = 0.243 \pm 0.003$
(stat) (Izotov \ea 1997). On the theoretical side, the recent
provisional value for primordial D/H = $(2.7 \pm 0.6) \times 10^{-5}$
(Tytler, Fan, \& Burles 1995) corresponds to $Y = 0.248 \pm 0.002$
(Schramm \& Turner 1997).  For our work, we adopt a value $Y = 0.239$
corresponding to $n_{\it He} / n_H = 0.0785$.
The electron density in fully ionized gas is then $n_e \approx 1.16
n_H$, and the column ratio is
\begin{equation}
   \label{eta}
   \eta \approx 1.70 \left( \frac {J\subH} {J\subHe} \right)  
   \left( \frac {3 + \alpha_4} {3 + \alpha_1} \right) T_{4.3}^{0.06} \; .
\end{equation}
In our radiative transfer calculations, we derive the photoionization
rates by integrating over the ionizing spectrum, and take into account
the ionization balance between H~I, H~II, He~II, and He~III.
The approximations above are usually quite good, except that
$J\subH$ and $J\subHe$ should be qualified to be the portion of the
background resulting from direct emission from ionizing sources.
As shown in \S~3, when the re-emission from absorbing clouds is included  
it would be inappropriate to simply picture
the shape of the metagalactic ionizing background as a 
piecewise power law with the two exponents $\alpha_1$ and $\alpha_4$.

The absence of a smooth Gunn-Peterson trough in the H~I \Lya\ line
suggests both that most of the baryons are clumped into distinct
structures at $z \lesssim 4$ and that the H~II regions around emitting
sources have overlapped by that time.  The observations of \HS\ give a
similar conclusion for helium at redshifts $z < 2.7$.  Combining the
Gunn-Peterson limits of $\tau_{\it HeI}(z=2.7) < 0.05$ for He~I
$\lambda$584 (Reimers \ea 1992) with the condition $\tau\subHe <
1.0$ for He~II, we see that $\lesssim 10^{-3}$ of the baryons can be
in the form of smoothly distributed He~I or He~II.  This implies
either an implausibly efficient clearing-out of the spaces between
clouds or the overlap of He~III spheres by $z = 2.7$.  The same
argument could be extended to redshift $z = 3.1$, if there is indeed
flux at all observed wavelengths below the He~II edge in the spectrum
of \Qj\ (Hogan \ea 1997).  However, the observed patchy He~II
absorption toward the $z=2.9$ quasar HE2347-4342 (Reimers \ea 1997) casts
some doubt on this simple picture of reionization.

We adopt the standard parameterization for the absorption-line
distribution in equations~(\ref{line-opt-depth}) and
(\ref{cont-opt-depth}),
\begin{equation}
\label{col-dist}
\fnz = (A / N_r) (\nhone / N_r)^{-\beta} (1 + z)^{\gamma} ,
\end{equation}
with a reference column of $N_r \equiv 10^{17}$~\cdu.  New
high-resolution spectra of \Lya\ clouds have enabled us to construct a
new, more accurate opacity model for the IGM. A simple, but reasonably
accurate single power-law fit to the data in \S~\ref{sec:col-dist} has
$A = 0.054$, $\beta = 1.51$, and $\gamma = 2.58$.  Better models with
breaks and multiple power laws are discussed in \S~\ref{sec:col-dist}.
The spectral filtering by the IGM produces an effective spectral index
of the background ($\alpha_b$) that is considerably larger (softer
radiation field) than that of the intrinsic sources ($\alpha_s$).  (As
we have stated, the background spectrum is not a power law, but since
the ionization balance of H~I and He~II is controlled by the flux at
the ionization edges, it is useful to define $J\subHe / J\subH \equiv
4^{-\alpha_b}$.)  The value of $\eta$ can be estimated in this simple
model by using the source-function approximation, in which $J_{\nu}
\approx S_{\nu} = j_{\nu} / \kappa_{\nu}$.  Here $j_{\nu}$ is the
proper volume emissivity and $\kappa_{\nu} \equiv (dz/dt)
(d\tau_{\nu}/dz)/c$ is the opacity per unit length.  It is easy to
show that the threshold opacities of He~II (at 4 Ryd) and H~I (at 1
Ryd) are in the ratio $\kappa\subHe / \kappa\subH = (\eta /
4)^{\beta-1}$.  The sources of ionizing emissivity have a spectral
index $\alpha_s$, so that $j\subHe / j\subH = 4^{-\alpha_s}$.
Combining this with equation (\ref{eta}), in which we take $\alpha_1
\approx \alpha_4$ to obtain $\eta = 1.70 \times 4^{\alpha_b}$, we
derive the relation $(2 - \beta) \alpha_b = \alpha_s - [1 - (\ln
1.7)/(\ln 4)] (\beta - 1)$.  For $\beta = 1.51$, we then have
$\alpha_b \approx 2.0 \alpha_s - 0.64$.  We shall see later that
$\alpha_s \approx 1.8$ (Zheng \ea 1997) is a plausible value for the
350--912~\AA\ spectra of quasars, implying a background spectral index
$\alpha_b \approx 3.0$, a continuum opacity ratio $\kappa\subHe /
\kappa\subH \approx 5.5$, and a column ratio $\eta \approx 110$. In
the remainder of this paper, we employ our new opacity models as well
as more accurate numerical models of radiative transfer to derive the
ionizing spectrum and the ratio $\eta$.  However, this simple analytic
model demonstrates both that the background should be softer than the
emitted flux and that its index $\alpha_b$ is quite sensitive to the
source index $\alpha_s$.

As the forms for equations~(\ref{line-opt-depth}),
(\ref{cont-opt-depth}), and (\ref{col-dist}) suggest, the clouds that
dominate either the line or the continuum opacity (if $1 < \beta < 2$,
as is observed) are those that are marginally saturated, with optical
depths $\tau \sim 1$.  However, the column densities at which this
occurs are quite different for line and continuum absorption and for
H~I versus He~II.  For example, the H~I continuum opacity at 1~Ryd is
dominated by clouds with $\nhone \sim 1/\sigma\subH \sim
10^{17}$~\cdu, while the H~I \Lya\ opacity is dominated by clouds with
$N\subH \sim 10^{13}$~\cdu.  The clouds that dominate the He~II
continuum opacity at 4~Ryd have columns $\nhone = \nhetwo / \eta \sim
(\sigma\subHe \eta)^{-1} \sim 10^{15}$~\cdu, while the absorption in
the He~II \Lya\ line is controlled by clouds with columns around
$\nhone \sim 10^{12}$~\cdu.  The most important column ranges for our
problem are thus N(H~I) $\approx 10^{12}$--$10^{13}$~\cdu\ and
$10^{15}$--$10^{17}$~\cdu.  Because the absorption-line distribution
is so crucial to understanding the He~II absorption, we examine it in
more detail in the next section and derive a new opacity model.

\subsection{Absorption-Line Models}
\label{sec:col-dist}

Although many authors have fitted the column density distribution over
a limited range in $\nhone$, the last attempt to find a global fit was
that of Petitjean \ea (1993).  Since then, spectra of higher quality
and greater redshift coverage have appeared, which should allow us to
achieve greater accuracy and to fit the distribution at lower columns.
In addition, by constructing a new sample, we can estimate the errors
in the column density distribution and the resulting opacities.
Because clouds of different column densities are found by different
observational methods, fits to the column density distribution in the
literature have generally contained artificial discontinuities at
certain columns. For example, our Paper~I relied on a combination of
the Press \& Rybicki (1993) distribution for low columns and the
Sargent, Steidel, \& Boksenberg (1989) distribution for Lyman-limit
systems (LLS).  This distribution had a mismatch of a factor of $\sim \!
5$ at $\nhone = 10^{17}$~\cdu\ and $z \sim 3$, which had drastic
effects on the continuum opacities.  The fits we obtain below should
give much more reliable results.

To estimate the distribution of absorbers in $\nhone$ and $z$, we have
taken line lists from the following sources: Lu \ea (1996), Kulkarni
\ea (1996), Hu \ea (1995), Giallongo \ea (1993), Cristiani \ea (1995),
Rauch \ea (1992), and Carswell \ea (1991) for \Lya\ forest lines, and
from Stengler-Larrea \ea (1995) and Storrie-Lombardi \ea (1994) for
Lyman-limit absorbers.  The objects in our sample overlap somewhat
with those of Petitjean \ea (1993).  However, about half the lines are
from new \Keck\ spectra, and thus many lines have much lower columns
(down to $2 \times 10^{12}$~\cdu) than were detectable a few years ago.

\begin{figure}
\plotone{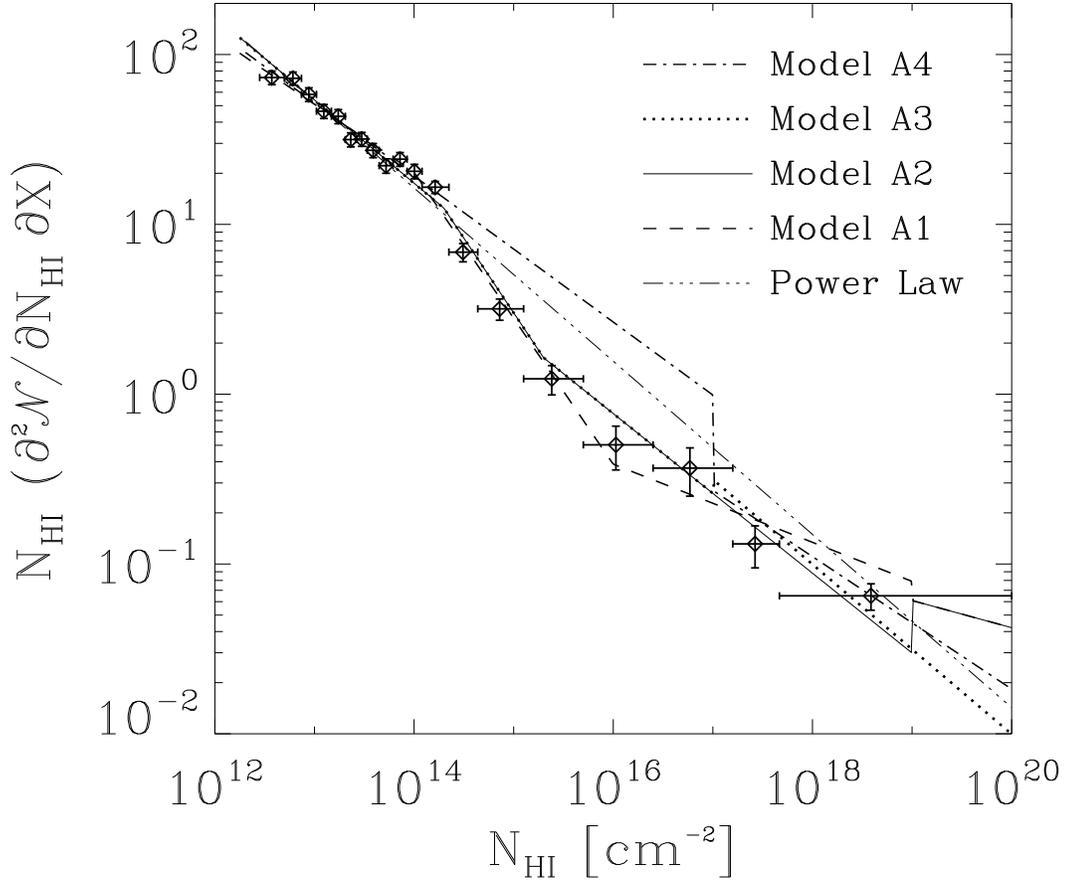}
\caption{
\label{fig:col-dist}
Data and models for the H~I column density distribution.  The sources for
the data are discussed in the text; error bars are based on $N^{1/2}$
statistics.  The variable $X$ refers to the ``absorption path
length'', defined as in Petitjean \ea (1993) by $dX = (1+z) dz$.  This
makes the distribution less sensitive to redshift, while the choice of
$y$-axis makes the structure in the spectrum easier to see and reduces
the effect of binning on the plotted points.  Our four absorption-line
models are plotted for the mean redshift of the sample, $\bar{z} = 3.0$.}
\end{figure}

By constructing several models for the absorption-line distribution,
denoted A1 through A4, we hope to illustrate the range of uncertainty in
$\eta$ due to the cloud distribution.  To arrive at our fits, we use a
maximum-likelihood method, which avoids binning the data in H~I
column.  To treat the Lyman limit systems with continuum $\tau_{\it H
I} > 3$, for which $\nhone$ is difficult to determine, we include in
the likelihood function the binomial probability of seeing the
observed number of these systems in the sample.  Completeness
corrections have been included for the low-column clouds seen in the
\Keck\ spectra.  Explicit formulae for this method are given in
Appendix~\ref{app:statistics}.  The column density distribution in our
sample is shown in Figure~\ref{fig:col-dist}.  We have included the
best single power-law fit, which was already mentioned in
\S~\ref{sec:phys}, but it is clear from the graph that this fit
is poor, with fit probability $P_{\rm fit} = 3 \times 10^{-5}$.
The poor fit is caused by a deficit of clouds with $N\subH \sim
10^{15}$~\cdu, in agreement with Petitjean \ea (1993).

In deriving models A1 and A2, we have assumed that $\nhone$ and $z$
are statistically independent.  As in equation~(\ref{col-dist}), we
assume an evolution in $z$ proportional to $(1+z)^{\gamma}$.  A
three-power-law model is required to get a good fit to the observed
column density distribution, which we require to be continuous.  Our
model A1 is designed to minimize, and A2 to maximize, the He~II line
and continuum opacity relative to that of hydrogen.  These models are
not reliable in the damped ($\nhone > 10^{20}$~\cdu) region, so there
we have switched over to the distribution described in
Storrie-Lombardi, Irwin, \& McMahon (1997), although this does cause a
small discontinuity in the column distribution.  In any case, this
region has little effect on our results.

In the past, it has been controversial whether the \Lya\ forest clouds
and LLS evolve at different rates, and whether this indicates a
difference in their origin.  The redshift exponent $\gamma$ derived in
Stengler-Larrea \ea (1995) is $\gamma = 1.55 \pm 0.39$, more consistent
with our estimate of $\gamma = 2.58 \pm 0.27$ for the total sample
than was the Sargent, Steidel, \& Boksenberg (1989) estimate of
$\gamma = 0.68 \pm 0.54$ we used in Paper~I.  A value of $\gamma=2.58$
for the LLS is acceptable if one considers only the limited redshift
range $2 < z < 4$, which contains most of the \Lya\ forest lines in
our sample.  However, this does not {\em rule out} a difference in the
evolution rates, which would produce faster redshift evolution of
$\eta$.  To see the strength of this effect, we use our model A3,
which is exactly the same as A2 up to $\nhone = 10^{17}$~\cdu\, but
switches over to the redshift evolution of Stengler-Larrea \ea (1995)
above this column density.  For purposes of comparison with Paper~I,
we have included that paper's absorption-line model (based on Press \&
Rybicki 1993), here labeled A4.  All of these models are summarized in
Table~\ref{table:col-dist} and displayed in Figure~\ref{fig:col-dist}.

A conversion of models for the average column density distribution of
intervening absorbers to estimates of the corresponding line optical
depth $\tau_s^{LF} (z)$ must also assume a distribution in line widths
$b$.  Our calculations assume the line width distribution inferred by
Hu \ea (1995), which is a gaussian with $\langle b \rangle  = 28$ \kms\ 
and $\sigma(b) = 10$ \kms\ but truncated below $b = 20$ \kms.   
Strictly speaking, this distribution is not based on the full data
sample we use to derive our column density models.  We have performed
calculations which assume different line width distributions,
including simulations in which we assume all absorbers have $b = 25$
\kms\ and $b=35$ \kms.  The variation in line optical depth
due to these differing line-width distributions is less than $5\%$.
Our calculation of the He~II line optical depth is dependent on the
relation of the measured H~I line widths to the corresponding He~II
line widths, since $b\subHe = (0.5-1) b\subH$ (see \S 3).

\placetable{table:col-dist}
\placefigure{fig:col-dist}

\begin{figure}
\plotone{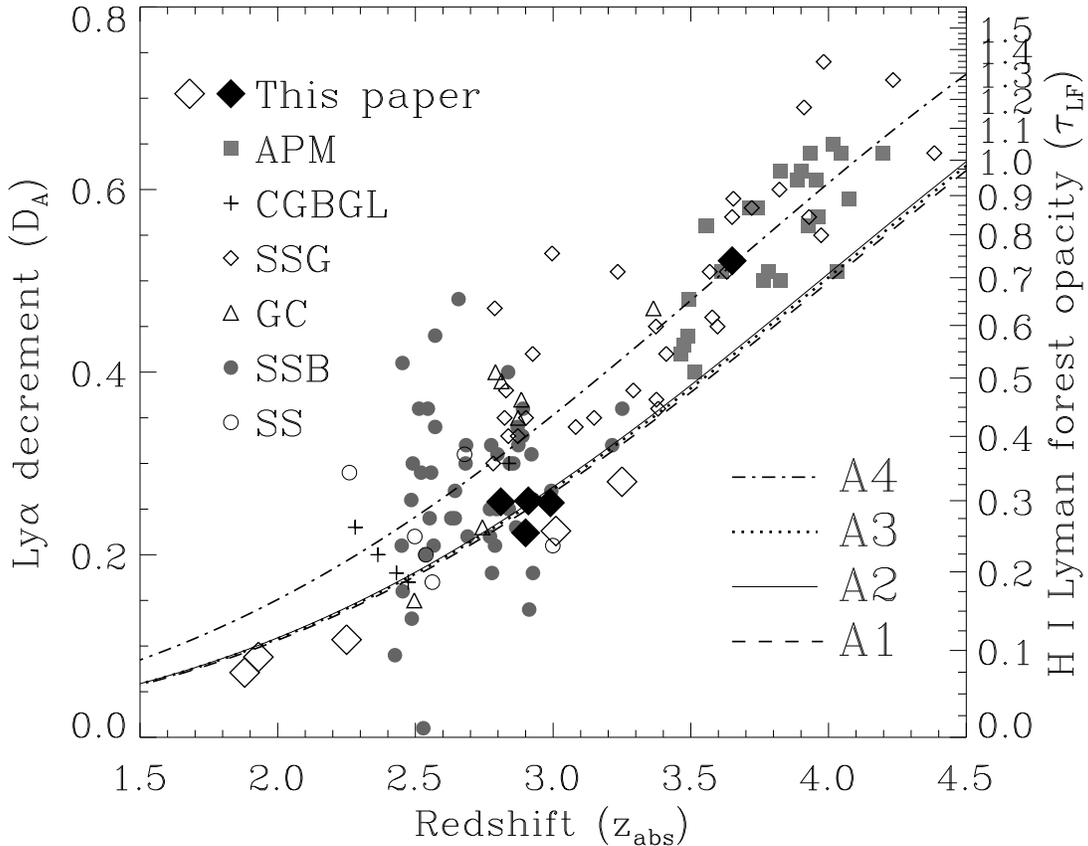}
\caption{
\label{fig:dsuba}
The H~I \Lya\ effective optical depth and \Lya\ decrement, $D_A$, 
as a function of mean absorption redshift. The quasar emission redshift
is somewhat higher, as the observations are taken typically from
1050--1170 \AA\ rest frame.  Data points marked ``this paper'' are
computed from reconstructed spectra, given the sample of lines discussed
in \S~\ref{sec:col-dist}.  Large filled diamonds are based on {\it Keck} 
spectra; large empty diamonds are based on other, lower S/N spectra.
Small data points are from samples of low-resolution observation.
The different samples are labeled in the legend: 
(APM) Automated Plate Measurement QSO Survey of Storrie-Lombardi, Irwin, \&
McMahon (1997); (CGBGL) Cristiani \ea (1993); (SSG) Schneider, Schmidt, 
\& Gunn (1991a,b); (GC) Giallongo \& Cristiani (1990);
(SSB) Sample of Sargent, Steidel, \& Boksenberg (1989) as analyzed by
Storrie-Lombardi \ea (1997); and (SS) Steidel \& Sargent (1987).
The results of our opacity models are also shown.  Model A4, which is
based on the Press \& Rybicki (1993) distribution and is a reasonable fit 
to the low-resolution measurements, lies far above the others.  Models
A1, A2, and A3  are based on the absorption-line sample in 
\S~\ref{sec:col-dist} and are a good fit to the data points measured by 
the high-resolution technique.  }
\end{figure}

Our models A1--A3 agree well with the observed number of Lyman limit
systems.  They are in poor agreement with model A4 and with the
observed H~I \Lya\ decrements, $D_A$, found in various low-resolution
surveys.  Here, $1- D_A$ is the ratio of the fluxes below 
(1050 -- 1170 \AA\ in the QSO rest frame) and above ($\lambda \geq 1216$ \AA) 
the H~I edge.  Figure~\ref{fig:dsuba} shows this quantity, which is
related to the effective optical depth by $D_A = [1 - \exp(-\tau_{\it eff})]$.  
We have included the $D_A$ calculated from our quasar
line lists, our models, and measurements from several low-resolution
samples.  Our models and line lists tend to show less absorption than
is measured in the low-resolution observations, 
although there is a large scatter
in the low-resolution measurements.

This discrepancy may indicate a shortcoming in the line-list sample,
such as an insufficient completeness correction at low H~I columns.
There are other interpretations as well.  The low-resolution
observations are based on extrapolating a power-law continuum from
longward of the \Lya\ emission line, while the high-resolution line
lists involve local continuum fits on the shortward side.  If the
spectra have significant curvature, as indicated by the composite
spectrum of Zheng \ea (1997), the low-resolution spectra will
overestimate the continuum and $D_A$ as well.  The discrepancy could
be also caused by a smooth Gunn-Peterson effect.  It will be important
to resolve this question.  The apparent presence of breaks in the H~I
column-density distribution (Fig.~\ref{fig:col-dist}) and the
resulting deficit of observed lines around $10^{15}$~\cdu\ have a
large impact on the He~II continuum opacity.  In Paper I, we
extrapolated the Press \& Rybicki (1993) distribution up to
$10^{17}$~\cdu, giving an unphysical discontinuity in the column
density distribution there.  Since the Press \& Rybicki distribution
is based on Ly$\alpha$-forest absorption, which is insensitive to the clouds
with $\nhone \sim 10^{17}$~\cdu, this extrapolation is doubtful, and the
H~I continuum opacities are almost certainly described better by our
models A1--A3.

\begin{figure}
\plotone{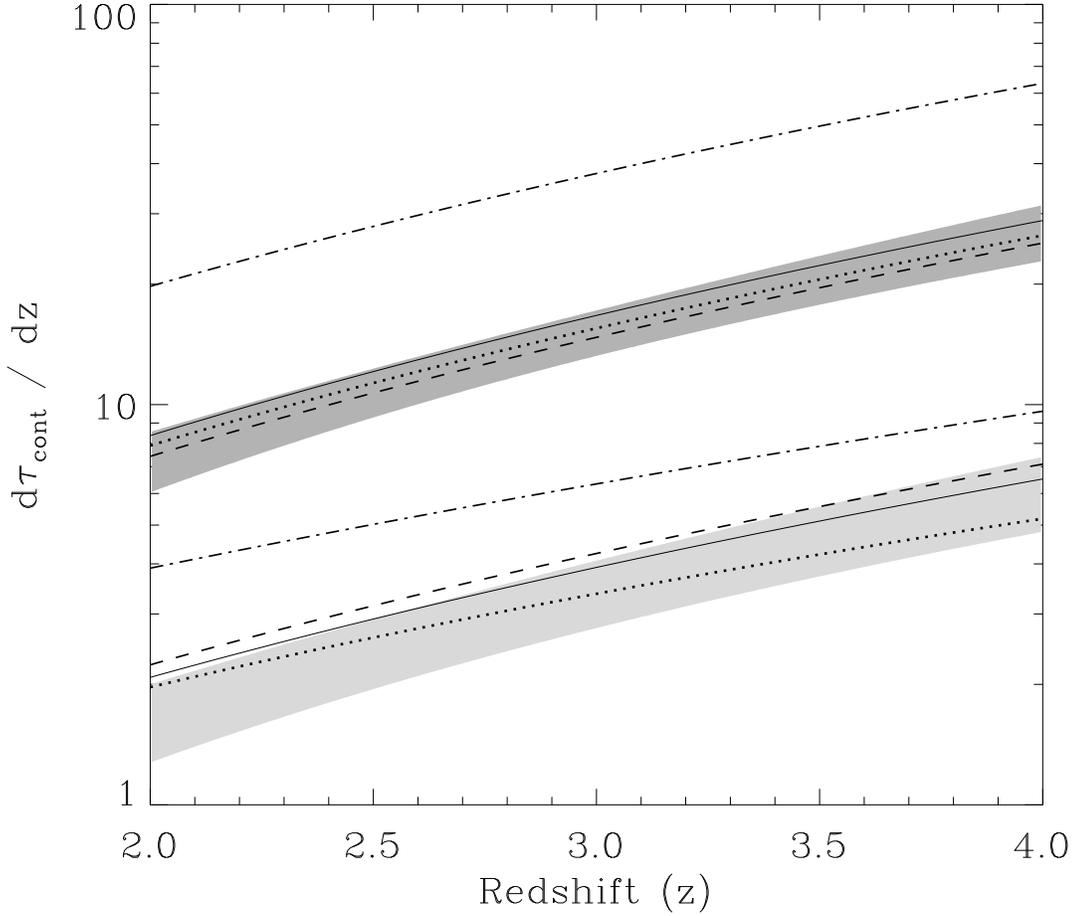}
\caption{
\label{fig:opac}
The H~I {\em (lower)} and He~II {\em (upper)} threshold continuum
opacities as a function of redshift, as derived from the
absorption-line sample directly {\em(shaded region)} and from the
model fits.  Line codes for opacity models (A1--A4) are same as in
Figure 3.  To make comparing the models easier, the He~II opacities
assume $\eta = 70$, instead of deriving $\eta$ from the simulations.
The shaded regions represent $1\sigma$ errors derived from the
bootstrap technique discussed in the text.  While models A1--A3 are
reasonable fits, model A4 (from Paper I) gives much more opacity.  }
\end{figure}

Although the spread among our models does serve to indicate the
observational errors in our opacity estimates, we can get a better
estimate of the errors involved by performing a ``bootstrap
resampling'', as discussed in Appendix~B.  This technique does not
take into account errors in the observed columns, or systematic
effects such as errors in the completeness functions or the results of
clustering.  As a result, this estimate is probably a minimal estimate
of the errors involved.  The H~I and He~II continuum opacities derived
from the sample are shown in Fig.~\ref{fig:opac}, together with the
opacities derived from our models.  The derived uncertainties are only
about 15\%, much smaller than the dispersion among estimates in the
literature.  Partly because our models A1 and A2 assume continuity of
the H~I column distribution, whereas our bootstrap error estimate does
not, models A1 and A2 are not quite $2\sigma$ apart from each other,
despite our best efforts to maximize the difference.  This
difference should be sufficient to indicate the observational errors.

\subsection{Diffuse IGM vs.\ Clouds}
\label{sec:diffuse-igm}

A framing question for the interpretation of the He~II observations has 
been whether the absorption occurs in low-column Ly$\alpha$ clouds or 
in a smooth IGM.  An implicit question is whether the distinction has 
any real meaning.  A picture in which physically distinct clouds
are embedded in a more uniform medium has been implied by models in which 
the clouds are pressure-confined by surrounding hotter gas
(Ostriker \& Ikeuchi 1983) or represent self-gravitating bodies within a more
tenuous medium (Rees 1986).  Amplifying the latter hypothesis,
Shapiro, Giroux, \& Babul (1994) defined the IGM as gas whose
thermal velocity exceeds the circular velocity of all of its
encompassing structures, and is thus smeared out.  Reisenegger \& 
Miralda-Escud\'e (1995) used a modified Zel'dovich approximation
to follow the evolution of structure, and suggested a division between
clouds and IGM based on the local deformation tensor.

A more recent view, inspired by recent N-body/hydrodynamic simulations 
(e.g., Cen \ea 1994; Zhang \ea 1995, 1997; \ME\ \ea 1996; 
Hernquist \ea 1996), is that there is no physical motivation for
a distinction.  The distribution in velocity space
and the physical state of the baryons determine the shape of the quasar
absorption spectra, which may include features that
have historically been associated with the ``uniform IGM'' and ``clouds.''  
Rather than counting and tabulating the properties of discrete lines and 
measuring any remaining absorption trough, 
the new challenge is to directly compare the absorption profile
of quasar spectra with that predicted by large-scale models for the 
evolution of density fluctuations combined with estimates
of the metagalactic ionizing background.  This has inspired promising 
new statistical methods (e.g., Croft \ea 1997; 
Miralda-Escud\'e \ea 1996).

A distinction between Ly$\alpha$ forest clouds and a ``smooth'' IGM
remains useful even within this latest view.  In this paper, we use
the terms ``cloud'' or ``absorber'' to denote the structures that
result in Ly$\alpha$ absorption lines (the ``Ly$\alpha$ forest'')
with H~I column densities ranging from about $10^{17}$~\cdu\ down to
$10^{12}$~\cdu\ and perhaps even lower.  High-resolution spectra taken
with the \Keck\ Telescope (Hu \ea 1995) show that discrete absorption
features exist whose equivalent widths 
correspond to column densities down to at least
$\nhone = 10^{12.3}$ \cdu.  This may be taken to represent a cloud
distribution whose number, line widths, and column densities are well
characterized, and whose properties may be extrapolated to somewhat lower
column densities.  

As discussed above, models for the averaged mean intensity and shape of 
the metagalactic background depend in part on the averaged distribution 
in column density and redshift of the high column density end of the 
Ly$\alpha$ forest.  It is likely that N-body/hydrodynamical simulations 
will continue to depend on these models, and hence on an empirical column 
density distribution, to translate their baryon distributions into 
absorption spectra.  Recent \Keck\ spectra at high S/N and high resolution 
(Fan \& Tytler 1995; Cowie et al. 1995; Songaila \& Cowie 1996) have shown 
that 50\%--75\% of the Ly$\alpha$ forest clouds with $\log N_{\rm HI}>14.5$
have undergone some chemical enrichment, as evidenced by weak but
measurable C~IV and Si~IV absorption lines.  The typical inferred 
metallicities range from 0.003 to 0.01 of solar values (see Giroux \& Shull 
1997). Lyman limit systems have somewhat higher metal abundances 
(Reimers \& Vogel 1993), with possible contributions from local sources 
of hot gas (Giroux, Sutherland, \& Shull 1994; Gruenwald \& Viegas 1993;
Petitjean, Rauch, \& Carswell 1994).  The presence of these metals
provides strong evidence for ``feedback'' from star formation.
Evidently, many of the Ly$\alpha$ absorbers have been affected by physical 
processes besides gravitational clustering, perhaps outflows from star-forming 
regions or mergers of star-forming galaxies.  These processes have not
been reliably incorporated into the current generation of large-scale 
simulations.

Although a fundamental theory of the IGM and \Lya\ forest is beyond
the scope of the paper, we do require a physical model for the
absorbers in order to infer, for example, the likely levels of He~II
present in an H~I absorber.  As a result, we summarize below the
physical properties of our absorbers.  We also show that they are
compatible with the properties ascribed to the corresponding
absorption lines by the results of large scale structure simulations.

One can show that absorbers down to $N\subH
\sim 10^{12}$~\cdu\ still have a density contrast with the mean
diffuse IGM.  By redshift $z \approx 3$--5, many of the baryons have
already collapsed into galaxies and clouds.  
We take the baryon density to be $\Omega_b h_{75}^2 = 0.038 \pm 0.007$, 
corresponding to primordial D/H $= (2.7 \pm 0.6) \times 10^{-5}$ 
(Schramm \& Turner 1997). 
This density is an upper limit to baryons in the diffuse IGM. 
Suppose that a fraction $f_{\rm IGM}$
of the baryons resides in the diffuse IGM. The mean hydrogen density
of the diffuse IGM at $z \approx 3$ is then
\begin{equation}
   \langle n_H \rangle = (1.16 \times 10^{-5}~{\rm cm}^{-3})
            \left( \frac {1+z} {4} \right)^3  f_{\rm IGM} \; .
\end{equation}
To produce an H~I absorption cloud with column density $N\subH =
(10^{14}~{\rm cm}^{-2}) N_{14}$, we first approximate the absorber as
a slab of thickness $L = (100~{\rm kpc}) L_{100}$ with
constant total hydrogen density $n_H$, H~I density $n\subH$, and
ionized hydrogen density $n_{\it HII}$.  We assume that the cloud
temperature is set by photoelectric heating at $T = (10^{4.3}~K)T_{4.3}$.

If then we set the baryon density $n\subH = N\subH / L$, the mean hydrogen
density in the cloud is (from \S~\ref{sec:phys})
\begin{equation}
   n_H = (5.9 \times 10^{-5}~{\rm cm}^{-3}) J_{-21}^{1/2} T_{4.3}^{0.38}
    N_{14}^{1/2}  L_{100}^{-1/2}     \; ,
\end{equation}
assuming an ionizing background of
$J\subH = 10^{-21} J_{-21}$~ergs~cm$^{-2}$ s$^{-1}$ Hz$^{-1}$ sr$^{-1}$ and
a spectral index near 1 Ryd of $\alpha_1 \approx 1.0$,
after filtering by the IGM.  The cloud contrast factor is
\begin{equation}
   \delta_c \equiv  \frac {n_H} {\langle n_H \rangle } = (5.1) J_{-21}^{1/2}
   T_{4.3}^{0.38}
    N_{14}^{1/2} L_{100}^{-1/2} \left( \frac {1+z}{4} \right)^{-3}
   f_{\rm IGM}^{-1}  \;,  
\end{equation}
and approaches 1 at a column density
\begin{equation}
   N\subH \approx (3.8 \times 10^{12}~{\rm cm}^{-2}) f_{\rm IGM}^2
   J_{-21}^{-1} T_{4.3}^{-0.76} L_{100} \left( \frac{1+z}{4} \right)^6 \; .
\end{equation}
Therefore, for a diffuse fraction $f_{\rm IGM} \lesssim 0.5$ and
standard values of the other scaling parameters, we see that clouds
with $N\subH$ at or below $10^{12}$ cm$^{-2}$ retain some contrast with
diffuse portions of the IGM.  This is consistent with the results of
numerical models (e.g., Zhang \ea 1997) which associate underdensities with
these column densities but which also assume a low background of
$J_{-21} \sim 0.1$--$0.4$.

In addition, many of the Ly$\alpha$ clouds probably have some
flattening in order to avoid problems of having volume filling factors
greater than 1 or exceeding the baryonic nucleosynthesis limit (Rauch
\& Haehnelt 1995; Madau \& Shull 1996).  We can address this
question with the opacity models of \S~\ref{sec:col-dist}.  Let us
suppose that almost all of the baryons are in clouds with
$10^{12}$--$10^{17}$~\cdu.  At $z=3$, the clouds have 
$L = $~(21, 17, 17, 6)~$J_{-21}^{-1}$~kpc for opacity models A1--A4, 
respectively.  These thicknesses are much smaller than the observed 
$\sim \! 200 h_{75}^{-1}$~kpc transverse sizes of the \Lya\ clouds 
(Bechtold \ea 1994)  consistent with the 
general flattening of such clouds in the numerical simulations.
However, they are comparable to the estimated sizes of Lyman limit systems,
consistent with evidence that suggests these systems are roughly spherical
(Steidel \ea 1997).   The volume filling factors are less than 1 in this 
model for columns down to $N\subH = $~(2.1, 3.0, 3.0, 1.6)~$ 
\times 10^{10}$~\cdu.  If fewer baryons are in the \Lya\ forest, 
the thicknesses become even smaller.

The total hydrogen densities obtained in the above model are
$n_H = $~(4.1, 4.6, 4.6, 7.7~$\times 10^{-3}~{\rm cm}^{-3})
J_{-21} (N\subH /10^{17}~{\rm cm}^{-2})^{1/2}$.  
Based on QSO luminosity functions Q1 and Q2 with $J_{-21} =$ 0.3--1.0, 
we take the density in the clouds to be
\begin{equation}
\label{hdens}
n_H = (2 \times 10^{-3}~{\rm cm}^{-3})(N\subH /10^{17}~{\rm cm}^{-2})^{1/2}
\end{equation}
throughout our calculations.  This assumption admittedly has large
uncertainties, but the effect of density in our radiative transfer 
calculations will turn out to be small.  This $n_H$--$N_{HI}$ relation 
is also favored by numerical simulations for 
$N_{HI} < 10^{14.5}~{\rm cm}^{-2}$ (e.g., Zhang \ea 1997).  
\subsection{Ionizing Emissivity}
\label{sec:ioniz-emiss}

By now, it is clear that quasars produce an appreciable fraction of
the H-ionizing radiation at high redshifts, and probably almost all of
the He~II-ionizing radiation.  The observed quasars are probably
sufficient to provide the flux deduced from the proximity effect at $z
\approx 2.5$, within the uncertainties (Haardt \& Madau 1996), as well
as ionize the IGM (\ME\ \& Ostriker 1992).  The actual spectral index
of quasars in the EUV has been determined only recently.  Based on
HST-FOC prism spectra of quasars with $z_{\it em} \sim 3$, Tytler
\ea (1996) suggested an average spectral index of $\alpha_Q = 1.5 \pm
0.2$ for 330--1300~\AA\ in the quasar rest frame, a softer index
than seen at higher wavelengths.  Using FOC spectra of 41 radio-quiet
quasars, most with $z_{\it em} < 1.5$, Zheng \ea (1997) find that the
slope steepens below 1000 \AA, producing a slightly softer index of
$\alpha_Q = 1.77 \pm 0.15$ for 350--1050~\AA, with no clear
redshift dependence.  The results are completely consistent, given the
different wavelength coverage.  Zheng \ea find that they can explain
the observed spectrum if photons are Compton-scattered above an H~I
Lyman continuum edge; this model implies that the frequency power-law
continues up to the X-rays, where it does indeed match the observed
spectrum (Laor \ea 1997).  

To compute the ionizing photon emissivity, we use the empirical quasar
luminosity function from Pei (1995).  This fit uses a Gaussian
dependence on redshift, implying a steep cutoff of the quasar
emissivity at high $z$.  The resulting comoving emissivity for $(\nu >
\nu_H)$ can be expressed as
\begin{equation}
  \label{emissivity}
j_{\nu} = j_{H\ast} \left(\frac{\nu}{\nu_H} \right)^{-\alpha_s}
  \exp\left[ - \frac{(z-z_{\ast})^2}{2 \sigma_{\ast}^2} \right]
  (1+z)^{\alpha_{BH}-1} .
\end{equation}
We assume a spectral slope of $\alpha_{BH}=0.83$ from the B-band 
(4400~\AA) to the H~I Lyman edge (912~\AA) to be consistent with the mean
quasar spectra discussed above.  Pei used values of $\alpha_{BH}=0.5$
and 1.0, but the value has little effect on the value of
$j_{\nu}$ because B-band observations of high-$z$ quasars fall fairly
close to the Lyman edge.  We refer to the emissivity model resulting
directly from Pei's luminosity function as model~Q1.  The parameters
for this model are given in Table~\ref{table:emissivity}.

\placetable{table:emiss-coeff}

We also consider a variant, model Q2, that has more quasars at high
redshifts.  There are several reasons to consider such a model.  It is
possible that dust in intervening galaxies reduces the observed number
of high-$z$ quasars.  Model~Q2 lies between the ``medium'' and
``high'' absorption cases of Fall \& Pei (1993).  In addition, there
may be systematic biases against the detection of high-$z$ quasars.
Finally, we shall see that the flux derived from model Q1 is
insufficient to match that estimated from the proximity effect at high
$z$, while model Q2 will be much more consistent.  The Gaussian form
assumed by Pei is well constrained at $z < 3$, but the falloff at
$z>3$ is less certain and even a constant comoving density of quasars
is not ruled out. On the other hand, a recent search (Shaver \ea 1996)
for radio-loud quasars found none with $z>5$, even though there should
be no bias from dust in this case.  This radio survey is consistent
with model Q1 but not with model Q2, unless radio-loud quasars behave
differently than the more numerous radio-quiet quasars.
\placetable{table:emissivity}

An important question is at what epoch the universe became ionized in
H~I and He~II (Shapiro \& Giroux 1987; Donahue \& Shull 1987).  The
average recombination rate in the universe is probably small enough
that one can estimate this epoch simply by integrating the number of
photons until it matches the number of atoms, assuming standard
nucleosynthesis and neglecting the fraction of atoms that are not in
the IGM or \Lya\ forest.  We have done this for our quasar emission
models.  The universe reionizes in H~I at $z=4.7$ in model Q2, but
only at $z=3.7$ in model Q1.  Thus, model Q1 may need to be
supplemented by Lyman continuum emission from massive stars to
produce the smooth evolution of H~I opacity seen to redshifts as high
as $z=4.7$ (Schneider, Schmidt \& Gunn 1991b).  For He~II, the results
depend slightly on the spectrum, but the ionized redshifts are not
necessarily lower than for H~I.  In fact, even for model Q1 the
spectrum must be as steep as $\alpha_s = 2.3$ for He~II reionization
to be delayed to a redshift of $z=3.2$, the highest redshift probed by
current He~II observations.  The critical slope for He~II ionization
fronts to lie inside of H~I fronts is $\alpha_{\it cr}=1.84$
for $Y = 0.239$ (Giroux
\& Shull 1997), close to our standard value of $\alpha_s = 1.8$.  This
slope gives reionization epochs both H~I and He~II of $z=3.7$ and
$z=4.7$ for models Q1 and Q2 respectively.  Since the Pei luminosity
function extends only to $z=4.5$, our calculation thus involves a
mild extrapolation. However, these luminosity functions are declining
rapidly at high $z$, and may underestimate the population
of quasars at $z > 4$.  It thus seems likely that most of the universe
has become ionized in He~II by the observed redshifts.

As noted above, hot stars are another possible source of hydrogen-ionizing
photons.  In our radiative transfer models, we use a fit to the
composite spectrum of an OB association undergoing a starburst (Madau
\& Shull 1996; Sutherland \& Shull 1998).  This spectrum is fairly
hard ($j_{st} \propto \nu^{-1.5}$) just above the H~I edge, but it
falls off rapidly past 45 eV with negligible radiation beyond 4 Ryd.
Thus, the effect of starlight is like that of the decaying neutrinos
in Sciama's (1994) model, in that it can ionize H~I effectively but
not He~II.  Suppose the stellar and AGN emissivities at the H~I edge
are in the ratio $\phi_{st} \equiv j_{st} / j_q$.  Then the spectral
index from 1 to 4 Rydbergs is given by $\alpha_s = \alpha_q + \ln(1 +
\phi_{\it st}) / \ln(4)$.  We would like to calculate an upper limit
to $\phi_{st}$ based on observations of $\tau\subHe$.  However, as
we shall see below, it is at present difficult to obtain an upper
limit on $\alpha_s$ from $\tau\subHe$, so this constraint is actually
much weaker than the constraint relating H~I Lyman continuum radiation
to metal production in stars (Madau \& Shull 1996). Future observations
should change this situation (see \S~\ref{sec:future-obs}).

One could also consider the possibility that the \Lya\ clouds are
photoionized by ``local'' sources---a class of sources (e.g., local
starbursts) that dominate the ionizing flux at the observed clouds
but contribute little to the background $J_{\nu}^{\it bg}$ at an
average point in space.  We can actually rule out this scenario
by the following argument. If clouds have a typical size $R_{cl}$,
their space density $n_{cl}$ is related to their redshift frequency by
$(d\NN / dz) = \pi R_{cl}^2 n_{cl} (dl / dz)$, where $dl$ is the
proper length increment.  Suppose that each cloud contains its own
ionizing source, which dominates the ionizing flux out to the edge of
the cloud.  This requires its luminosity $L_{\nu}^{\it loc}$ to
satisfy $L_{\nu}^{\it loc} / 4 \pi R_{cl}^2 \gtrsim 4 \pi J_{\nu}^{\it
bg}$.  These sources collectively produce an ionizing emissivity of
$j_{\nu}^{\it loc} = n_{cl} L_{\nu}^{\it loc} / 4 \pi \gtrsim 4
J_{\nu}^{\it bg} (dz/dl) (d\NN / dz)$.  In the source-function approximation
discussed above, the additional background due to these ``local''
sources is $J_{\nu}^{\it loc} \approx j_{\nu}^{\it loc} / \kappa_{\nu}
\gtrsim 4 J_{\nu}^{\it bg} (d\NN / dz) / (d\tau_{\nu} / dz)$.
Considering the case of the H~I \Lya\ forest clouds, which are easily
visible for $\nhone > 10^{13.5}$~\cdu, we find $(d\NN / dz) \sim 220$
and $d\tau\subH / dz \sim 3$ at $z \sim 3$.  One can see that
$J_{\nu}^{\it loc} \gg J_{\nu}^{\it bg}$, violating our original
assumption that $J_{\nu}^{\it loc}$ is a minor part of the background.
This {\it reductio ad absurdum} shows that the abundant
\Lya\ clouds really are a probe of the metagalactic background.

The above reasoning therefore provides strong justification for
assuming a general metagalactic background of ionizing radiation
that controls the ionization state of \Lya\ forest clouds.   
The argument is not quite as strong for Lyman-limit systems, which at
$z \approx 3$ have $(d\NN / dz) \approx 2$.  Neither does it rule
out the presence of a population of absorbers with N(H~I) $\geq
10^{15}$ cm$^{-2}$ that are dominated by their ``local'' radiation
field.  Giroux \& Shull (1997) have argued that the high Si~IV/C~IV 
and low C~II/C~IV ratios found in a sub-sample of \Lya\
clouds with N(H~I) $\geq 10^{15}$ cm$^{-2}$ (Songaila \& Cowie 1996)  
may be compatible with photoionization by a locally dominated
radiation field.

\subsection{Radiative Transfer Calculations}

In Paper I, we used the source function approximation to relate the
emitted ionizing spectrum to the \hetohyd\ ratio.  As a rule of thumb,
the source function approximation is good only when emitted photons
are absorbed within a Hubble length.  The universe becomes less opaque
as cosmic time increases, however, so this approximation is not as
good for the case of \HS\ ($\zem = 2.74$) as for \Qj\ ($z_{\it em} =
3.29$).  In addition, Haardt \& Madau (1996, hereafter HM96; see also
Fardal \& Shull 1993) have shown that several new processes alter
$\eta$ slightly, including emission from \Lya\ clouds in He~II and H~I
lines and continua as well as self-shielding effects in high-column
clouds.  We have thus carried out radiative transfer calculations to
derive $\eta(z)$.

The spatially-averaged specific intensity at a given redshift and
frequency (in \fluxu) is given by
\begin{equation} 
  \label{avg-flux}
  J_{\nu}(z_{\it obs}) =
  \int_{z_{\it obs}}^{\infty} \left( \frac{1 + z_{\it obs}}{1 + z} \right)^3
  \, j_{\nu}(\nu_{em}, z) \, \frac{dl}{dz} \, \exp(-\tau_{\it eff}) \, dz 
\end{equation} 
(Bechtold 1993), where $j_{\nu}$ is the proper volume emissivity (erg
s$^{-1}$ cm$^{-3}$ Hz$^{-1}$ sr$^{-1}$) equal to $\epsilon(\nu, z)/4\pi$
in the notation of HM96, 
$dl/dz = (c / H_0) (1+z)^{-2} (1+\Omega_0 z)^{-1/2}$ 
is the proper length
increment, and $\tau_{\it eff}$ is the effective continuum optical
depth from $z$ to $z_{\it obs}$, as given by
equation~(\ref{cont-opt-depth}).  We note here that, although the flux
depends on many parameters, it is at least independent of the
cosmological parameters $H_0$, $\Omega_0$, and the cosmological
constant, $\Lambda$, as long as
the emissivity is actually determined observationally, as in a quasar
survey, and not assumed {\it a priori}.  First of all, the optical
depth $\tau$ is independent of these parameters.  The emissivity
$j_{\nu}$ depends on the quasar luminosities and densities as $j_{\nu}
\propto n_Q L_{Q,\nu}$, where an integration over the luminosity
function is implied.  In turn, $L_{Q,\nu} \propto {r_A}^2$ where $r_A$
is the angular size distance.  The quasar density is $n_Q \equiv d N_Q
/ dV$ with $N_Q$ the number of quasars seen in a survey, and $V$ the
proper volume sampled, or $n_Q = (dN / dz) / (dl / dz) / r_A^2$.  The
ionizing flux is proportional to $j_{\nu} (dl / dz) \propto n_Q
L_{Q,\nu} (dl / dz) \propto dN / dz$, an observed quantity, and thus
is independent of $H_0$, $\Omega_0$, and $\Lambda$.  

If we use the two luminosity functions from Pei (1995), 
with $\Omega_0 = 1$ and $\Omega_0 = 0.2$, and 
correct for the quasar spectral slope that differs as
well, we find that the low-$\Omega_0$ model gives a higher
photoionization rate by about 40\%.  This difference cannot be rooted
in the original observations.  Instead, the difference arises from the 
method by which the QSO luminosity function is propagated
from low-$z$ to high-$z$ assuming pure luminosity evolution. 
This parameterization depends only on $z$ and
not on appropriate cosmological distances and luminosity 
evolution parameters.  Thus, this introduces a dependence on $H_0$ 
and $\Omega_0$.  

In our computer code, we compute the ionizing flux by setting up a
grid in $(z, \nu$) space and propagating the flux according to
equation~(\ref{avg-flux}) along photon trajectories, $\nu \propto
(1+z)$.  The column ratio $\eta$ is adjusted to match the
photoionization rates at each epoch.  The radiative transfer models
incorporate emission from the H~I Lyman continuum and the He~II Lyman,
Balmer, and 2-photon continua, as well as He~II \Lya\ emission from
the clouds.  In order to find the opacity of He~II, we compute $N\subHe$ 
in an approximate manner as a function of $N\subH$, the
photoionization rates $\Gamma\subH$ and $\Gamma\subHe$, and the cloud
density (see \S~\ref{sec:diffuse-igm}).  Details of this treatment are 
explained in Appendix~\ref{app:re-emiss}.  We neglect the opacity of He~I.
\ME\ \& Ostriker (1992) and Haardt \& Madau (1996) have considered the 
effects of He I theoretically and conclude that if the sources
of ionization are quasars or stars, the net effect is small, but not 
completely negligible.  For our new models for the distribution and 
physical properties of absorbers, our conclusions are similar, and we 
discuss the small effects of He~I opacity in Appendix~\ref{app:small-effects}.
In the few absorbers at $z > 2$ for which measurements of He~I absorption 
exist (Reimers \ea 1992; Reimers \& Vogel 1993), our
assumed density and photoionization models are consistent
with observations (see Appendix~\ref{app:small-effects}).

\section{RESULTS}
\label{sec:results}

\begin{figure}
\plotone{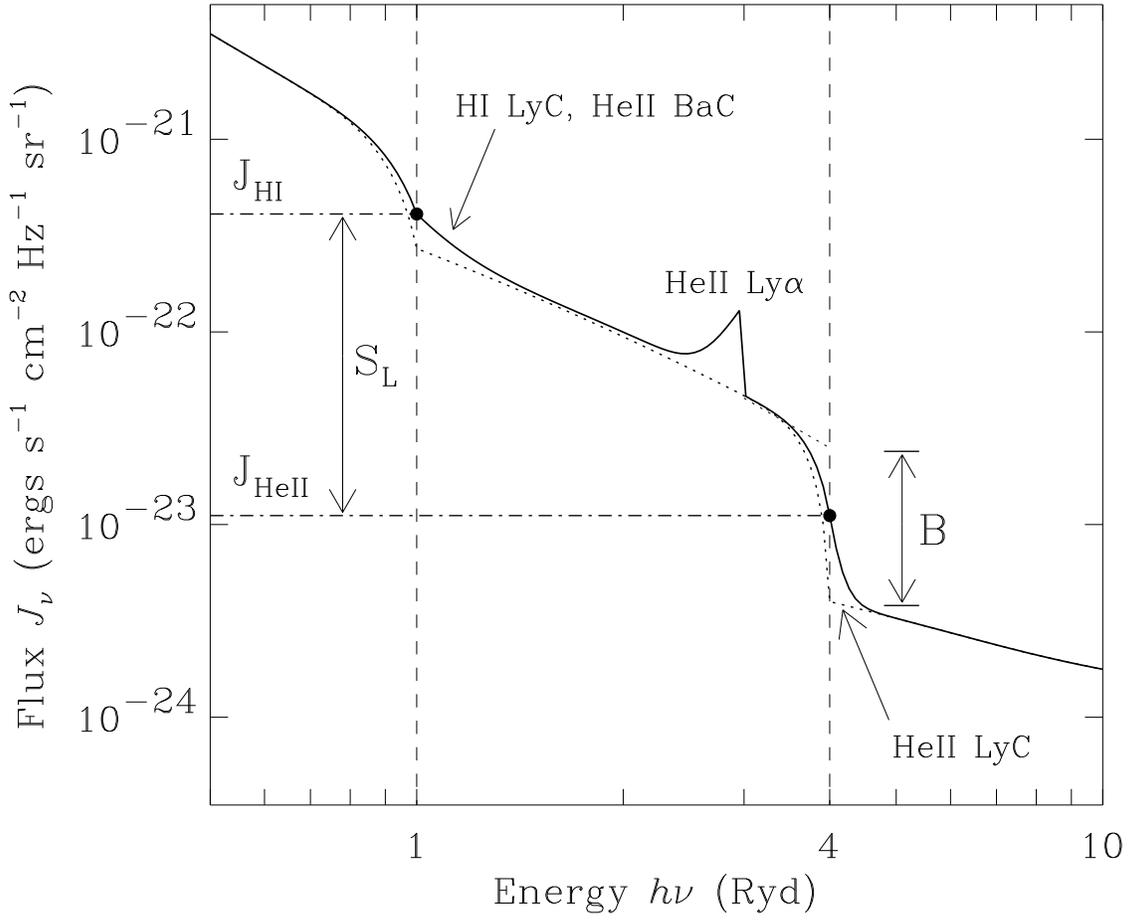}
\caption{
\label{fig:schematic}
Schematic picture of the ionizing background.  The dotted line shows
the spectrum with the cloud re-emission omitted. The parameter $S_L$
is the ratio of intensities at 1 Ryd and 4 Ryd (see eq. [7]), while 
$B$ is the 4 Ryd break.   }
\end{figure}

A typical spectrum resulting from our radiative transfer code is
shown in Figure~\ref{fig:schematic}.  Parameters commonly used
to describe the spectrum are: $J\subH$ and $J\subHe$, the fluxes
at the ionization thresholds of H~I (1 Ryd) and He~II (4 Ryd); 
$S_L \equiv J\subH / J\subHe$; and $B$, the strength of the break at 4 Ryd.
The effect of the line and continuum emission can clearly be
seen in the He~II 304 \AA\ line and the recombination continua, although 
the contribution from He~II 2-photon emission is spread
over a large frequency range and is thus difficult to discern.
The ionization edges are not sharp, but blurred by redshifting
below the threshold and by recombination radiation above it.
The spectrum is poorly approximated by a power law above the 
ionization thresholds, unless the cloud re-emission is omitted.
We {\em define} the effective threshold
power-law indices $\alpha_1$ and $\alpha_4$ by the relations
$\Gamma\subH = (2.53 \times 10^9 \, {\rm sr} \, {\rm cm}^2 \, 
{\rm erg}^{-1} \, {\rm s}^{-1} ) \, J\subH [5/(3 + \alpha_1) ]$ and 
$\Gamma\subHe = (6.33 \times 10^8 ) J\subHe [5/(3 + \alpha_4) ]$,
which hold true for pure power-law spectra.
For $z=3$, absorption model A2, and quasar model Q2 with spectral index
$\alpha_s = 2.1$, these effective indices
are $\alpha_1 = 1.4$ and $\alpha_4 = 1.0$ omitting cloud re-emission and
$\alpha_1 = 2.6$ and $\alpha_4 = 6.6$ including re-emission.
Using the metagalactic background from our simulations, we have
calculated the equilibrium temperatures in photoionized
clouds with and without re-emission.  In general, temperatures
are $10\%$ lower if re-emission is included.

The qualitative appearance of the spectrum is similar to that
in HM96.  However, as discussed in Appendix~\ref{app:re-emiss},
the differences in our treatment of self-shielding and
cloud re-emission lead to values of $\eta$ that are $\sim \! 70\%$
higher than theirs.  This difference is small compared to the
differences induced by uncertainties in the observational parameters. 

\begin{figure}
\plotone{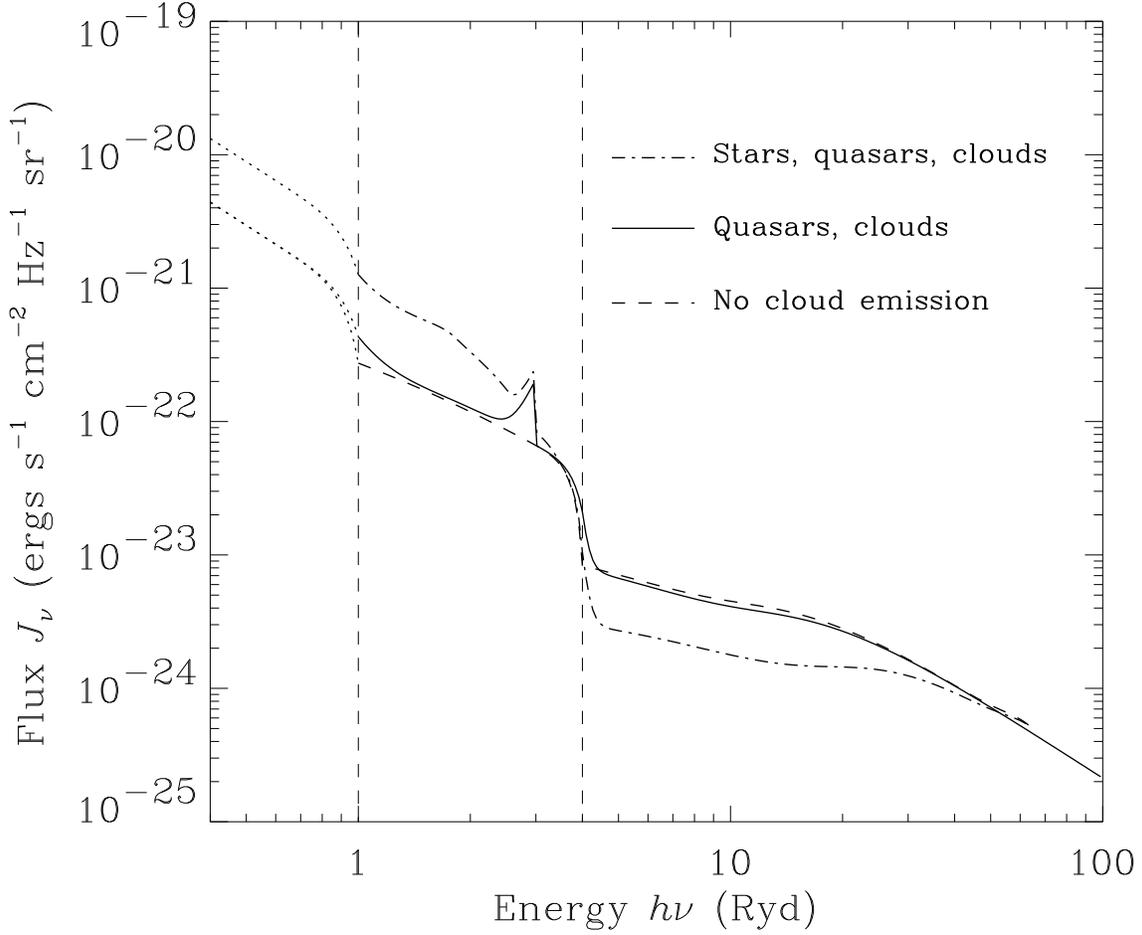}
\caption{
\label{fig:spectra}
Typical spectra produced by the radiative transfer simulations, for
$z=3$.  These curves assume quasar model Q1, opacity model A2, and an
ionizing slope of $\alpha_s = 1.8$.  The upper curve includes a stellar 
contribution, fixed at 1 Ryd to have an emissivity twice that of the 
quasars.  The effect of re-emission from the clouds can be seen in the 
bump at 3 Ryd and the enhanced radiation just above the H~I and He~II 
thresholds.  The region below 1 Ryd affects neither the H~I nor He~II 
ionization rates, and so we have not bothered to include processes that 
affect it, such as H~I \Lya\ emission; hence it is unrealistic and shown 
with dotted lines.  }
\end{figure}

\begin{figure}
\plotone{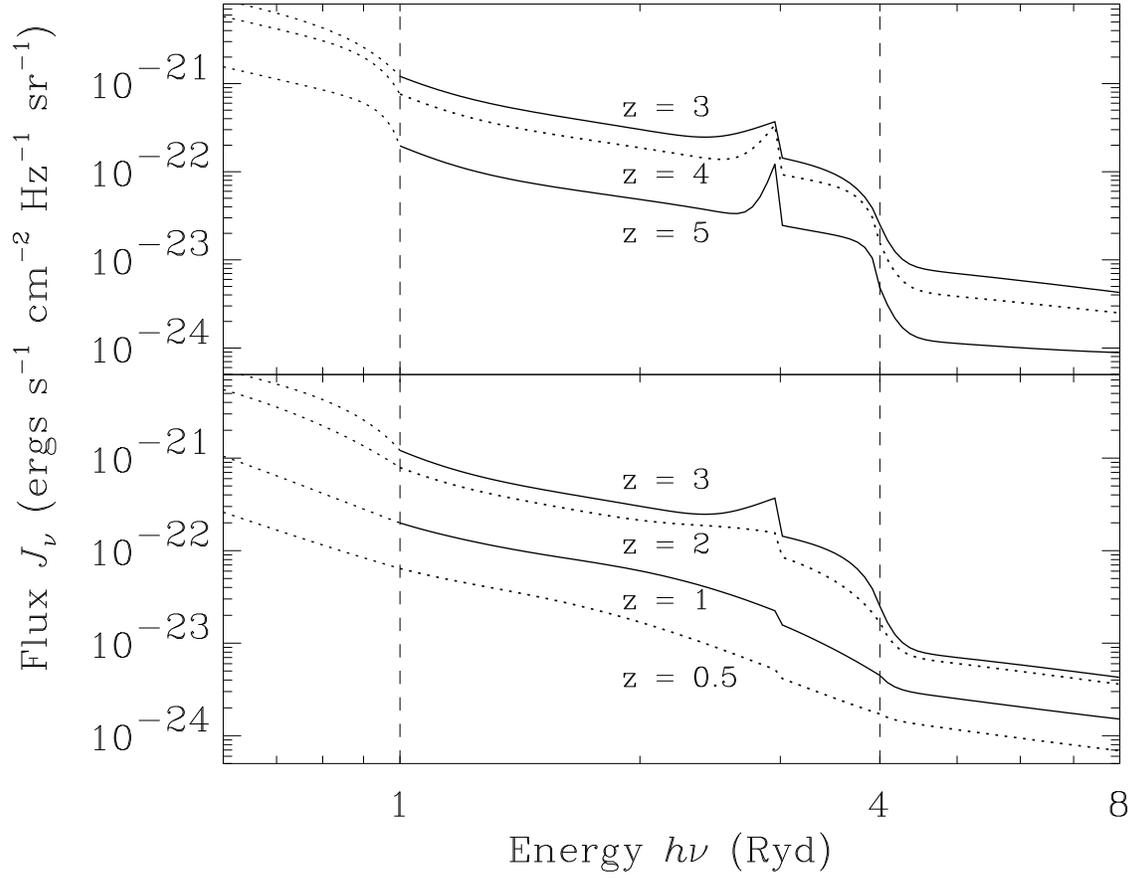}
\caption{
\label{fig:fevolve}
Evolution of the ionizing background with redshift, assuming quasar
model Q2 with spectral index $\alpha_s = 2.1$ and absorption model
A2.  Top panel: $3 \le z \le 5$.  Bottom panel: $0.5 \le z \le 3$.
See discussion in \S~3.  }
\end{figure}

The effect of stellar emission is shown in Figure~\ref{fig:spectra},
and the evolution of the spectrum with redshift in
Figure~\ref{fig:fevolve}.  The intensity of the spectrum peaks at $z
\sim 3$.  The features in the spectrum gradually disappear with cosmic
time as the universe becomes less opaque.  The break parameter in
Figure~\ref{fig:fevolve} evolves approximately as $B \approx 3.2 z$
over the plotted redshift range ($0.5 < z < 5$), although $B$ at a
given redshift depends strongly on the assumed $\eta$ and opacity
model.

The intensity of the ionizing background in these models
spans a large range, depending on the assumed quasar luminosity
function, the stellar contribution, and the opacity model.
The H~I proximity effect gives an independent estimate of
the intensity, although many systematic uncertainties could be
involved.  Recently, Giallongo \ea (1996) found a value of $J\subH
\approx (5\pm1) \times 10^{-22}$ \fluxu, while Cooke, Espey, \& Carswell
(1997) found $J\subH \approx 10^{-21 \pm 0.4}$ \fluxu, with very
little evolution over the redshift range of $2.0 < z < 4.5$.  These
estimates assume that the quasar has the same spectrum as the ionizing
background; in view of the bumps and wiggles in the background shown
in Figure~\ref{fig:spectra}, this is unlikely.  
The photoionization rate $\Gamma\subH$ is a better measure
of the integrated background. Assuming
that the clouds see the ``bare'' spectrum of the quasar observed in
the proximity effect, and that this bare spectrum typically has a
spectral index $\alpha_s = 1.8$ as argued above, 
the conversion from the estimated fluxes to the photoionization rate
is $\Gamma\subH = (2.64 \times 10^9 \, {\rm sr} \, {\rm cm}^2 
\, {\rm erg}^{-1} \, {\rm s}^{-1}) \,  J\subH$, based on the photoionization
rate equation above.

An additional constraint on the photoionizing background radiation
comes from N-body hydrodynamical simulations of the  
Ly$\alpha$ forest (e.g., Cen \ea 1994; Hernquist \ea 1996).
For these optically thin clouds, the column density distribution
determines, via recombination theory, the ratio of $n_H^2/\Gamma_{HI}$
times $dl/dz$, or equivalently $\Omega_b^2 h^3/ J_{HI}$.  One recent 
simulation (Zhang \ea 1997) finds that,
in order to reproduce the Ly$\alpha$ column density distribution, 
the ionization rate must lie in the 
range $0.3 < \Gamma_{-12} < 1.0$, somewhat lower than estimates 
$\Gamma_{-12} \approx 2-3$ ($J_{-21} \approx 1$) from the proximity effect.
For $\alpha_s \approx 1.8$, the range $\Gamma_{-12} = 0.3 - 1.0$ 
corresponds to $0.1 < J_{-21} < 0.4$ (see Fig. 6).  Similar conclusions 
are found from other simulations (\ME\ \ea 1996; Hernquist \ea 1996).
Thus, one must conclude that an inconsistency still exists for 
the ionizing radiation field inferred from the proximity effect,
from Ly$\alpha$ simulations, and from cosmological radiation transfer 
of QSO sources.  In the most extreme case, $\Gamma_{-12} = 0.3$
from simulations and $\Gamma_{-12} = 2.6$ from the proximity effect, 
the discrepancy could be almost a factor of 10.  
Moreover, the discrepancy is larger when one adopts the ``standard'' value 
$\Omega_b h_{75}^2 = 0.022 \pm 0.018$ from Big Bang Nucleosynthesis
(Copi, Schramm, \& Turner 1995) instead of the somewhat higher value
$\Omega_b h_{75}^2 = 0.038 \pm 0.007$ implied by deuterium measurements 
in two high-$z$ QSO absorption systems (Tytler, Fan, \& Burles 1995). 
Thus, many details of these estimates remain to be pinned down.   

\begin{figure}
\plotone{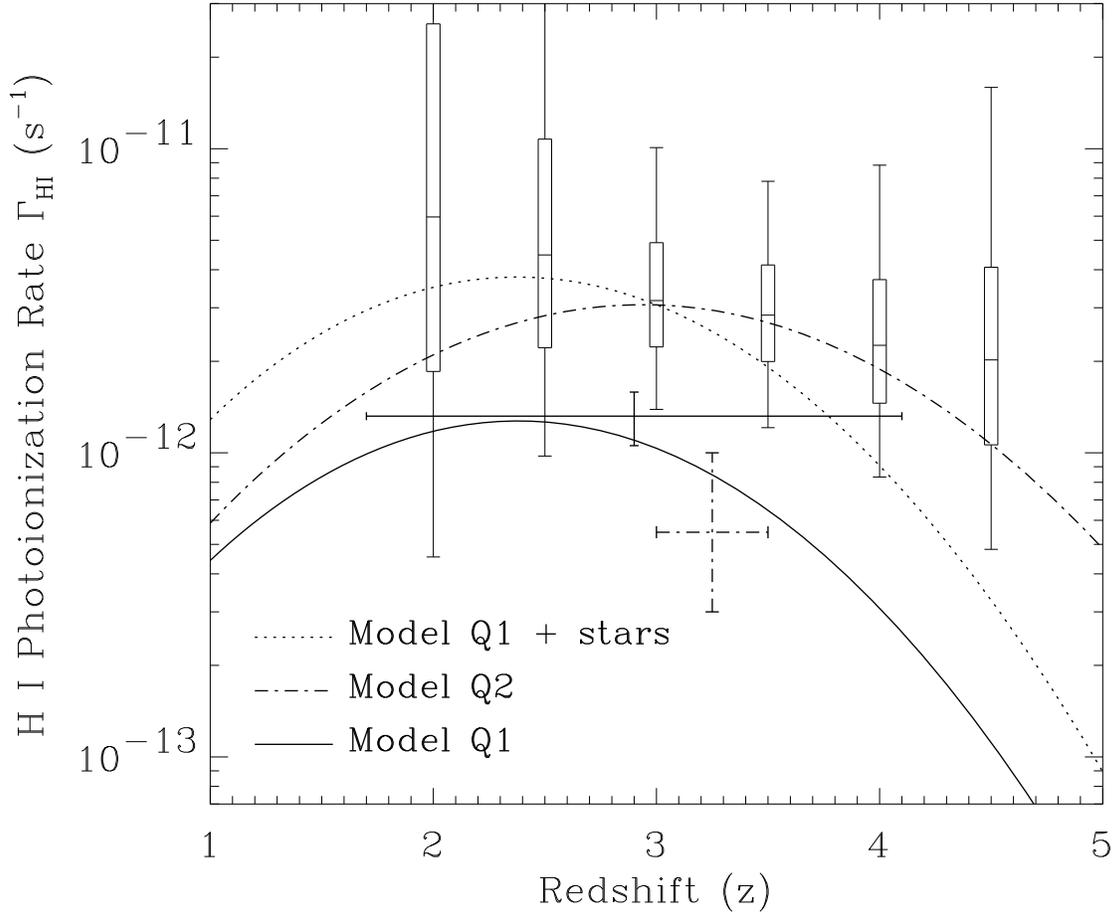}
\caption{
\label{fig:phrate}
Photoionization rate of H~I atoms from our models, assuming absorption
model A2 and quasar spectral index $\alpha_s = 1.8$.  Analytic fits
to $\Gamma_{HI}(z)$ are provided in text (see eq. [16]).
Data points, based on the proximity effect, are converted from $J_{\nu}$
as discussed in \S~3.  Box plots from Cooke, Espey, \& Carswell (1997)
show the median, quartiles, and 95\% limits.  The single point with solid 
lines is from Giallongo \ea (1996).  The dot-dashed point is the preferred 
range from hydrodynamical simulations by Zhang \ea (1997).   The 
``Model Q1 + stars'' curve includes a stellar contribution with emissivity 
twice that of the quasars at the H~I Lyman edge.  Models assume quasar 
spectral index $\alpha_s = 1.8$.  }
\end{figure}

Estimates of the photoionization rates are shown in Figure~\ref{fig:phrate},
along with the results of our models.  These rates
are well described by the analytic form of HM96, 
\begin{equation}
    \label{photoioniz}
    \Gamma_{HI} = A_\Gamma (1+z)^{B_\Gamma} \exp[ -(z-z_c)^2 / S] \; .
\end{equation}
This is not surprising, since we are both using versions of the Pei (1995) 
luminosity function.  With spectral slope $\alpha_s = 2.1$, we find
$A_\Gamma = 0.56 \times 10^{-12}$~s$^{-1}$, 
$B_\Gamma = 0.60$, $z_c = 2.22$, and $S = 1.90$ for model Q1, and
$A_\Gamma = 1.26 \times 10^{-12}$~s$^{-1}$, 
$B_\Gamma = 0.58$, $z_c = 2.77$, and $S = 2.38$ for model Q2.
If the proximity effect really
gives an accurate estimate of $J\subH$, the unmodified Pei (1995)
luminosity function (model Q1) fails to produce enough photons at high
redshifts ($z > 3.5$).  The dramatic fall-off in quasar flux past this
point also might tend to produce a sharp increase in the opacity of
the \Lya\ forest, which clashes with the observed smooth behavior 
seen in quasars with redshifts up to
$z_{em}=4.9$ (Schneider, Schmidt, \& Gunn 1991b).
Our model Q2 is in agreement with the proximity
effect estimates, as is model Q1 if it is supplemented by a large
contribution from star-forming galaxies.  In view of the objections to
model Q2 raised by the radio observations of Shaver \ea (1996), this
constitutes circumstantial evidence for a significant stellar
contribution to the ionizing background at high $z$.  

\begin{figure}
\plotone{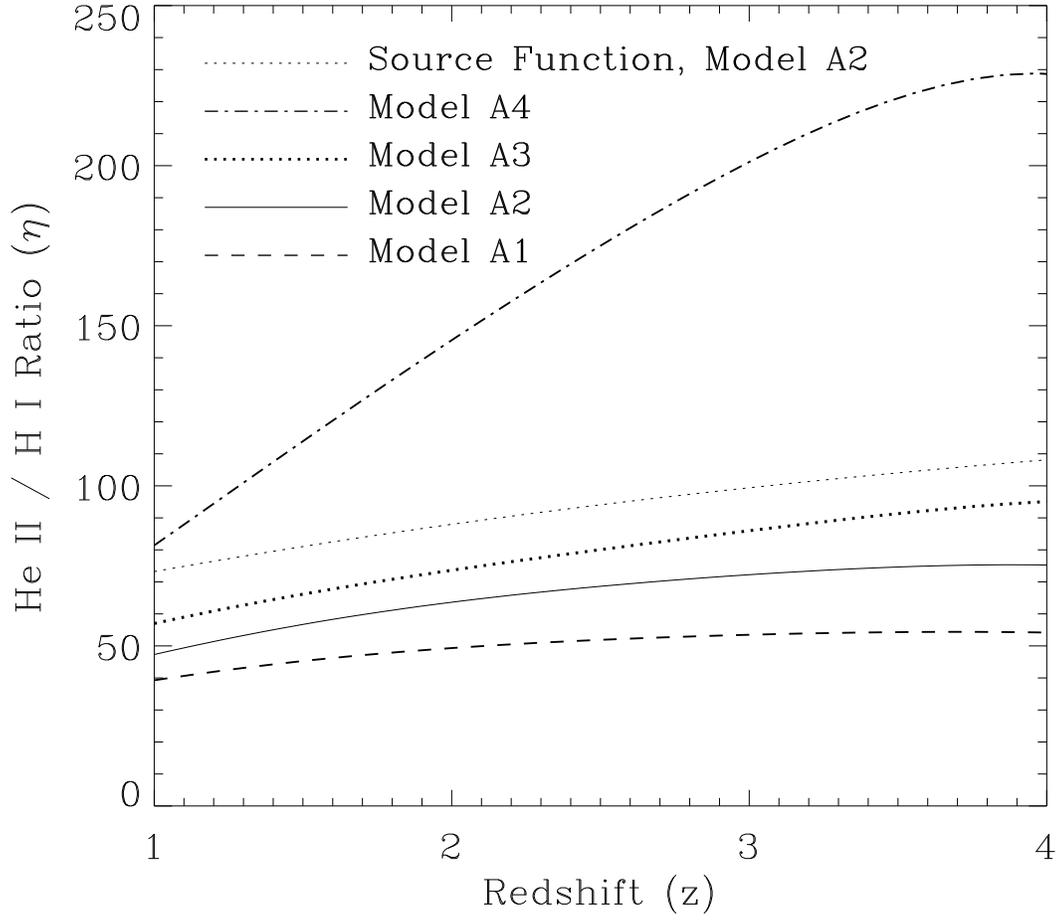}
\caption{
\label{fig:eta-z}
The ratio $\eta \equiv \nhetwo / \nhone$ as a function of redshift.
(a) Effect of the absorption-line model, assuming quasar model Q2 with
spectral index $\alpha_s = 1.8$ and including cloud re-emission.  
The results of the source-function approximation are shown as well.  }
\end{figure}

In Figure~\ref{fig:eta-z} we show the column ratio $\eta(z) = N\subHe
/ N\subH$ that results from our photoionization models.  This value is
only valid for the optically thin clouds, as self-shielding affects
$\eta$ in the thicker clouds. For comparison with Paper I, we have
shown the results of the source-function approximation.  Since the
universe is less opaque in H~I than in He~II, more of the H~I-ionizing
flux is redshifted away, and $\eta$ is smaller than in the 
source-function approximation.  

Figure~\ref{fig:eta-z} also shows the effects of our main parameters
upon $\eta$.  It can be seen that the most important variables are the
intrinsic spectral index $\alpha_s$ of the ionizing sources and the
continuum opacity model.  The quasar luminosity function, the cloud
density, and the cloud emissivity are less important.  The large
difference between A4 and the other absorption models is explained
simply by Figure~\ref{fig:opac}, where the continuum opacities of both
species but especially that of He~II are much higher than in the other
models. The parameter $\eta$ depends on the ionizing emissivity in two 
distinct ways.  The smaller the He~II-ionizing intensity, the easier it 
is to make almost all the He into He~II in the clouds that dominate the
He~II continuum opacity, saturating $d\tau\subHe/dz$ and lowering
$\eta$.  This effect is generally confined to large redshifts ($z
\gtrsim 3.5$), where the ionizing flux is low, with our assumption
for the density within the clouds.  For $z \sim 3$, the ionizing
intensity is large; while it drops again at low redshifts, the
continuum opacity is not as important as cosmological redshifting
there.  The change with redshift of the ionizing emissivity also
affects $\eta$.  The ionizing photons seen at a given point in space
come from a range of redshifts; because the universe is less opaque in
H~I than in He~II, this range is larger for H~I.  Consequently, an
increase in ionizing emissivity with redshift tends to increase the
H~I-ionizing flux relative to the He~II-ionizing flux and thus
increase $\eta$.  Both of these effects make $\eta$ higher in model Q2 
than in model Q1.

With our QSO emission models, which peak at $z \sim 3$, $\eta$
increases slightly with redshift up to $z \sim 4$. Beyond this
redshift He~II becomes a majority species and equation (\ref{eta}) is
no longer valid.  The increase before this point is slight over the
observable range of redshifts, because we have assumed the shapes of
the column density distribution and ionizing spectrum to be constant
with redshift.  Between $z = 2.4$ (\HS) and $z = 3.15$ (\Qj), $\eta$
increases by only 5--10\%.

We find that the cloud re-emission contributes a fair fraction, $\sim
\! 20\%$, of the total ionizations of both H~I and He~II at $z \approx
3$.  However, the fractional increase is roughly the same for both
species, so the net effect on $\eta$ is not large.  For the purpose of
calculating $\eta$ to within $\sim 5\%$, we could leave out all cloud
emission except for the Lyman recombination continuum.  However, these
diffuse emission processes may have a larger effect on photoionization
of metal ions.

In Figure~\ref{fig:alphas-alphab} we have plotted the intrinsic
(emitted) index $\alpha_s$ versus $\eta$ and the spectral index
$\alpha_b$ of the background at a redshift of $z = 3$.  This plot
clearly displays the strong dependence of $\alpha_b$ on $\alpha_s$
discussed in \S~\ref{sec:phys}.  If quasars are the principal source
of ionizing photons, our models predict a plausible range for the
column ratio of $25 < \eta < 400$.  We would like to use the He~II \Lya\
absorption measurements to constrain the models further.

The dependence of the $\tau\subHe$ on the various parameters is shown
in Figures~\ref{fig:cum-opt-depth} and \ref{fig:optdepth-contour}.  For
the parameters chosen, $\tau\subHe$ has mostly converged by a lower
limit of $N^{\rm min}_{HI} \sim 10^{11}$--$10^{12}$~\cdu.  Mildly damped
He~II lines, with $N\subHe \gtrsim 10^{18}$~\cdu, make a minor
contribution to $\tau\subHe$, as can be seen by the turn-down around
$N\subH \sim 10^{17}$~\cdu\ in Figure~\ref{fig:cum-opt-depth}.  
The He~II \Lya\ optical depth depends not only upon $\eta$ and the
absorption-line model, but also on the line velocity widths,
parameterized by Doppler width $b\subHe = \xi\subHe b\subH$, and on the
IGM density.  If we simply try the combinations of absorption models,
spectra, and luminosity function shown in Figure~\ref{fig:eta-z},
choose $N_{\it min} = 10^{11}$ or $10^{12}$~\cdu, and consider the
extremes of either pure thermal ($\xi\subHe = 0.5$) or pure
bulk-velocity line broadening ($\xi\subHe = 1$), we find that {\em
none} of the individual parameters are constrained by the He~II
measurements.  A value of one parameter that minimizes $\tau\subHe$
can be traded off against other parameters that maximize it.  (Our
criterion for a successful value is that it lie within the 2$\sigma$
boundaries of the observed quasars; in practice, the models are
constrained mainly by \HS.
It is worth keeping in mind, though, that fluctuations 
in the observed optical depth are likely to be substantial, as
explored in the next section, and the D96 limits may be an
underestimate of the true uncertainty.)  
For the successful models, $\eta$ falls
in the range 70--400.  However, most of these successes come with 
the deprecated absorption model A4, which gives a higher $\eta$
by a factor 2--3 than other models.

For example, one can satisfy the optical depth constraint with bulk
motions, $N_{\it min} = 10^{11}$~\cdu, a spectral index of
$\alpha_s=2.1$, and the quasar model~Q2, giving $\eta=125$ and
$\tau\subHe=1.12$ at $z=2.4$; we will refer to this as our
``standard'' model below.  If we define the Lyman forest to consist
only of clouds with $N\subH > 10^{12}$~\cdu, it is still possible to
satisfy the observed limits without invoking a smooth Gunn-Peterson
effect.  Absorption model A4 and quasar model Q2 with $\alpha_s=1.8$
give $\eta=120$, and, with bulk motions, produce $\tau\subHe = 1.07$ at
$z=2.4$.  Thus, pure line-blanketing can explain at least the mean
level of absorption.  As shown in Figure~\ref{fig:optdepth-contour},
however, this requires some combination of high $\alpha_s$, low
$N_{\it min}$, maximal Doppler widths, and a favorable opacity model,
particularly if we restrict our attention to the favored opacity
models A1--A3.  There is thus some weak evidence for a diffuse
Gunn-Peterson effect and/or a stellar contribution to the ionizing
background.  

The above results are even less restrictive if a diffuse Gunn-Peterson
effect is allowed, since we now have an additional free parameter.
Nevertheless, we can look for a lower limit on the column ratio $\eta$
by considering the maximal contribution to $\tau\subHe$ from the \Lya\
forest and diffuse IGM at a given $\eta$.  Fang \& Crotts (1995) found
evidence for a small Gunn-Peterson effect in H~I ($\tau_{\it GP} =
0.115 \pm 0.025$), although as we have argued above this could also be
induced by a turnover in the quasar spectrum.  We take $\tau\subH^{\it
(GP)} < 0.19$ as an upper limit, which corresponds to $\tau\subHe^{\it
(GP)} < 0.0475 \eta$.  The argument of Fang \& Crotts includes an
extrapolation of the \Lya\ forest down to $N_{\it min} =
10^{12}$~\cdu.  By considering model A2 with this lower limit and
$\xi\subHe=1$, we get a \Lya\ forest contribution of $\tau\subHe^{\it (LF)}
= 0.30$ at $\eta = 10$.  Combined with the Gunn-Peterson opacity, this
is enough to account for the observations of D96 within 2$\sigma$.  We
thus find a lower limit of $\eta > 10$.  

A corresponding upper limit on $\eta$ {\em cannot} be set this way.
Assuming that He~II has not yet reionized, so that all the He in the 
clouds is in the form of He~II, and
using the density model of equation~(\ref{hdens}), we obtain
$\eta_{\it max} = 144 \, \Gamma_{-12} \, (N\subH /10^{17}~{\rm
cm}^{-2})^{-1/2}$ up to $N_{\it max} = 10^{17}$~\cdu, where the H~I
becomes self-shielded and $\eta_{\it max}$ decreases sharply.  
If we now adopt unfavorable assumptions of
thermal linewidths, $N_{\it min} = 10^{12}$~\cdu, and absorption model
A1, we obtain $\tau\subHe = 1.0$ at $z=2.4$, in perfect agreement with
the observations of D96.  This is because the high $\eta$ simply
pushes the lines further up the saturated portion of the curve of
growth.   In this case $\tau\subigm\suplf \lesssim 0.2$,
showing that a fraction $f\subigm < 3 \times 10^{-4}$ of the baryons
are in structures with column density less than $10^{12}$~\cdu. This
seems physically implausible, but we would not venture to calculate
an upper limit to $\eta$ based on this argument.

It has in fact been suggested that the high He~II optical depths
towards \Qj\ (Jakobsen \ea 1993) and the redshift dependence of the
${\rm Si~IV} / {\rm C~IV}$ ratio (Songaila \& Cowie 1996) both
constitute evidence that ionization bubbles around individual quasars
may have not yet overlapped at a redshift of $z \approx 3.2$. Reimers
\ea (1997) suggests a similar explanation for the patchy He~II
absorption observed at $z \approx 2.9$ towards \HE.  We consider this 
explanation unlikely for two reasons mentioned above: the total number 
of 4 Ryd continuum photons emitted by this redshift is likely
to be sufficient to fully ionize helium, while the space between the
\Lya\ lines must be extraordinarily clear of gas at $z \lesssim 3.2$
according to the flux measurements of Hogan \ea (1997). Although these
statements depend on the poorly known QSO luminosity function
at $z > 4$ and uncertain ionizing spectra above 4 Ryd, we have 
nevertheless considered what would happen to $\eta$ if the 
non-overlapping I-front explanation was correct.
If we mimic the overlap of He~III spheres by fixing $\eta$ to an
arbitrary high value for $z \geq 3.2$ and then releasing it, we find
that $\eta$ recovers by $z=2.7$ to the value it has in our standard
models.  Observations of \HS\ thus cannot help us to decide this
question; future observations should preferably include some quasars
at $2.7 < z < 3.2$.

Our results show that the mean He~II \Lya\ opacity $\tau\subHe$ and
the He~II$ /$H~I ratio $\eta$ are not strongly constrained by theory
at this point, as they are too sensitive to the observational
parameters.  In addition, $\tau\subHe$ is not uniquely determined by
$\eta$, as the Doppler widths and the distribution of low-column gas
are also involved.  However, if it were possible to determine $\eta$
accurately from observations that resolve the He~II lines, its
sensitivity to the background spectrum would tightly constrain the
spectrum, as discussed in \S~\ref{sec:ioniz-emiss}.


\section{HE~II ABSORPTION FLUCTUATIONS}
\label{sec:abs-fluct}

We have argued for a picture in which the He~II absorption
occurs in density structures where helium is mostly photoionized to
He~III by quasars.  To explain the mean level of the absorption, it
appears that $N\subHe / N\subH \equiv \eta \sim 100$. As a result, the
universe is optically thick to He~II-ionizing radiation at high
redshifts.  A consequence of this picture, however, is that there
should be fluctuations in the He~II absorption, from both intrinsic
structure in the gas and fluctuations in the level of the ionizing
background.  Some questions then arise: how large should these
fluctuations be? If we wish to find the mean level of absorption, must
we observe many different quasars at a given redshift?  How can one
distinguish between fluctuations in the gas density and in the
ionization level?  What can these fluctuations tell us about the
properties of quasars and the process of structure formation?  In this
section we address these questions.

\subsection{Intrinsic fluctuations}

The very existence of discrete absorbers (``clouds'') implies
fluctuations in the gas density at high redshifts.  Since the
clustering of \Lya\ clouds with $\nhone \sim 10^{13}$~\cdu\ is small
(Cristiani \ea 1997), it is probably sufficient to regard the random
fluctuations from these clouds as a Poisson process.  However, the
clustering of clouds with lower columns, which produce a large
portion of the He~II \Lya\ absorption, is not known observationally.
We will assume that Poisson statistics apply to these clouds as well.
We also assume that the smooth IGM component, if present,
does not fluctuate.
In our Poisson idealization, the absorption loses correlation over
velocity scales of $\sim 30 \xi\subHe^{-1}$~\kms, since each absorption line
is independent.  Observations (Cristiani \ea 1997) and hydrodynamical
simulations (\ME\ \ea 1996) support the notion that the clouds are
very weakly clustered beyond $\sim 150$~\kms.

This point of view is, of course, naive.  In hydrodynamical
simulations of the IGM, as in the real universe, no distinction is
made between the clouds and the IGM.  Local density maxima in the
simulations are generally fairly sharp, which gives rise to the
appearance of ``clouds'', while local density minima are generally
less sharp.  If one chooses to define an IGM as the matter outside
detectable clouds, it will have large fluctuations in optical depth.
Since the density structure reflects the large-scale structure of the
universe, there is no {\it a priori} reason to expect the clouds to obey
Poisson statistics.  In our defense, we can offer two reasons for
proceeding: first, our method at least gives an estimate for the
fluctuations that can be improved upon in future work.  Also, an
estimate based on the observed almost-Poisson nature of the H~I \Lya\
is in the same spirit as semi-analytic treatments 
(Bi \& Davidsen 1996), which assume a simplified global
model for density fluctuations and reproduce many of
the qualitative properties of the IGM and absorbers
seen in  the output of hydrodynamical simulations. 

The level of fluctuations expected from variation in cloud numbers
is easy to estimate.  We have found that half of the He~II absorption is due 
to clouds with $N\subH \gtrsim 5 \times 10^{12} \eta_{100}^{-1}$~\cdu, which
(in model A2) have a line density $d\NN/dz \sim 590 [(1+z)/4]^{2.6}
\eta_{100}^{0.5}$.  Using our Poisson idealization, we can guess that,
over an observed redshift interval $\Delta z_{\it obs}$, the
fluctuations in optical depth would satisfy 
$(\Delta \tau / \tau)\subHe \sim [\Delta z_{\it obs}(d\NN/dz)]^{-1/2} 
\approx 0.041 \eta_{100}^{-0.25} \Delta z_{\it obs}^{-0.5}
[(1+z)/4]^{-1.3}$.  From Monte Carlo simulations with our model column 
density distributions, we have found a slightly lower result,
\begin{equation}
  \left( \frac{\Delta \tau }{\tau} \right) \subHe \approx  
 0.03 (\xi\subHe \eta_{100})^{-0.1} \Delta z_{\it obs}^{-0.5}
   \left[ \frac {1+z}{4} \right]^{-1.0} \; .  
\end{equation}
Here we have defined $\tau \equiv -\log(\bar{T})$ where $\bar{T}$ is
the mean transmission within the redshift interval.
The exponents differ slightly from their expected values because the
absorption is beginning to be nonlinear ($\tau\subHe \gtrsim 1$).
For the observations of D96, which measured He~II absorption with
$\tau\subHe = 1.00 \pm 0.07$ in the redshift range $2.23 < z < 2.60$,
we expect intrinsic fluctuations in optical depth of about
$\delta \tau\subHe = 0.06 (\xi\subHe \eta_{100})^{-0.1}$, of the same order as
the observational error from photon statistics.

\subsection{Ionization fluctuations}

Thus far, our computations have assumed the ionizing flux to be uniform
in space.  However, as pointed out by Zuo (1992), much of the ionizing flux
comes from bright but rare quasars, and there should actually be
substantial variation in the flux level from point to point in space.
Fardal \& Shull (1993, henceforth FS93) calculated the consequent
statistical effect on the H~I \Lya\ forest, predicting a weak,
large-scale correlation.  It is intriguing that such a correlation was
found in the data analysis of Press \& Rybicki (1993), although the authors
caution that the correlation is at the expected level of systematic
errors.

In the case of He~II, the ionizing flux is even more likely to come
from a few discrete sources.  Star-forming galaxies may produce a
large fraction of the H~I-ionizing photons but are unlikely to emit
appreciable He~II-ionizing photons (Sutherland \& Shull 1998).  Even
if quasars are the main sources of both types of photons, the universe
is more opaque in the He~II continuum than in H~I (see
Fig.~\ref{fig:opac}).  Thus, far fewer quasars contribute to the flux
at a given point, and the resulting variations in the He~II ionization
level are much larger than those in H~I.  Effects on the He~II
absorption can show up in three ways: as the proximity effect, where
the quasar providing the background source also provides the enhanced
ionization; as the influence of known quasars on neighboring lines of sight;
and as a statistical effect, independent of knowledge of the positions
of any observed quasars.  We discuss the latter here.

How large are these statistical fluctuations in the He~II line
absorption expected to be?  To answer this, we adapt the methods and
notation that FS93 used to discuss the analogous case of ionization
fluctuations in the H~I forest.  We assume for now that quasars emit
ionizing radiation isotropically and are not variable except on
cosmological time scales.  As in FS93, we conduct simple Monte Carlo
experiments, placing a sample of randomly distributed quasars in a
three-dimensional box and seeing what effect this has on the optical
depth along a line of sight.  Here, we consider only the {\em mean} 
optical depth at a given wavelength.  For now, we ignore the intrinsic 
fluctuations in the line absorption, just as we ignored the ionization 
fluctuations in \S~\ref{sec:abs-fluct}.  Later we will give examples that
combine the two effects.

We draw our sample of quasars from the Pei luminosity function with
$\Omega_0 = 1$ (our model Q1), with a cutoff imposed three magnitudes
fainter than the characteristic luminosity $L_z$.  The comoving
density of the quasars above this threshold is $\phi = 2.4 \times
10^{-5} h_{75}^3 {\rm Mpc}^{-3}$, independent of redshift, since pure
luminosity evolution is assumed.  Emission from fainter quasars could
smooth out the flux but is not well-constrained observationally.
However, we do account for the re-emitted radiation, representing it
as a perfectly smooth background radiation field.  We found above that
it typically contributes $\sim 20\%$ of the He~II ionizations, so we
take the quasar flux fraction to be $f_Q = 0.8$.

In FS93, the quantity calculated was the correlation function of the
number density of H~I clouds.  In the He~II case, we would instead
like to know the statistical properties of the optical depth
$\tau\subHe$.  There are two cases to consider, depending on whether
the absorption is dominated by discrete lines or by a diffuse IGM.  In
both cases the density of He~II ions at a given point $\vec{x}$ goes
like $[J\subHe(\vec{x})]^{-1}$, barring any {\em dynamical} effect of
the ionizing flux on the absorbing gas. However, equation
(\ref{igm-opt-depth}) shows that the optical depth of a diffuse IGM
scales as $\tau\subHe^{\rm GP} \propto [J\subHe(\vec{x})]^{-1}$.  The
optical depth of discrete lines is less sensitive to the flux,
$\tau\subHe^{\rm LF} \propto [J\subHe(\vec{x})]^{1-\beta}$ (equation
\ref{line-opt-depth}).  We found that $\beta \approx 1.5$ for the
column densities that contribute the most to the He~II line opacity
(\S~\ref{sec:col-dist}).  In reality, the behavior may be intermediate
between these two cases.  These cases are equivalent to the formalism
of FS93 for the number density of H~I \Lya\ clouds, with $\beta = 2.0$
and $\beta = 1.5$ respectively.

The most important difference from the H~I fluctuations is a shorter 
attenuation length, defined as the inverse of the continuum opacity.
For example, if we evaluate opacity model A2 with
$\alpha_s = 2.1$ and $\eta \approx 100$ at the mean energy, 
$h \bar{\nu} = [(\alpha_s+3)/(\alpha_s+2)] h \nu\subHe = 5.0$~Ryd,
of a He~II-ionizing photon, we find a comoving attenuation length of
\begin{equation}
  r_0 \equiv  \frac{dl/dz}{d\tau\subHe/dz}
  =  \frac{(500 h_{75}^{-1} {\rm Mpc}) \left(\frac{1+z}{4}\right)^{-1.5} }
      {13.9 \left(\frac{1+z}{4}\right)^{2.6}}  
 \approx (36~{\rm Mpc}) h_{75}^{-1} \left(\frac{1+z}{4}\right)^{-4.1} \, .
\end{equation}
This gives only $\bar{N} = 1.1 [(1+z)/4]^{-12.3}$ 
quasars brighter than our luminosity threshold within a
``flux sphere'', or one attenuation length.  
The rapid diminution of this number with
redshift suggests that the spatially-averaged models used in
\S~\ref{sec:results} may break down as $z$ increases beyond 3.

In our Monte Carlo simulations, we assume that the flux drops off as
$r^{-2} e^{-r/r_0}$ where $r$ is the distance from the quasar (a more
sophisticated treatment would treat the opacity in a Monte Carlo
fashion as well, and integrate the flux over frequency to get the
ionization rate).  We then adjust the ionization level and resulting
optical depth at each point for the flux observed at that point, as
discussed above.  As expected, we see substantial variation in the mean 
absorption, which we can describe statistically in several ways.  

The main statistic used in FS93 was the correlation function of the
normalized absorption.  The correlation function at zero separation is
the variance, denoted by $\xi_0$ in FS93.  The 1$\sigma$ level of the
fluctuations in the local mean optical depth, 
which we call $(\delta\tau / \tau)\subHe$, is the square
root of the variance.  In our Monte Carlo simulations, we find that
the local fluctuations in the He~II \Lya\ optical depth occur at a level
\[ (\delta\tau / \tau)\subHe = \left\{   
\begin{array}{ll}
\label{ioniz-fluct}
   0.19 \left[(1+z)/4\right]^{2.1} &  (\mbox{cloud-dominated});  \\
   0.33 \left[(1+z)/4\right]^{2.2} &  (\mbox{IGM-dominated}).
\end{array} 
\right. \] 
We have depicted this level of fluctuations by the shaded bands in
Figure~\ref{fig:obs-opt-depth}.  However, this is a somewhat theoretical
quantity, since the small-scale variations are dominated by the
intrinsic fluctuations in the gas properties.  We also calculate
the fluctuations within a fixed redshift interval, as in 
\S~\ref{sec:abs-fluct}:
\[ (\Delta\tau / \tau)\subHe = \left\{  
\begin{array}{ll}
\label{ioniz-fluct-box}
    0.078 \, \Delta  z_{\it obs}^{-0.4} \left[(1+z)/4\right]^{2.9} &
           (\mbox{cloud-dominated});  \\
    0.15 \, \Delta  z_{\it obs}^{-0.4} \left[(1+z)/4\right]^{3.2}  &  
           (\mbox{IGM-dominated}).
\end{array}
 \right. \] 
In both cases, the fluctuations with IGM-dominated absorption are nearly
twice as strong as in the line-dominated case, because a smooth IGM
is more sensitive to flux variations than discrete clouds.  
The ionization fluctuations within large redshift intervals dominate the 
expected intrinsic fluctuations, especially at higher ($z \gtrsim 3$) redshifts.

The correlation length for H~I fluctuations---the separation at which
the correlation function falls off appreciably, say by $1/2$---was 
found in FS93 to be a
substantial fraction of the attenuation length $r_0$.  For He~II, we
estimate a correlation length in redshift space of 
$\Delta z \approx 0.06 [(1+z)/4]^{1.1}$,
corresponding to a comoving length of 
$\Delta r \approx 30 h_{75}^{-1} [(1+z)/4]^{-0.4}$~Mpc 
or a velocity difference of 4000~\kms.  The fluctuations
due to quasar ionization thus occur on larger
scales than fluctuations induced by gravitational clustering.  Indeed, the
scale is so large that it may interfere with attempts to measure the
average He~II line opacity.  A large redshift path is necessary to
make sure the derived average is accurate.  It is interesting to note
that in the published spectrum of \HS\ (D96), there are two apparent
emission peaks (at $\lambda_{\rm obs} \approx$ 960 \AA\ and 1070 \AA)
extending over at least 10 \AA.  The former peak may correspond to
He~II Ly$\beta$ emission (256 \AA\ rest frame), but we know of no line
corresponding to the latter feature.  One possibility is that this is
a region of enhanced transmission induced by photoionization from a
quasar about $7'$ away from the line of sight, with $z \approx 2.5$.
We estimate that an isotropically emitting quasar capable of ionizing
this region would need to have a blue magnitude of only $B \approx
20$.

Of course, our model of isotropically, steadily emitting quasars may
be too simple. If quasars emit light in two oppositely
oriented cones (Dobrzycki \& Bechtold 1991)  
with total solid angle $4 \pi f_{\Omega}$, 
there would need to be $f_{\Omega}^{-1}$ more quasars than
in the luminosity functions used here, but each quasar affects a
volume smaller by $f_{\Omega}$.
The regions strongly influenced by the quasars thus have
the same filling factor as before, and the local fluctuation level
$(\delta\tau / \tau)\subHe$ is unchanged.
The correlation length, however, goes down by $\sim \! f_{\Omega}^{1/2}$.
If the cone opening angle is large, there is a
quantitative but not qualitative change in the fluctuations.
However, the best way to test this idea would be to look for the
influence of quasars near the line of sight.  If the cone emission
idea is correct, the ionization zone should always be either
completely absent or offset from the quasar redshift.

Quasar variability could also affect these results.  
The He~II ions will reach equilibrium with the local ionization rate
on a timescale given by the inverse of the photoionization rate,
$\Gamma\subHe^{-1} \sim 10^5$~yrs. 
Structure would be induced in the He~II absorption if individual quasars vary
more slowly than this timescale, but more quickly than the timescale
$\sim \! 3 \times 10^7$ yrs for photons to cross 
the quasar ionization zones we have discussed.
Quasar lifetimes are not well known; one might invoke 
the Eddington accretion time ($4 \times 10^8$ yrs) or 
the dynamical time for bright galaxies ($\sim \! 10^8$ yrs).  
If nearly every galaxy once hosted a bright quasar, then
quasar lifetimes may be as short as $10^7$ yrs.  On the other hand,
the existence of the H~I proximity effect argues that typical quasar
lifetimes are at least $10^7$ yrs.  
As in the case of anisotropic emission, the variance in the ionization level 
is unchanged by the fluctuations, since any reduction in the duty cycle
or lifetime of quasars is compensated by an increase in the number
of quasars deduced from observations. 
Again, however, the correlation length is potentially shortened.
If a quasar emits a burst of radiation, it causes a shell of enhanced
ionization to travel outwards.  Because of the time needed for light
to travel to the observer, this shell would appear not as a sphere but
as a parabola cupped towards Earth, although we can think of no
practical way to observe this curious pattern.

\subsection{Influence on Observations}

Because fluctuations in He~II absorption could be quite
large, how does this affect our interpretation of the observations?
The most obvious effect is that it makes the true optical depth at a
given redshift more uncertain and weakens any conclusions about
$\eta$.  Consider a thought experiment in which one measures the
absorption along a line of sight many times, each time with a
different realization of absorption lines and ionizing quasars.  In
Figure~\ref{fig:tau-histogram} we show the results of such an
experiment for a line of sight with He~II absorption observed in the
quasar rest wavelength range 262~\AA$ < \lambda < $293~\AA, 
similar to that of \HS\ (D96).  Even if ionization fluctuations are 
suppressed, as would be the case if quasars are short-lived, we find 
significant variation ($\pm20$\%) in the observed
opacity.  Ionization fluctuations due to isotropically emitting,
constant-luminosity quasars contribute additional variation to the 
results, with the amount increasing rapidly with redshift.  
The uncertainty in the average optical depth could therefore 
be reduced by obtaining a larger redshift coverage.

In summary, fluctuations in He~II absorption can come about on $\sim
\! 30$~\kms\ scales because of random variations in the number of
clouds (and also clustering of these clouds, on scales corresponding
to large-scale structure of galaxies), and on $\sim \! 3000$~\kms\
scales because of variations in the He~II-ionizing flux.  The amount
of ionization fluctuations depends on the quasar luminosity function
and He~II continuum opacity, as well as the quasar lifetime and
beaming angle.  Observations of the absorption fluctuations thus have
the potential to constrain these parameters, which are difficult to
measure by other means.  To do this, it will be necessary to measure
the absorption in the spectra of numerous quasars and to look for
ionized zones from quasars close to the line of sight.

\section{FUTURE ULTRAVIOLET OBSERVATIONS}
\label{sec:future-obs}

Considerable scientific interest was generated by the first 
measurements of He~II absorption, even those at low
spectral resolution.  Future UV instruments with greater throughput
offer the possibility of increasing the scientific information.
First, from a single sightline, one can glean information about
the He~II opacity over a range of absorption redshifts. 
With $\sim \! 5$ sightlines, one can look for the expected variations 
arising from Poisson fluctuations in ionizing sources and absorbers.  
Toward the brightest targets, one can hope to resolve individual 
absorption lines, patches of absorption, and voids in the absorption 
that may correspond to large-scale structure and voids in the gas 
distribution. 	Extending the wavelength coverage down to 912 \AA\ 
opens up the IGM down to $z = 2$, 
allowing the study of evolution in the He~II absorption
and increasing the odds of finding a suitable target.

To date, there have been detections of He~II absorption toward
four QSOs, at emission redshifts $z_{\rm em}$ = 2.74, 2.90, 3.18, and 3.29.
These spectroscopic experiments are exceedingly difficult,
even at low resolution, for two reasons.  First, 
viable background targets with clear lines of sight for He~II Gunn-Peterson
studies are rare (M\o ller \& Jakobsen 1990; Jakobsen \ea 1994),
owing to the strong absorption by the IGM at the redshifts ($z \geq
2.9$) needed to bring the He~II $\lambda304$ absorption into the HST band.  
Observation at lower redshifts ($z = 2$--3) is potentially more fruitful
if one can observe down to the 912 \AA\ limit, as was done with HUT
(Davidsen \ea 1996). 
Second, most high-redshift QSOs are
faint; the four observed targets have fluxes at the He~II edge ranging
from around 0.3 to 3 $\times 10^{-15}$ ergs cm$^{-2}$ s$^{-1}$ \AA$^{-1}$.
Thus, faint-object UV spectroscopy at moderate resolution
(30 km~s$^{-1}$) has been nearly impossible, even with the powerful 
spectrographs (FOS, GHRS, STIS) aboard the {\it Hubble} Space Telescope.

However, the future prospects for UV spectroscopy are somewhat brighter.
In addition to the He~II absorption studies planned with the
Space Telescope Imaging Spectrograph (STIS), two future instruments will
contribute to this subject. 
The first is the Far Ultraviolet Spectroscopic
Explorer (FUSE), scheduled for launch in October 1998.  
The second is the Cosmic Origins Spectrograph (COS), 
planned for the HST 2002 refurbishment mission.
We discuss the potential of FUSE in detail here as
an example of what can be learned in future He~II observations.

Because the FUSE satellite operates in the wavelength range 915--1196 \AA, 
it will be able to detect He~II \Lya\ absorption in the range 
$2.01 < z < 2.94$. Its effective area of 20--100 cm$^2$ is
sufficient to perform moderate-resolution 
spectroscopic studies of QSOs to fluxes $F_{\lambda} > 10^{-14}$
ergs cm$^{-2}$ s$^{-1}$ \AA$^{-1}$. Although the limiting fluxes
accessible to FUSE are comparable to those with HST/STIS, the 
access to shorter wavelengths opens up He~II studies at
redshifts ($2.1 < z < 2.9$) with considerably less H~I and He~II 
absorption from the IGM. Its resolution of 0.03 \AA\ is much better 
than that of the He~II absorption spectra taken so far, although
STIS and COS have similar resolution. However,
FUSE has an estimated dark count of $\sim \! 0.02 \,
{\rm counts}\, {\rm s}^{-1} \, {\rm \AA}^{-1}$.  
This dark count, together with uncertainties in the calibration and
scattered light, may be difficult to characterize.  
Separating continuous He~II absorption from that arising in the
Ly$\alpha$ forest would require binning of order 0.3--0.5 \AA, 
and exposure times $\sim \! 10^6$ s.  Even if one takes shorter
observations and bins the data
to 3--10 \AA\ resolution, the variable dark count may prevent
measurements of $\tau\subHe$ to the 10--20\% accuracy  
needed to detect the predicted fluctuations (Fig. 1, Fig. 13).
FUSE will clearly require long
integration times to observe faint, high-redshift objects such as 
quasars.  Thus, it pays to think creatively about the sorts of 
physical inferences possible with such observations.

A basic question is whether FUSE is capable of determining the
evolution and fluctuations of $\tau\subHe$ in the range $2.1 \lesssim
z \lesssim 2.9$.  Such measurements would help to determine the
evolution of low-column gas and the ionizing spectrum over this
redshift range.  As suggested by Figure~\ref{fig:obs-opt-depth}, the
evolution of $\tau\subHe$ is probably quite rapid.  The number of
potential targets probably depends most on the efforts of observers on
the ground; Picard \& Jakobsen (1993) estimate that there may be
several hundred quasars with He~II threshold fluxes $F_{-15} > 1$ in
the range $2 < z < 3$, where $F_{-15} \equiv F_{\lambda} /
(10^{-15}$~erg s$^{-1}$ cm$^{-2}$ \AA$^{-1})$.  However, only six such
quasars are known at present.  The dark counts dominate the FUSE
signal unless the quasar is  bright, $F_{-15} > 10$; hence
the necessary exposure time is very sensitive to the quasar flux.  For
a quasar with redshift $z \approx 2.5$, we expect an optical
depth $\tau\subHe \sim 0.8$.  To determine the transmission in a
10~\AA\ interval to an accuracy of 10\% would require $\sim \! (5 \times
10^4 \, {\rm s}) \,(1 + 0.15 F_{-15}) / (1.1 F_{-15}^2$).  This is
feasible for $F_{-15} \gtrsim 1$, but it rapidly becomes impractical
for fainter targets.

FUSE will also be able to make a contribution to understanding
the large-scale fluctuations in the He~II absorption.  The top panel
of Figure~\ref{fig:fuse-spectrum} shows a plausible spectrum in a
model where quasars radiate isotropically and statically, as assumed
in \S~\ref{sec:abs-fluct}.  The regions of increased ionization due to
quasars along the lines of sight are visible as broad bumps in the
spectrum.  We estimate that $\sim \! 10$ such exposures would
constrain the optical-depth correlation function well enough to be
able to test the assumptions of isotropy and constant emission,
although much depends on the contribution of unsaturated gas 
(because of its sensitivity to the ionization level) and on
the brightness of the available targets.

FUSE will be able to distinguish between absorption from Ly$\alpha$ forest
clouds and the diffuse IGM, even in highly binned spectra, 
by determining the amount of
absorption at the rest $256.31$~\AA\ He~II Ly$\beta$ edge.  For discrete
clouds, this absorption has an optical depth $\tau_{\beta} \approx
0.47 \tau_\alpha$, whereas for a diffuse IGM the optical depth is
$\tau_{\beta} \approx 0.16 \tau_\alpha$; the difference arises from
saturation effects in the discrete lines.  Thus, a measurement of the
Ly$\beta$ decrement could strongly constrain the relative
contributions of these two sources of absorption.  FUSE is capable of
measuring this decrement for quasars with $z_{\it em} > 2.7$, with
exposure times $\lesssim 10^4$ s for a quasar with $F_{-15}
\sim \! 3$ in the unabsorbed continuum.

Besides measuring the evolution of gross absorption properties for a
large sample of quasars, FUSE will be capable of
providing more spectral detail in a few lines of sight.  Even for
bright targets such as \HS\ ($z = 2.74$) and \HE\ ($z = 2.90$), which
have fluxes $F_{\lambda} = 1.2 \times 10^{-15}$ and $3.6 \times
10^{-15}$ erg~s$^{-1}$~cm$^{-2}$~\AA$^{-1}$, respectively, above the
He~II break, these observations will require very long exposures.
Consider He~II lines with central optical depth $\tau_0 \approx 4$,
which make the median contribution to the line absorption in most of
our models.  Just to ensure an average significance level of 3$\sigma$
in these lines will require an exposure of $2 \times 10^5$~s.  Binning
the spectra at the full resolution will result in extremely noisy
plots, with no apparent features.  However, statistical methods may be
able to pull out a great detail of information from these spectra.

We have carried out simulations of a ``mock'' \HS\ spectrum.  We have
chosen not to use the actual lines in the quasar spectrum, because of
the lack of a published \Keck\ spectrum (though a portion of such a
spectrum is reproduced in Bi \& Davidsen 1997); however, the redshift,
flux level, and optical depth match that of the real \HS.
Figure~\ref{fig:fuse-spectrum} shows a segment of a simulated spectrum
with 0.15~\AA\ pixels, in the case of thermally broadened lines and a
substantial IGM.  Although the spectrum looks quite noisy, due to the
small pixel size, it is possible to identify most of the saturated
He~II lines in this spectrum.  Binning to the HUT resolution of 3~\AA\
would throw away most of this small-scale information but would retain
important information about large-scale fluctuations such as those
resulting from peaks in the photoionizing radiation, some of which are
visible in the figure.

Such a high-resolution spectrum is probably most useful in combination
with a high-resolution H~I spectrum, such as those returned from
\Keck.  Starting with a simulated H~I spectrum of similar quality to
those of Hu \ea (1995), we have computed the He~II absorption spectrum
attained in a FUSE exposure of $3 \times 10^{5}$~s.  We then find a
global fit to the H~I and He~II spectra using the three parameters
$\eta$, $\xi$, and $\omigm$.  In this idealized case, $\eta$ is
determined to an accuracy of 30\% and $\xi$ to within 10\%.  Thus, the
high resolution of FUSE should enable observers to break the parameter
degeneracy that plagues the low-resolution $\tau\subHe$ measurements.
The method of comparing H~I and He~II spectra should determine $\eta$
to high accuracy even if the spectrum is produced by a fluctuating IGM
rather than discrete lines (Bi \& Davidsen 1997).

Two instruments aboard HST are also expected to make advances in the
study of He~II absorption.  Many of the issues discussed in relation
to FUSE apply to these instruments as well.  The recently installed
STIS will be able to perform
better background subtraction in the far-UV with its 2-D detector than
previous HST spectrographs.  Whereas FUSE can observe He~II absorption
in the redshift range $2.1 < z < 2.9$, the HST optics are limited to
$z > 2.8$, so the two instruments complement each other.  The
effective area, dark count, and energy resolution of STIS are all
superior to the corresponding parameters for FUSE.  On the other hand,
at higher $z$ it is more difficult to find suitable targets.  The
quasars are obviously farther away, and the larger He~II opacity at
higher $z$ will make detection of flux below the He~II \Lya\ edge more
difficult.  In addition, in most quasars the rest-frame flux at 304
\AA\ is obscured by Lyman limit systems, eliminating many
of the intrinsically brightest targets from consideration.  These
problems will be partially alleviated by the Cosmic Origins
Spectrograph (COS), which is scheduled for the 2002 HST refurbishment
mission.  The COS instrument will be much more powerful than any of
the other instruments discussed here; it is designed to deliver $\sim
\! 1500$ cm$^2$ effective area in the range 1150--1400 \AA.  This
capability, some 10--20 times greater than that of STIS, will allow 30
km~s$^{-1}$ spectroscopy of sources down to $10^{-15}$ ergs cm$^{-2}$
s$^{-1}$ \AA$^{-1}$, and lower-resolution studies of even fainter
sources.  Astronomers may then be able to measure He~II optical-depth
fluctuations toward a sufficiently large number of sources to
characterize the expected fluctuations.

We have argued above that He~II absorption measurements have several
important uses, but constraints on Big Bang nucleosynthesis are 
unfortunately not among them.  To be useful, the $^4{\rm He} / ^1{\rm H}$ 
ratio must be obtained to high accuracy. As seen in this paper, however, 
the ionization corrections are large and uncertain.  
Ionization corrections would not affect the \hethree/\hefour\ 
ratio, if it could be derived from the $\lambda$304 line.
However, this measurement will be nearly impossible, 
owing to saturation in the \hefour\ line. The detection of deuterium in
the blueward wing of H~I \Lya\ is already difficult,
because the isotopic shift is only 82~\kms\ and because there is a 
large risk of contamination by other H~I \Lya\ lines.  
In He~II, both of these problems are
exacerbated. The isotopic shift of \hethree$^+$ \Lya\ is only
14~\kms. In order to detect a \hethree\ line with central optical
depth $\tau_c \sim 1$, the corresponding \hefour\ line must have
$\tau_c \gtrsim 1000$ ($^3{\rm He} / ^4{\rm He} < 10^{-3}$ for ${\rm
D} / {\rm H} < 10^{-4}$ ) and will completely cover the \hethree\ line
for any Doppler width $b > 5$~\kms.  Also, the problem of
contamination by ordinary \Lya\ lines is far worse for the \hethree\
forest.  We reluctantly conclude that this experiment is impractical.

However, there are several practical observations that could help to 
maximize the significance of future He~II observations.  Any quasar that is
observed in He~II with FUSE should also be observed in H~I 
with \Keck\ or another telescope capable of high-resolution, 
high-S/N spectra.  In addition, concentrated
surveys for neighboring quasars in a $\sim \! 30'$ radius around primary
targets would help to interpret ionization fluctuations in the He~II
absorption.  A survey for low-optical-depth ($\tau_{LL} < 1$) Lyman
limit systems would help to determine the column density distribution for
$N\subH < 10^{17}$~\cdu\ and thus estimate the continuum opacity of H~I and
He~II.  Finally, the study of weak metal lines (C~IV, Si~IV, O~VI) in
\Lya\ clouds will continue to be relevant to the He~II problem 
(Songaila \& Cowie 1996; Giroux \& Shull 1997) as they constrain
the spectrum of ionizing radiation. However, the metal-line ratios often depend
more strongly on the temperature and density than the He~II/H~I ratio.

\section{SUMMARY}

We have used a radiative transfer code and a new absorption-line
sample to relate the spectrum of ionizing sources in the early
universe to the ionization conditions in the \Lya\ forest and diffuse
IGM.  We find that a model with only line-blanketing and ionization
by quasars is sufficient to explain the observed level of absorption,
although contributions from stars or the presence of a diffuse IGM
would help to explain the data.  We show that the H~I and He~II
absorption really does probe the metagalactic background radiation,
and that the observed quasars are probably sufficient to ionize both
H~I and He~II by redshift $z \sim 3.5$.

By considering fluctuations both in the density of clouds and the ionizing
background, we find that a long redshift path is required to obtain
the mean level of absorption.  This is not entirely a drawback, however,
as these fluctuations carry interesting information in themselves.  
Observations by FUSE, STIS, and future high-throughput spectrographs
should allow us to determine the spectral 
shape and sources of the ionizing background with high accuracy
and to disentangle the properties of the ionizing sources from those of the
absorbers, leading to a much better understanding of the latter.

\acknowledgments

We thank Arthur Davidsen, Sally Heap, Craig Hogan, Susanne Koehler,
Jerry Kriss, Limin Lu, Piero Madau, and Michael Rauch for useful 
conversations, and Lisa Storrie-Lombardi for communicating the APM 
survey data on $D_A$ in advance of publication.  We also thank Ed Murphy 
for the use of the FUSE (FSIM)
simulator.  This work was supported by the Astrophysical
Theory Program at the University of Colorado (NASA grant NAGW-766 and
NSF grant AST-96-17073).

\newpage

\appendix

\section{Treatment of Cloud Re-emission and Self-Shielding}
\label{app:re-emiss}

In this Appendix, we discuss two effects on the ionizing background.
The first effect is the emission of diffuse radiation from the
absorbing clouds, referred to as ``cloud re-emission''.  As was noted
in Fardal \& Shull (1993) and explored in detail in HM96, this diffuse
radiation has a significant effect on the high-$z$ ionizing flux,
because the universe as a whole is optically thick in the continuum
for a wide range of redshifts.

The second effect is the self-shielding of the absorbing clouds.  In
the gas that produces most of the He~II and H~I \Lya\ absorption, the
relationship between the ionizing background and $\eta$ is probably
quite simple (see equation \ref{eta}).  As the column density of an
absorber increases, the gas becomes optically thick in the continuum,
first in He~II (as long as $\eta > 4$) and then in H~I, meaning that
$\eta$ becomes a function of column density.  This effect, also
studied by HM96, increases the He~II continuum opacity
slightly at 4 Ryd and more at higher energies.  The He~II \Lya\
opacity can also be affected, because of the damping wings that set in
at $N\subHe \sim 10^{19}$~\cdu.

Both of these problems can be treated by considering an ensemble of
slabs of different column densities.  As discussed in
\S~\ref{sec:phys}, the clouds that dominate the continuum opacity are
moderately optically thick.  The same is true for the cloud
re-emission, since the number of photoionizations saturates as the
slab becomes optically thick.  
The absorbing clouds become optically thick in H~I at H~I column densities
$\eta / 4$ times larger than those in clouds that are optically thick
in He~II,
and the ionizing photons above 4 Ryd contribute little to the
ionization of H~I.  The radiative transfer in the relevant H~I and
He~II clouds are thus nearly separate problems, and we have treated
them as such.  

The ionization structure of these clouds depends on the transfer of
LyC radiation within them.  We choose to model clouds as slabs since
this is specific, tractable, and the sheet or pancake is a generic
feature of numerical simulations.
We have treated the radiative transfer by solving an integral
equation (the Milne solution for a grey atmosphere) for the number of
photoionizations at any optical depth in a given slab.  If
$\Gamma(\tau')$ is the total photoionization rate at an optical depth
$\tau'$ within a slab of total optical depth $\tau$, and $\Gamma_{\rm
dir}(\tau')$ is the photoionization rate resulting from external
photons only, then
\begin{equation}
  \Gamma(\tau') = \Gamma_{\rm dir}(\tau') + \frac{p\subHLy}{2}
    \int_{-\tau\subH/2}^{\tau\subH/2} 
    E_1(|\tau'' - \tau'|) \, \Gamma(\tau'') \, d\tau''.
\end{equation}
In this equation, the recombination radiation is assumed to be emitted
isotropically; the dependence of $\sigma\subH$ on $\nu$ is ignored,
since the recombination photons have energies close to 1 Ryd for
the temperatures under consideration.  Our calculation of $\Gamma_{\rm
dir}$ assumes an isotropic distribution of the incident flux and a
power-law incident spectrum, although we could improve on this
slightly by using the example spectra derived in this paper.  We solve
this equation for slabs of varying optical depth.  Assuming constant
density within the slabs, we can solve for the ionization fractions at
every point.  Integrating over the slab, we find $N\subHe$ as a
function of $N\subH$.  In general, $\eta(N\subH)$ 
increases smoothly but rapidly once $N\subHe \gtrsim 3 \times 10^{17}$~\cdu, 
only to turn over and fall nearly as $N\subH^{-1}$
when $N\subH \gtrsim 3 \times 10^{17}$~\cdu\ (HM96).  

These columns are perpendicular columns, where the slab is viewed
face-on.  For consistency, we must consider inclined clouds, where the
projected area goes down by a factor $\mu \equiv \cos \theta$ while the
column density goes up by the same factor.  The cloud distribution
that would be seen if the clouds were all face-on is related to the
observed distribution by
\begin{equation}
  \frac{\partial^2\NN}{\partial \Nobs \partial z} (N_{\it obs}) = 
  \int_0^1 \left( 
  \frac{\partial^2\NN}{\partial \Nperp \partial z} \right)_{\rm face-on}
  (\Nperp) \mu^2 d\mu \; ,
\end{equation}
with $\Nperp = \mu \Nobs$.  There are two approximate methods we
consider to relate these two distributions.  If the observed
distribution is a pure power law in $\Nobs$, as in equation
(\ref{col-dist}), the face-on distribution is simply a factor of
$(3-\beta) \approx 1.5$ higher.  It is also useful to consider the
clouds as being at a characteristic angle of $\theta_1$ where
$\cos{\theta_1} = \mu_1 = 1/\sqrt{3}$ according to
the model of 2-beam ($n=1$) Gaussian quadrature.
We actually use both approximations, as discussed below.

The above model is too complicated to compute within our radiative
transfer code, especially for the sake of effects that will not
change $\eta$ radically.  As did HM96, we need to find approximations
that can be computed quickly.  We begin with the cloud emissivity.
The processes we consider are those that can contribute to the ionization
of H~I or He~II, namely H~I and He~II LyC recombination radiation, 
He~II BaC recombination radiation, and He~II 2-photon and Lyman line
emission.

Consider the H~I Lyman recombination radiation from \Lya\ forest and
Lyman limit clouds in ionization equilibrium.  For clouds that are
optically thin at the Lyman limit ($\tau\subH \ll 1$), a fixed
fraction of recombinations $p\subHLy$ go directly to the ground state
and emit a Lyman continuum photon that escapes from the cloud
($p\subHLy = 0.43$ at a temperature of $2 \times 10^4$~K). Thus, the
H~I atoms in the cloud emit recombination photons at an average rate
$\equiv f\subHLy \Ghthin$, where $\Ghthin$ (written simply as
$\Gamma\subH$ in the rest of the paper) is the ionization rate of
hydrogen atoms in optically thin clouds and $f\subHLy$ is the
normalized emission probability; clearly $f\subHLy \approx p\subHLy$
in this case.  In optically thick clouds ($\tau\subH \gg 1$), almost
all of the incident photons are absorbed, and most of the ionizations
and recombinations take place within a few optical depths of the cloud
edge, no matter how large $\tau\subH$ becomes.  The average emission
probability of atoms in these clouds is $f\subHLy \rightarrow {\rm
const} / \tau\subH$.  Note that the escape probability formalism would
erroneously predict that the emission rate scales as $f\subHLy \propto
\tau\subH^{-2}$, a product of the average
ionization rate ($\propto \tau\subH^{-1}$) and the escape probability
($\propto \tau\subH^{-1}$).

The dependence of $f\subHLy \, \tau\subH $ on $\tau\subH$ is thus
similar to the factor $(1 - e^{-\tau\subH})$ contained in the integral
for the H~I continuum opacity.  The space-averaged volume density of
H~I atoms in clouds with columns in the range $[\nhone,\,\nhone +
d\nhone]$ at a redshift $z$ is $\nhone (\fnzt) (dz / dl) \, d\nhone$.
Multiplying this density by the emission rate, $\equiv
f\subHLy(\nhone) \, \Ghthin$, we find that the total re-emission of
Lyman continuum photons from the clouds is
\begin{equation}
 j\subHLy(\nu, z) = \frac{h\nu}{4\pi} \, 
    \frac{\phi_{\it em}(\nu)} {\sigma\subH} \,
    \frac{dz}{dl} \, \Ghthin \, 
    \int \fnz \, \left[ f\subHLy(\tau\subH) \, \tau\subH
    \right] \, d\nhone   \, ,
\end{equation}
where $\tau_{HI} = \sigma_{HI} N_{HI}$ and where 
$\phi_{\it em}(\nu)$ is the frequency distribution of the
recombination radiation.  Compare the form of the integral here
to that in equation~(\ref{cont-opt-depth}) for the continuum opacity.
As long as the column density distribution has a fixed shape, these
two integrals are proportional and close in magnitude.  Our strategy
is to exploit this similarity to express the cloud emissivity as
\begin{equation} 
  j\subHLy(\nu, z) = q\subHLy \, \frac{h\nu}{4\pi} \,
  \frac{\phi_{\it em}}{\sigma\subH} \, 
  \frac{dz}{dl} \, \Ghthin \,  
  \frac{d\tau\subH}{dz} \, ,
\end{equation} 
where the ensemble-averaged emission probability is
\begin{equation}
 q\subHLy \equiv 
    \left( \frac{d\tau\subH}{dz} \right)^{-1}
    \int \fnz \, \left[ f\subHLy(\tau\subH) \, \tau\subH
    \right] \, d\nhone   \; .
\end{equation}

We make use of the radiative transfer models discussed above to calculate
$q\subHLy$.  The
emission rate is given by the difference between emitted and destroyed
recombination photons, integrated over the slab, and is equal to
\begin{equation}
 f\subHLy(\tau\subH) \, \Ghthin = \frac{1}{\tau\subH}  \left(
 p\subHLy \int_{-\tau\subH/2}^{\tau\subH/2} \Gamma_{\rm dir}(\tau') \, d\tau' 
 - \int_{-\tau\subH/2}^{\tau\subH/2} 
  \left[ \Gamma(\tau') - \Gamma_{\rm dir}(\tau')  \right] 
  \, d\tau' \right)  \;   . 
\end{equation}
The results do not depend on the assumed density.
We can then integrate over a power-law distribution of slabs of
different $\nhone$ to find $q\subHLy$.  The value of $q\subHLy$ turns
out to be insensitive to the exact input spectrum, cloud temperature
(which determines $p\subHLy$), or column density distribution.
Because many of the Lyman continuum photons are
unable to escape from moderately thick clouds, $q\subHLy < p\subHLy$.
Our method is similar for the He~II Lyman recombination radiation.
The H~I continuum opacity is generally negligible for the clouds with
$\tau\subHe \sim 1$ that dominate the emission, so the radiative
transfer models can be reused, with the minor difference of a slightly 
different $p_{\it Ly}$.  

Photons in the He~II Balmer continuum, Lyman 2-photon continuum, and
\Lya\ line escape from clouds with $\tau\subHe \sim 1$ (although the
\Lya\ photons resonantly scatter off He~II atoms many times first).
We use Case B recombination coefficients, since if a cloud has any
significant optical depth in the continuum the Lyman lines will be
optically thick, and the only Lyman line photons that escape are
\Lya.  The emission from these processes is simply proportional to the
total number of ionizations in the cloud.  The He~II Balmer emission
rate is given by
\begin{equation}
f_{\it HeBa}(\tau\subHe) \, \Ghethin = \frac{p_{\it HeBa}}{\tau\subHe} 
    \int_{-\tau\subHe/2}^{\tau\subHe/2} \Gamma(\tau') \, d\tau' \, ,
\end{equation}
with similar expressions for the other processes.  Since $\Gamma\subHe >
\Ghethin$ for moderately thick clouds, due to the He~II LyC radiation,
we find that $q > p$ for these processes.  The coefficients for all of
these processes are listed in Table~\ref{table:emiss-coeff}.

\placetable{table:emiss-coeff}

Our treatment of the re-emission differs substantially from HM96.
Their method is based on an approximate integration over column
density and energy, performed at each redshift step.  In contrast, we
perform the integration exactly but only for one model, and then
parameterize the re-emission rates in terms of the continuum
opacities.  In addition, HM96 use an escape probability formalism, and
they ignore the distinction between $\Nobs$ and $\Nperp$ and the
distribution of incident and re-emitted radiation in angle.  Despite
these differences, in the end our results agree fairly well.  Our
He~II \Lya\ line intensity, for example, agrees with theirs to better
than 20\%.

The fraction of photoionizations due to photons emitted from the
clouds depends on the number of photons that are absorbed before they
can be redshifted below the ionization threshold.  This fraction
naturally drops at lower redshifts as the opacity decreases.  At
$z=3$, we find that $\sim \! 20\%$ of H~I and He~II ionizations are
due to cloud re-emission.  Usually the contribution of LyC emission to
the ionization of each species is nearly equal, but the other
processes give a slight extra ionization of H~I, which is then
amplified non-linearly by the dependence of the opacity on $\eta$.  As
shown in Figure~\ref{fig:eta-z}, the effect of the re-emission is to
increase $\eta$ by 5--15\% at $z\sim3$.  Our results differ in this
respect from HM96, who found a 15\% {\em decrease} in $\eta$ from
re-emission for $z \approx 3$, changing to an increase in $\eta$ of
about the same magnitude for $z < 1.5$.  

We now turn to the self-shielding problem.
In our model, self-shielding occurs only in He~II; any effect of
self-shielding on the H~I column distribution is assumed to be
reflected in the distribution we are using.  (For discussion of
whether its effect can actually be seen in the H~I distribution, see
Fardal \& Maloney 1995).  Its effect on a slab is not as simple a
function of column density as the re-emission rate, so we choose to
model it in a higher level of detail.  Our radiative transfer models
above give us the run of $N\subHe$ with $N\subH$.  We can obtain an
approximation to this function by considering the equations of
ionization balance for H~I and He~II.  Neglecting He~I and assuming
$n\subH \ll n_{H}$,
\begin{eqnarray}
  \Gamma\subHe n\subHe &=& \alpha\subHe n_e (n_{He} - n\subHe) \, , \\
  \Gamma\subH  n\subH  &=& \alpha\subH  n_e n_H \, .
\end{eqnarray}
Solving for $n\subH$ and $n\subHe$, integrating over the
slab length, and equating the slab lengths in the two equations gives
\begin{equation}
  \frac{\sigma\subHe}{\sigma\subH}  \frac{n_{\it He}}{n_H} 
  \int_{-\tau\subH/2}^{+\tau\subH/2} \tilde{\Gamma}(\tau\subH') d\tau\subH'
    I\subH = \tau\subHe + I\subHe 
  \int_{-\tau\subHe/2}^{+\tau\subHe/2} \tilde{\Gamma}(\tau\subHe') 
         d\tau\subHe' \, ,
\end{equation}
where the normalized photoionization rate is $\tilde{\Gamma} \equiv
\Gamma(\tau') / \Gamma^{\rm thin}$ and where $I\subH \equiv \Ghthin / n_e
\alpha\subH$ and $I\subHe \equiv \Ghethin / n_e \alpha\subHe$ as in HM96.
The point of this exercise is that the integrals depend only on the shape
of the incident spectrum, and fairly weakly on that.  We have found that,
for realistic conditions, $N\subHe$ can be calculated from the quadratic
equation,
\begin{equation}
  \label{etacol-approx}
  \frac{1}{4} \frac{n_{\it He}}{n_H} 
  \left( \frac{\tau\subH}{1 + 0.5 \tau\subH} \right)
    I\subH = \tau\subHe + I\subHe 
  \left( \frac{\tau\subHe}{1 + 0.7 \tau\subHe} \right) \, .
\end{equation}
This approximation follows the numerical results closely, as shown in
Figure~\ref{fig:column-eta}.  When the opacity is integrated over
$N\subHe$, the approximation is usually accurate to within a few
percent.  Again, this equation applies to $\Nperp$ not $\Nobs$.  In
this case it is convenient to take $\Nperp = \mu_1 \Nobs$, as
discussed above.  We find that for the opacity models discussed in
this paper, this is accurate to within 3\% over all frequencies,
compared with an exact angular convolution.  To get further
computational speed, we break up the integral in
equation~(\ref{cont-opt-depth}) in \S~\ref{sec:phys} into segments
where $N\subH$ increases by a factor of 3 or so.  We interpolate
$N\subHe(N\subH)$ in each segment with a power law.  The integral can
then be done semi-analytically.

The effect of self-shielding on $\eta$ is also shown in
Figure~\ref{fig:eta-z}c.  
The threshold opacity is increased only slightly ($\sim \! 5\%$) by
the self-shielding at a fixed $\eta$, but because the strength of
the effect increases with energy, and because the He~II ionization
fraction calculation is nonlinear, $\eta$ typically increases by $\sim
\! 15\%$ at $z=3$.  This increase
is nearly model-independent, since the onset of self-shielding is
determined by the radiative transfer.  The onset of the turnover in
$\eta(N\subH)$ is, in contrast, dependent on the density in the
clouds, which is in turn dependent on the poorly known cloud
thicknesses.  The spectrum at energies much higher than 4 Ryd, which
probes the high columns where this turnover occurs, is thus more
model-dependent, but it has a negligible effect on $\Gamma\subHe$.

In contrast to the case of cloud emission, our results for the
self-shielding on the He~II continuum opacity differ somewhat from
those of HM96.  Their use of face-on columns, the perpendicular (1-beam)
approximation, and the escape probability formalism gives results that
nearly match ours, but only if we leave out their final step of
doubling $N\subH$.  For optically thick slabs, this step effectively
doubles the flux incident on a slab and thus reduces the strength of
self-shielding.  In addition, in our treatment $\eta(N\subH)$ rises
smoothly, whereas in theirs it remains constant until it jumps at
$\tau\subHe = 4$.  The optical depth of this jump is too high to make
any difference in the He~II threshold opacity.  We thus find a higher
$\eta$ if all other assumptions are held constant.  Combined with the
difference in our results for the re-emission, we find that $\eta$ is
nearly 70\% higher in our models compared to the results of HM96.

We use the same approximations for $N\subHe(N\subH)$ to calculate the
He~II line opacity.  The form of the curve of growth, 
as shown in Figure~\ref{fig:column-eta}, means that the
self-shielding is almost irrelevant until $N\subHe \sim 10^{19}$~\cdu,
where the damping wings set in (at a lower column than in
H~I).  In future observations these damping wings may be seen at
wavelengths corresponding to lines with $N\subH \sim 10^{16}$~\cdu,
depending on the typical density in Lyman limit systems.  We find that
these mildly damped systems contribute
 $\Delta \tau\subHe \sim 0.1$ to the \Lya\ optical
depth $\tau\subHe$ at $z=3$.

\newpage

\section{Statistical Methods for Cloud Samples}
\label{app:statistics}

Many current opacity models for the high-$z$ IGM are based
on low-resolution spectra and uncertain column density 
distributions.  One of the primary checks of the opacity models
is the flux deficit shortward of the \Lya\ emission line (the $D_A$
parameter discussed in \S~2.3) which assumes a constant power-law
extrapolation of the continuum longward of \Lya.  
However, new data from high-resolution
spectrographs on large telescopes allows one to construct 
more accurate line lists with better column density distributions.
One clear result of these studies is the turnover in \Lya\
clouds with N(H~I) $\geq 10^{14.5}$ cm$^{-2}$.  
It is also possible to quantify the spectral curvature of
the QSO continuum, which falls off in flux relative to a power-law
extrapolation at wavelengths
below 1216 \AA\ (Zheng \ea 1997).   For these reasons, we
have developed a new statistical model for the column density
distribution and the resulting IGM opacity.   

Wherever possible, we have avoided binning the data in column density.
To derive our fits to the absorption-line distribution, we use the
maximum-likelihood method.  By first fitting the number of clouds per
unit redshift, we derive a redshift exponent $\gamma$, although we
have chosen to use a different value for some parts of the models (see
Table~\ref{table:col-dist}).  The probability of finding a cloud with
column $N_i$ in the sample is, according to equation~(\ref{col-dist})
integrated over redshift,
\begin{equation}
  p_{\it obs}(N) = \sum_q \frac{1}{\gamma_i + 1} (1+z)^{\gamma_i + 1}
            \Bigr]_{z_{\it min}^{(q)}}^{z_{\it max}^{(q)}}
            \frac{A_i}{N_r} \left( \frac{N}{N_r}\right)^{-\beta_i} \, ,
\end{equation}
where $N$ lies in column range $j$ and $(q)$ denotes the quasars
in the sample.  We maximize the likelihood function 
by summing over the columns in the sample,
\begin{equation}
        L = \prod_i p_{\it obs}(N_i) \, ,
\end{equation}
subject to the constraint that the expected number of lines in the sample
equal the number actually observed.

We would also like to have a means of estimating the errors in the
quantities drawn from the column density distribution.  The bootstrap
procedure (Lupton 1993) is a means of calculating 
statistical quantities about a
variable $x$ without making assumptions about the form of its
probability distribution $p(x)$.  The central idea is to turn the
observed dataset $\{ x_i, i=1...N \}$ into an ``observed probability
distribution'' $p\substar(x) = (1/N) \sum_i \delta(x-x_i)$.  By
drawing from this observed distribution, one may calculate statistical
quantities in a Monte Carlo fashion.  This process is usually carried
out until the errors from the Monte Carlo process become small enough.
The bootstrap is no guarantee of success.   Some quantities, such
as the sample mean and other linear functionals of $p(x)$, are 
well estimated by this procedure, while others, such as the two-point
correlation function of $x$, are not.  

The bootstrap procedure seems useful in the case of quasar absorption
line samples, where one wants to know quantities such as the line and
continuum opacities in H~I and He~II and their associated errors.
These errors have often been estimated by having several people 
make fits to different datasets and then comparing  the dispersion
among the fits. This procedure is suspect for several reasons, 
notably that the datasets usually overlap significantly!

However, to use the bootstrap here one must make some slight
modifications to the standard procedure (discussed in Lupton 1993, for
example).  In our case, we have not a probability distribution but a
{\em frequency} distribution, where the total number of data points is
a Poisson variable.  We can represent this as follows.  Suppose the
frequency distribution of our parameter $x$ (counts per unit $x$ per
unit time) is $f(x)$ and we observe for a ``time'' $T$.  Then our
bootstrap frequency distribution is $f\substar(x) = (1/T) \sum_i
\delta(x-x_i)$.  In other words, the $m^{\rm th}$ Monte Carlo realization
will be specified by $f_m(x) = (1/T) \sum_i w_i^{(m)} \delta(x-x_i)$ where
the $w_i^{(m)}$ are Poisson variables with mean 1.

This procedure is especially useful in the case of linear functionals
of the underlying distribution, i.e. a quantity $G = \int g(x) f(x)
dx$ where $g(x)$ is some function.  For example, $G$ might represent 
the H~I continuum opacity and $g(x) = 1 - e^{-\tau}$.  These functionals
are so simple that there is actually no need to run any Monte Carlo
program.  The bootstrap estimate of $G$ is 
\begin{equation}
\langle G\rangle \substar = \langle  \int g(x) f_m(x) dx \rangle \substar 
= \langle  \frac{1}{T} \sum_i w_i g(x_i) \rangle \substar = 
\frac{1}{T} \sum_i g(x_i)  \; , 
\end{equation}
where the star denotes an average over the bootstrap ensemble. Of course,
one might have naturally made this estimate of $G$ without ever regarding it
as a bootstrap procedure.  

To estimate the uncertainty in this bootstrap estimate, we can calculate
the bootstrap variance:
\begin{eqnarray}
\langle \sigma^2_G\rangle \substar & = & \langle G^2\rangle \substar -
\langle G\rangle \substar^2 = \frac{1}{T^2} \sum_{ij} g(x_i) g(x_j)
\left[ \langle w_i w_j\rangle \substar - \langle w_i\rangle \substar
\langle w_j\rangle \substar \right] \nonumber \\
 & = & \frac{1}{T^2} \sum_{i} \left[ (g(x_i) \right]^2 .
\end{eqnarray}
It is not hard to show that this has an expectation value exactly the
same as the variance in $\langle G\rangle \substar$ resulting from
estimates drawn from different datasets.  In other words, the error
bar calculated by the above equation is, on average, the correct error
bar.  It is also simple to calculate and requires no arbitrary
binning.

The observational problem at hand is rarely so simple, and one
must extend these results in several ways.  An important application
is the case where one does not know the $\{x_i\}$ exactly, but instead
each has an error bar of its own.  If we can suppose the true value of
each $x_i$ to have a distribution $p_i(x_i)$, then the bootstrap procedure
is to draw a value from this distribution $w_i^{(m)}$ times.  If 
$\langle g(x_i) \rangle_{p_i}$ represents the value of $g(x)$ averaged
over the distribution of $x_i$, the
resulting bootstrap mean and variance are
\begin{eqnarray}
\langle G\rangle \substar  
 & = & \frac{1}{T} \sum_i \langle g(x_i) \rangle_{p_i}, \\
\langle \sigma^2_G\rangle \substar & = & 
 \frac{1}{T^2} \sum_{i} \left( \langle g(x_i) \rangle_{p_i} \right)^2 .
\end{eqnarray}

Another complication is that frequently the detection rate of the
events $\{ x_i \}$ is not perfect, and instead one has some detection
rate $F_{\rm det}(x)$.  To correct for this, we divide the ``observed''
bootstrap distribution by $F_{\rm det}(x)$.  The result is that $g(x_i)$
gets replaced by $g(x_i) / F_{\rm det}(x_i)$ in the equations above.

In addition, the observed frequency of events is often not constant
with time $t$, as was assumed above, and different ranges of $x$ will
often be observed for different amounts of time.  If one {\em knows}
that the $t$ and $x$ dependence are separate factors, it is not
difficult to adjust the equations for the $t$ dependence to derive a
smooth function $G(t)$.  If this is not known, but is consistent with
the data set, one can assume a separable dependence (at the cost of an
unquantified uncertainty).  Suppose that $f(x,t) = U(t) \tilde{f}(x)$.
Changing to the variable $\tau \equiv \int U(t) dt$, the 
distribution $f(x,\tau)$ is constant in $\tau$, so we can just apply the
theory above.  If each value $x_i$ was observable for a ``time'' $\tau_i$,
and had a detection rate $F_{\rm det}(x_i)$, 
we can write the bootstrap distribution as  $f\substar(x, \tau) 
= \sum_i (w_i \delta(x-x_i)) / (\tau_i F_{\rm det}(x_i))$.  
Upon performing the bootstrap average and converting back to $t$, we get
\begin{eqnarray}
\langle G\rangle \substar & = & 
  U(t) \sum_i \frac{g(x_i)}{\int_i U(t) dt \cdot F_{\rm det}(x_i)} \\
\langle \sigma^2_G \rangle \substar & = & 
(U(t))^2 \sum_i \left(\frac{g(x_i)}{\int U(t) dt \cdot F_{\rm det}(x_i)} \right)^2.
\end{eqnarray}

In the case of quasar absorption line distributions, the variable $x$
becomes column density $N_{\rm HI}$, the time $t$ becomes redshift
$z$, and the function $G$ may be the continuum opacity 
$d\tau_{HI}/dz$ or the line opacity $\tau_{\rm HI}$.
If the frequency of detecting absorption-line clouds is proportional
to $(1+z)^{\gamma}$, and there are several quasars in the sample,
the continuum opacity estimates become
\begin{eqnarray}
\langle \frac{d\tau_{HI}}{dz} \rangle\substar &=& 
  (1+z)^{\gamma} \sum_i \frac{ 1 - \exp\left(-\tau(N_i)\right) }
  { \sum_{(q)} \int^{(q)} (1+z)^{\gamma} dz \, 
  \, F^{(q)}_{\rm det}(N_i) }  ,                      \\
\langle \sigma^2_{d\tau_{HI}/dz} \rangle\substar &=&
  (1+z)^{2\gamma} \sum_i \left( \frac{1 - \exp\left(-\tau(N_i)\right)}
  { \sum_{(q)} \int^{(q)} (1+z)^{\gamma} dz \, 
  \, F^{(q)}_{\rm det}(N_i)} \right)^2 .
\end{eqnarray}
If $\Wbar_\lambda^{\rm rest}(N)$ is the average equivalent width at a
given column, the line opacity and its error estimate are calculated
from the formulae
\begin{eqnarray}
\langle \tau_{HI} \rangle\substar &=& 
  (1+z)^{\gamma+1} \sum_i 
  \frac{ \Wbar_{\lambda}^{\rm rest}(N) / \lambda^{\rm rest} }
  { \sum_{(q)} \int^{(q)} (1+z)^{\gamma} dz \, 
  \, F^{(q)}_{\rm det}(N_i) }  ,                      \\
\langle \sigma^2_{\tau_{HI}} \rangle\substar &=&
  (1+z)^{2(\gamma+1)} \sum_i \left( 
  \frac{ \Wbar_{\lambda}^{\rm rest}(N) / \lambda^{\rm rest} }
  { \sum_{(q)} \int^{(q)} (1+z)^{\gamma} dz \, 
  \, F^{(q)}_{\rm det}(N_i) } 
  \right)^2 .
\end{eqnarray}

\newpage

\section{Small Effects on $N\subHe / N\subH$}
\label{app:small-effects}

In \S~\ref{sec:results} we examined the major effects on
$\eta$ resulting from uncertainties in parameter space.  We list here
some of the less significant effects, some of which we have ignored
entirely in our calculations.  
We have tried varying the Pei (1995) luminosity function within the
uncertainties given in the paper.  The result was a variation in $\eta$
of $\sim \! 5 \%$, much less than the differences between our quasar
models 1 and 2.  Clearly, the uncertainty in the quasar luminosity
function is mostly due to systematic effects such as dust obscuration.

The ionizing spectrum could also be affected by the line opacity of
He~II.  A photon with energy slightly below 4~Ryd must run a
gauntlet of He~II absorption lines, many with significant opacity,
before being redshifted below the \Lya\ energy of 3 Ryd.
However, the mean spectrum is only affected by line splitting---e.g.,
the absorption of a Ly$\beta$ photon and subsequent re-emission of
Ly$\alpha$ and Balmer-$\alpha$ photons.  This splitting takes place
only rarely, about 1 time in 8 for the process just described.  We
have modeled this process as repeated tosses of a photon against a
screen that represents the Ly$\beta$ barrier.  Each time the photon
hits the screen, it can either bounce back, pass through, or undergo
line-splitting.  By solving the equation for the fraction of photons
that eventually undergo line-splitting, we find that this process
becomes significant for $z \gtrsim 4.0$.  The splitting of higher
Lyman-series lines will become significant at even higher redshifts.
Since the energy region 3--4~Ryd has little effect on the H~I or He~II
photoionization rates, we have ignored line-splitting processes in
this paper.  They may be of more significance for metal-line ratios
like Si~IV/C~IV.

We also ignore the He~I continuum opacity in this paper.  \ME\ \&
Ostriker (1992) showed that the optical depth is quite small unless
the ionizing background is dominated by decaying neutrinos. In the
standard picture of highly photoionized clouds, the He~I/He~II ratio
is $\sim \! 0.2 n_e J_{-21}^{-1}$ for $\alpha_b \approx 1.8$ between 1.0
and 1.8 Ryd.  Since the electron densities, $n_e$, in standard QSO
absorption systems are expected to be quite small, the continuum
opacity of He~I is substantially less than that of He~II and H~I.  
Observational indications of the presence of He~I in the IGM at 
$z \gtrsim 2$ are limited.  Reimers \ea (1992) and Reimers \& Vogel (1993) 
find He~I/H~I = 0.026--0.05 in partial Lyman Limit systems
with $N\subH = 10^{16.6}-10^{16.9}~{\rm cm}^{-2}$.
Since He~I/H~I = $\eta$(He~I/He~II)$ = 0.2 J_{-21}^{-1} n_e \eta$,
if $n_e \approx n_H = (2 \times 10^{-3}~{\rm cm}^{-3}) 
(N\subH / 10^{17} {\rm cm}^{-2})^{1/2}$, He~I/H~I $= 0.04 J_{-21}^{-1} 
\eta_{100} (N\subH / 10^{17} {\rm cm}^{-2})^{1/2}$, where 
$\eta_{100} = \eta / 100$.  Thus, the few known measurements are 
consistent with this density model, but our knowledge of the 
appropriate run of He~I/H~I with $N\subH$ is still very tentative.
If He~I/H~I $\approx 0.03$ in all absorbers at $z \gtrsim 3$,
the radiation at 2--3 Ryd is reduced by less than 10\% over
the case when He~I opacity is neglected, and less at later
times.  For the more generous He~I/H~I ratio implied by our 
$N\subH$ dependent density model, the reduction may approach 20--30\% 
at high redshift.  We have also ignored the
continuum opacity of metals in absorption-line clouds,
an approximation which has negligible effect on our results.

The column ratio $\eta$ is proportional to the primordial abundance of
He, which is probably uncertain by about 5\% (Schramm \& Turner 1997).
It also depends on the temperature of the \Lya\ clouds, but even for
temperature uncertainties of a factor $\sim \! 2$, equation~(\ref{eta})
shows that this introduces an uncertainty $\lesssim 5\%$.  It is clear
that these small uncertainties do not present obstacles to progress on
this problem.

\clearpage

\begin{deluxetable}{crrrrrr}
\footnotesize
\tablecaption{Absorption Line Distributions$^a$ \label{table:col-dist}}
\tablewidth{0pt}
\tablehead{
   \colhead{Model} &
   \colhead{$N_{\it min}$ (\cdu)\tablenotemark{b}} & 
   \colhead{$N_{\it max}$ (\cdu)\tablenotemark{b}} &  
   \colhead{$A$ \tablenotemark{c}} &
   \colhead{$\beta$\tablenotemark{c}} &
   \colhead{$\gamma$\tablenotemark{c}} &   
   \colhead{$P_{\it fit}$\tablenotemark{d}}
} 
\startdata
A1 &   ...         & $1\times 10^{14}$ & $1.45\times 10^{-1}$ & 1.40 & 2.58 & 0.15 \nl
   & $1\times 10^{14}$ & $1\times 10^{16}$ & $6.04\times 10^{-3}$ & 1.86 & 2.58 &      \nl
   & $1\times 10^{16}$ & $1\times 10^{19}$ & $2.58\times 10^{-2}$ & 1.23 & 2.58 &      \nl
   & $1\times 10^{19}$ & $1\times 10^{22}$ & $8.42\times 10^{-2}$ & 1.16 & 1.30 &      \nl
A2 &  ...          & $2\times 10^{14}$ & $6.69\times 10^{-2}$ & 1.49 & 2.58 & 0.04 \cr
   & $2\times 10^{14}$ & $2\times 10^{15}$ & $5.93\times 10^{-3}$ & 1.88 & 2.58 &      \cr
   & $2\times 10^{15}$ & $1\times 10^{19}$ & $2.95\times 10^{-2}$ & 1.47 & 2.58 &      \cr
   & $1\times 10^{19}$ & $1\times 10^{22}$ & $8.42\times 10^{-2}$ & 1.16 & 1.30 &      \cr
A3 &  ...          & $2\times 10^{14}$ & $6.69\times 10^{-2}$ & 1.49 & 2.58 & 0.04 \cr
   & $2\times 10^{14}$ & $2\times 10^{15}$ & $5.93\times 10^{-3}$ & 1.88 & 2.58 &      \cr
   & $2\times 10^{15}$ & $1\times 10^{17}$ & $2.95\times 10^{-2}$ & 1.47 & 2.58 &      \cr
   & $1\times 10^{17}$ & $1\times 10^{22}$ & $1.58\times 10^{-1}$ & 1.50 & 1.50 &      \cr
A4 &  ...          & $1\times 10^{17}$ & $1.32\times 10^{-1}$ & 1.43 & 2.46 &$<10^{-24}$   \cr
   & $1\times 10^{17}$ & $1\times 10^{22}$ & $4.22\times 10^{-1}$ & 1.39 & 0.68 &      \cr
\enddata
\tablenotetext{a}{ Models of opacities from different H~I column density 
distributions (see \S~2.3). 
}
\tablenotetext{b}{ Limits on H~I column density for subrange.  
The minimum column for the lowest subrange is left blank, since we 
treat it as a free parameter.  
}
\tablenotetext{c}{
  Parameters used in equation~\ref{col-dist}.
}
\tablenotetext{d}{
  Formal fit probability from combined K-S/binomial test.
}
\end{deluxetable}


\begin{deluxetable}{cccc}
\footnotesize   
\tablecaption{  
  Quasar emission models   
  \tablenotemark{a}
     \label{table:emissivity} 
}
\tablewidth{0pt}
\tablehead{
  \colhead{Model} & 
  \colhead{$j_{H\ast}$} & 
  \colhead{$z_{\ast}$} & 
  \colhead{$\sigma_{\ast}$}
}
\startdata
Q1  & $2.7 \times 10^{24}$   & 2.75  & 0.93 \nl
Q2  & $6.2 \times 10^{24}$   & 3.37  & 1.03 \nl
\enddata
\tablenotetext{a}{    
  The parameters are used in \S~2.5 [eq.~(\ref{emissivity})] of the text. 
  The comoving emissivity, $j_{H\ast}$, is given in units of
  ($h_{75}$ erg s$^{-1}$ Mpc$^{-3}$ Hz$^{-1}$ sr$^{-1}$).  Corrections
  have been applied for the different Hubble constant and spectral slope
  $\alpha_{\rm BH}$ assumed by Pei (1995).
}
\end{deluxetable}

\begin{deluxetable}{lcc}
\footnotesize
\tablecaption{
  Coefficients of continuous emission.  
   \label{table:emiss-coeff} 
}
\tablewidth{0pt}
\tablehead{
  \colhead{Process}     & 
  \colhead{$p_{\it process}$\tablenotemark{a}} & 
  \colhead{$q_{\it process}$\tablenotemark{b}}
}
\startdata
H I Lyman continuum     & 0.43  & 0.22 \nl
He II Lyman continuum   & 0.33  & 0.16 \nl
He II Balmer continuum  & 0.17  & 0.22 \nl
He II 2-$\gamma$        & 0.20  & 0.26 \nl
He II \Lya              & 0.45  & 0.58 \nl
\enddata
\tablenotetext{a}{ 
 $p_{\it process}$ denotes the probability of this
 process occuring in a single recombination. 
}
\tablenotetext{b}{
 $q_{\it process}$ denotes a dimensionless factor
  that accounts for its likelihood integrated over clouds of all
  columns.  (Since the integral converges at both its upper and
  lower limits, we have neglected the finite cutoffs used in our
  paper.)
}
\tablecomments{
  A temperature of $T = 2 \times 10^4$~K is assumed. 
}
\end{deluxetable}

\clearpage

\addtocounter{figure}{-1}         

\begin{figure}
\plotone{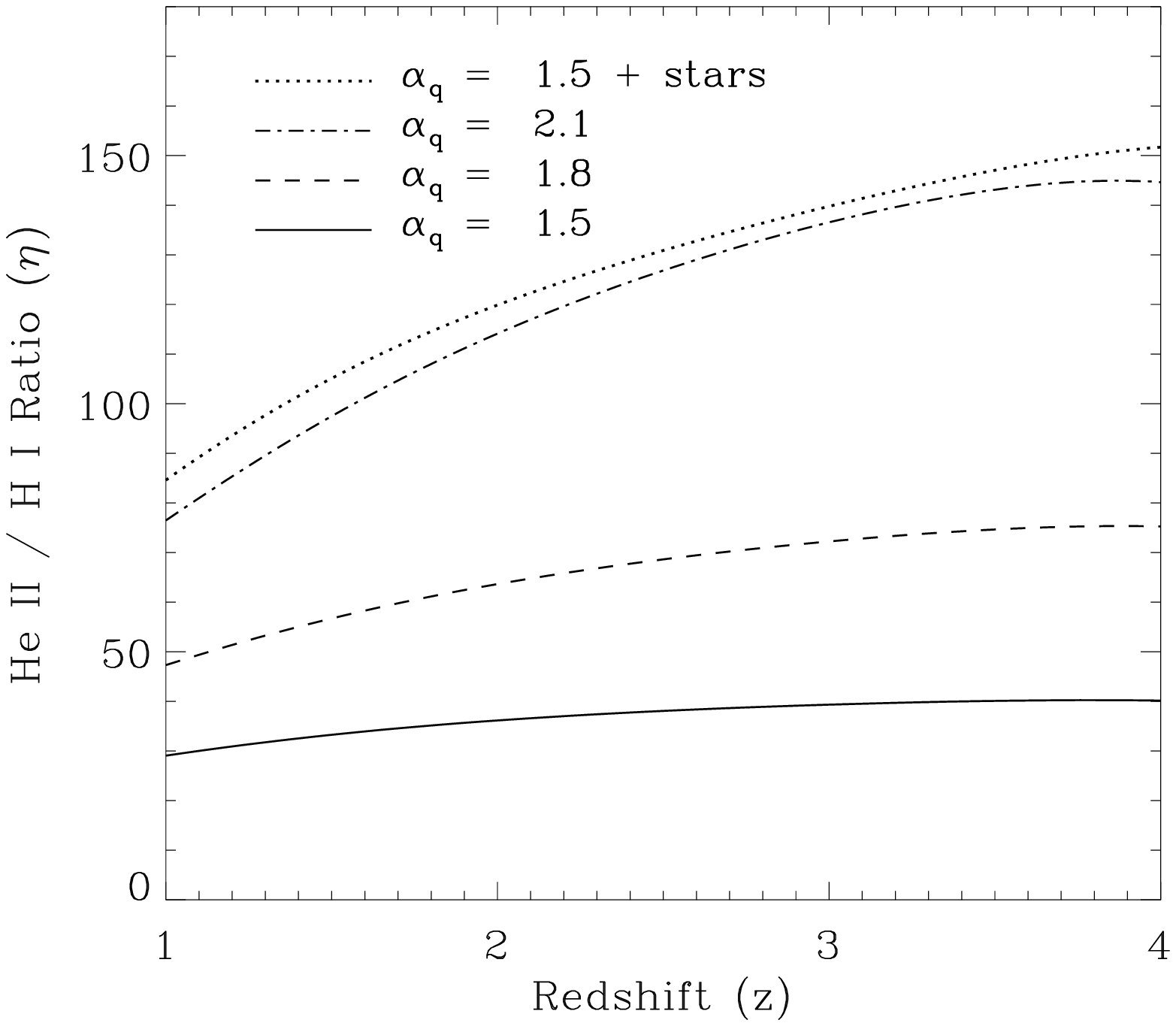}
\caption{
(b) Effect of the ionizing spectrum.  Uses
absorption model A2 and quasar model Q2.  Three curves
spanning the uncertainty in the mean quasar spectrum are shown.  In
addition, we include a model that includes a stellar component, of
magnitude 1.3 in relation to the quasar contribution at 1 Ryd.  As
argued in the text, this should produce results comparable to making
the quasar spectral index larger (softer) by $\ln(1+1.3) / \ln(4) = 0.6$, 
and this is indeed seen in the plot.  The agreement is not exact because the
stellar and quasar spectral shapes are different near 1 Ryd.  }
\end{figure}

\addtocounter{figure}{-1}         

\begin{figure}
\plotone{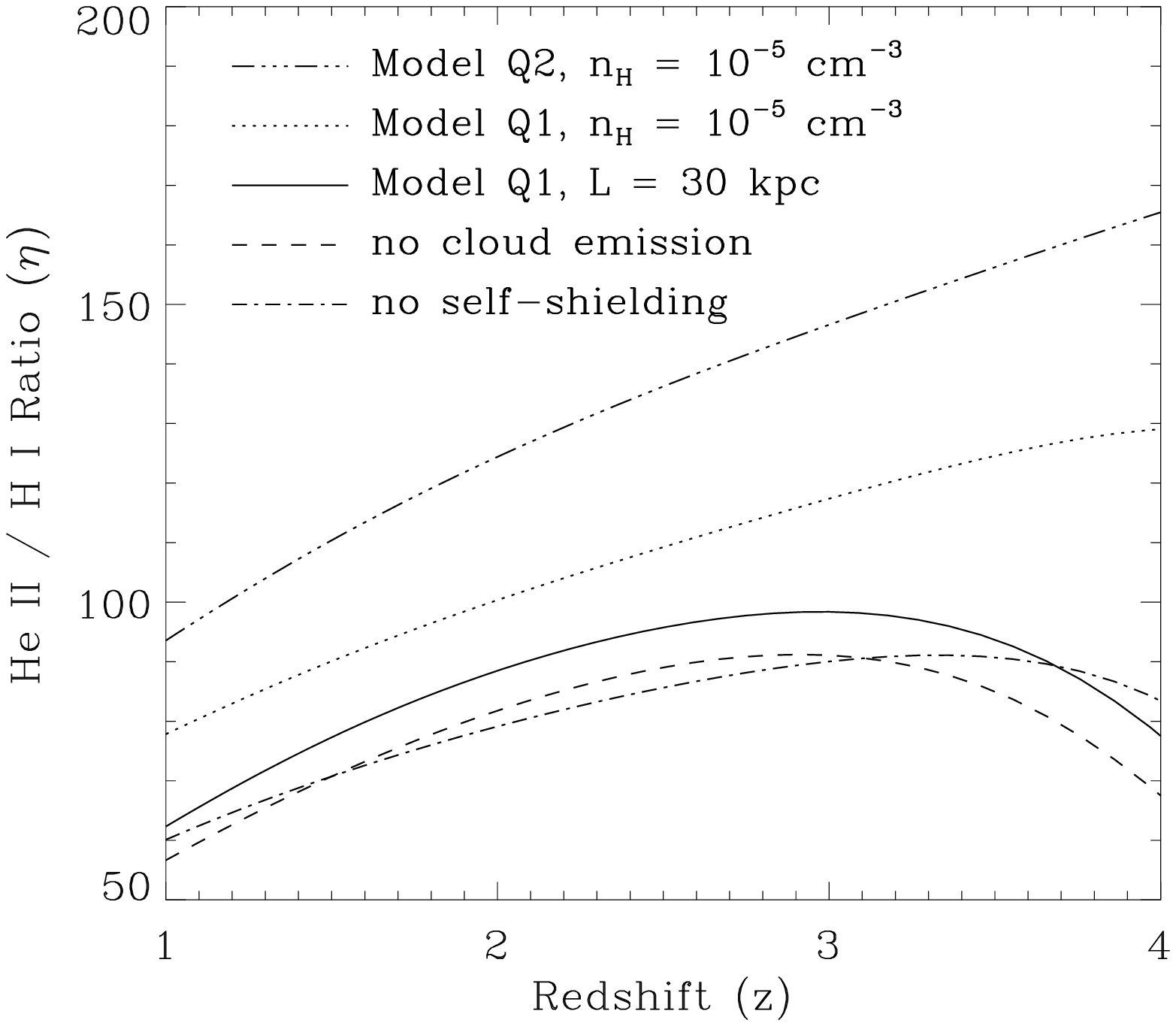}
\caption{ 
(c) Influence of the density within the absorbing clouds,
the quasar luminosity function, and the emission and self-shielding
processes.  All curves use opacity model A2 with a quasar spectral
index $\alpha_s = 2.1$.  The bottom three curves all use a density of
$n_H = 10^{-4}$~cm$^{-3}$ within the clouds and the Q1 luminosity
function, but we have omitted the cloud re-emission in the dashed curve 
and the self-shielding in the dot-dashed.  }
\end{figure}

\begin{figure}
\plotone{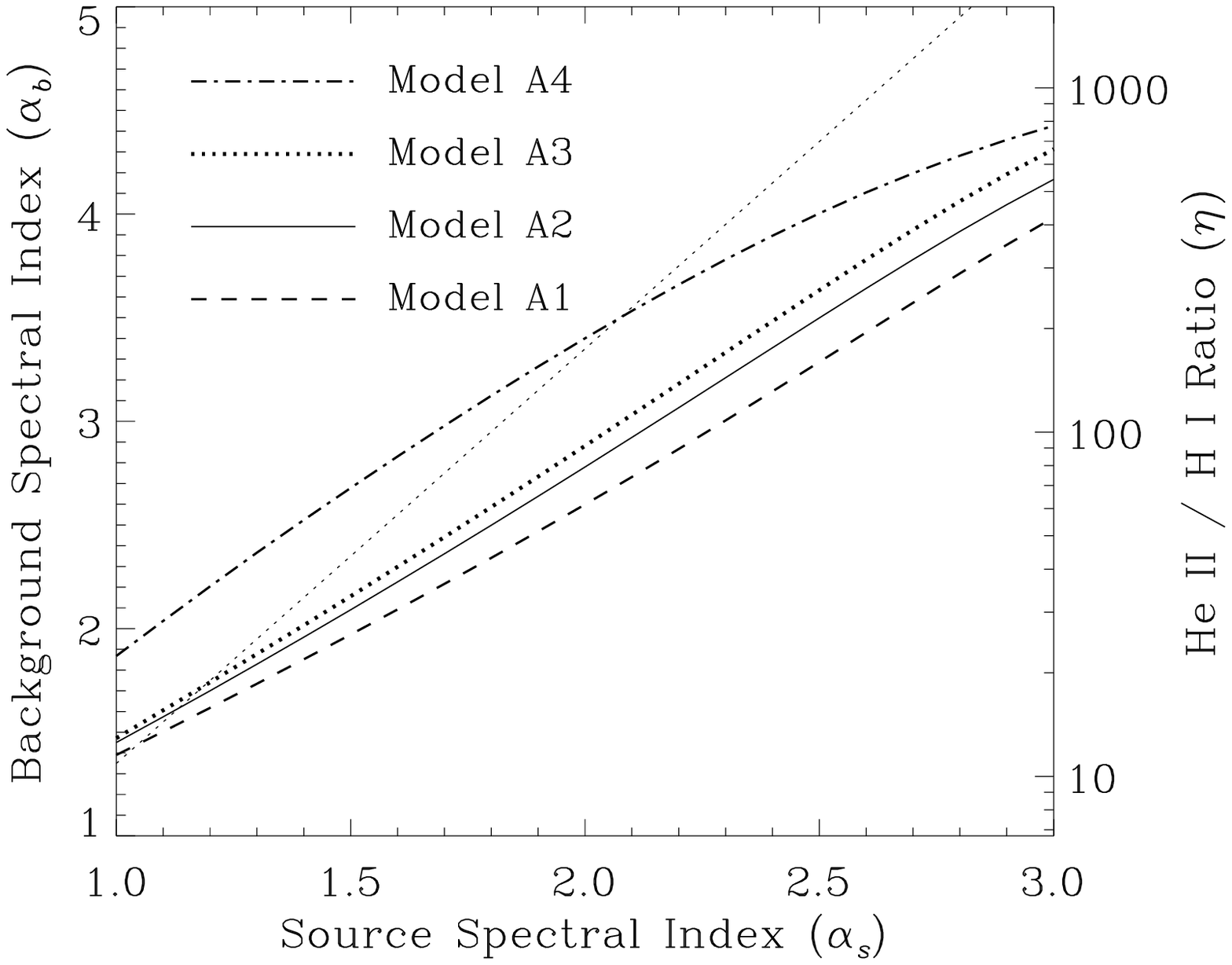}
\caption{
\label{fig:alphas-alphab}
Dependence of the background spectral index $\alpha_b$ upon the
intrinsic spectral index $\alpha_s$ of ionizing sources, for the
four opacity models in the text, for quasar model Q1 and  $z=3$.
The light dotted line shows the analytic model, 
$\alpha_b = 2.0 \alpha_s  - 0.64$, of \S~\ref{sec:phys}.
It is somewhat steeper than the numerical models, partly because it ignores 
the finite-density effects that affect models with soft spectra.  }
\end{figure}

\begin{figure}
\plotone{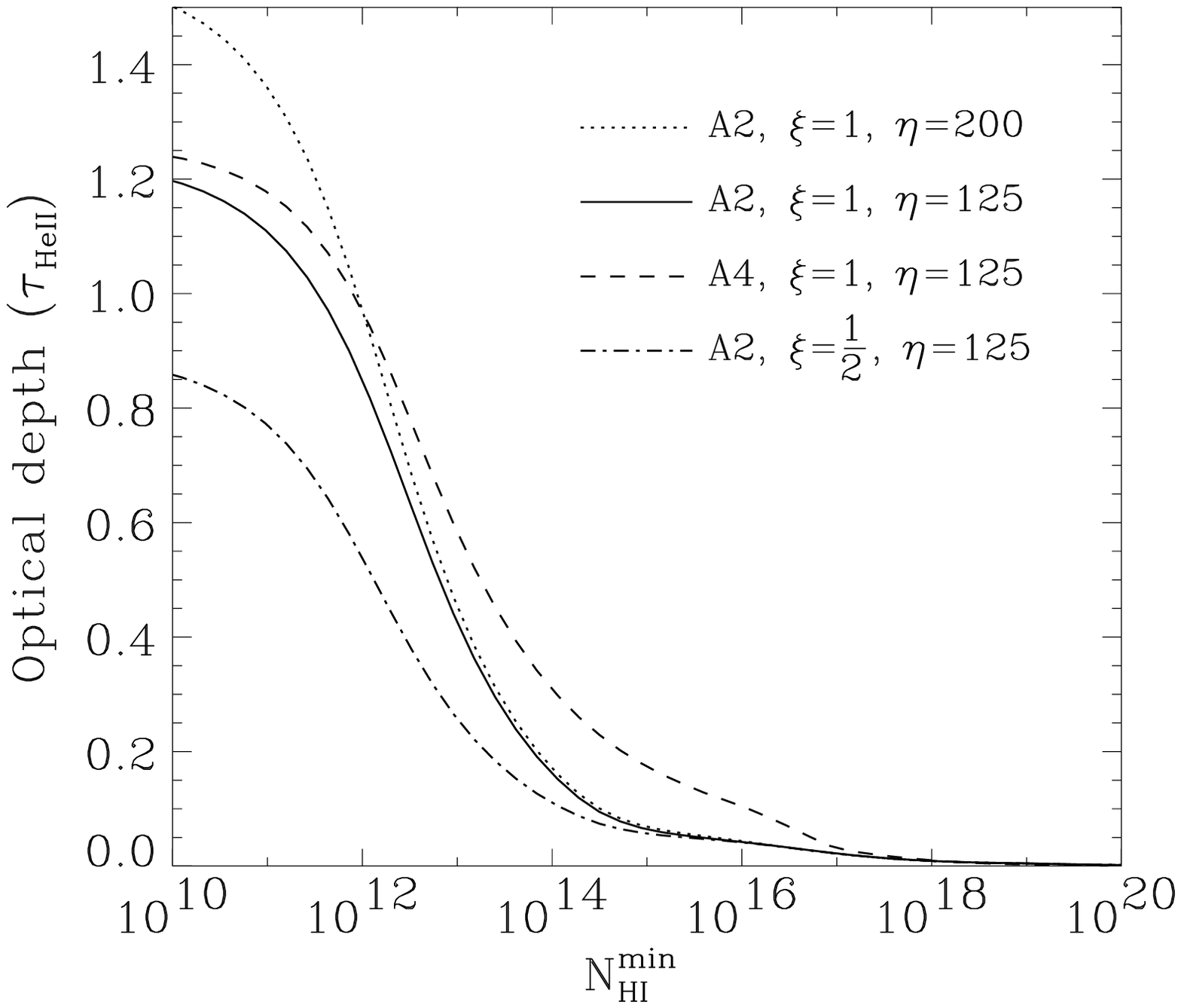}
\caption{
\label{fig:cum-opt-depth}
Cumulative value of $\tau\subHe$ with decreasing N$^{\rm min}_{HI}$ in
distribution of column densities, for several opacity models at $z = 2.4$. 
The solid line shows the results of our ``standard'' model without diffuse 
IGM.  This uses opacity model A2, bulk He~II Doppler widths, quasar model 
Q2, source spectrum $\alpha_s=2.1$, resulting in $\eta=125$.  Three variants 
are also shown: with a softer source spectrum of $\alpha_s=2.3$ giving 
$\eta=200$; with thermal He~II Doppler widths; and with the opacity model 
A4 (with $\alpha_s=1.6$ to give $\eta=125$ as in our standard model).  }
\end{figure}

\begin{figure}
\plotone{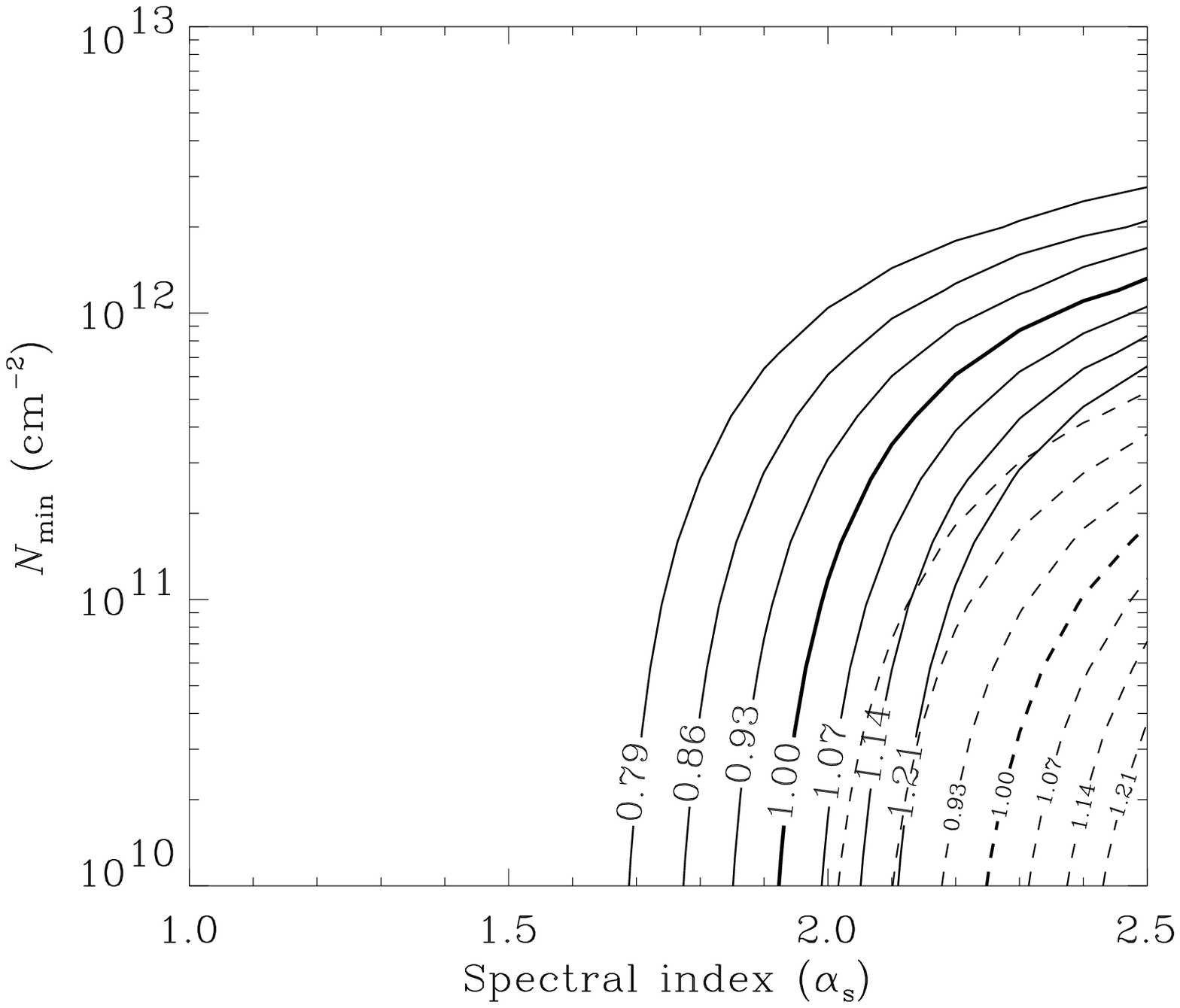}
\caption{
\label{fig:optdepth-contour}
Contours of the cumulative optical depth, $\tau\subHe(N\subH > N_{\it min})$,
as a function of the quasar spectral index $\alpha_s$, at $\langle z_{\it abs}
\rangle =2.4$ to match D96's observations of \HS.  Contours are in 1$\sigma$ 
intervals of 0.07 around their best value of $\tau\subHe = 1.00$. 
Curves are shown for bulk Doppler widths ($\xi\subHe = 1$, solid lines), 
and thermal widths ($\xi\subHe = 0.5$, dotted lines).  We assume negligible
ionizing radiation from galaxies and adopt opacity model A2 and quasar
model Q2. }
\end{figure}

\addtocounter{figure}{-1}         

\begin{figure}
\plotone{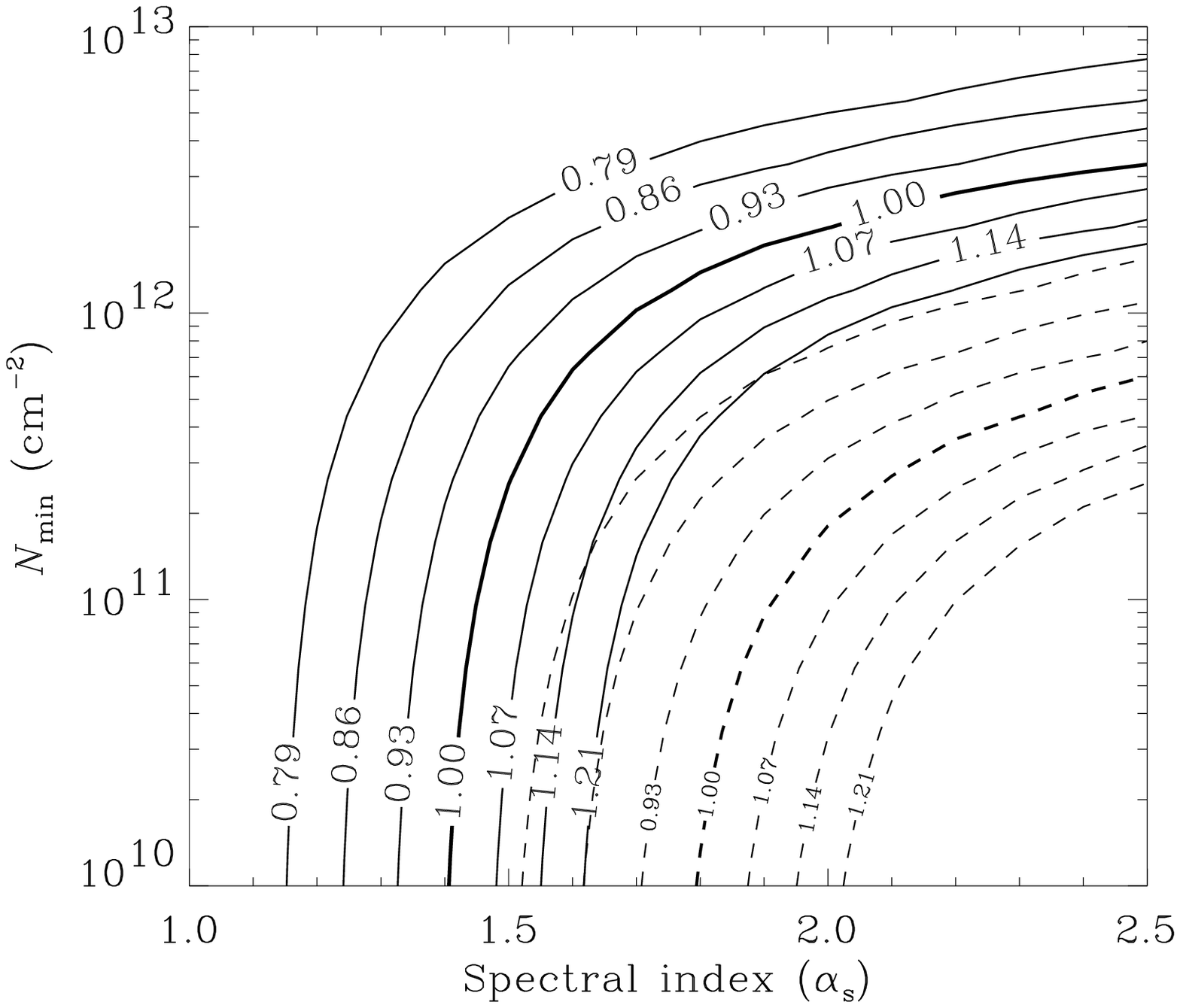}
\caption{
(b) Same, but for absorption model A4.  }
\end{figure}

\begin{figure}
\plotone{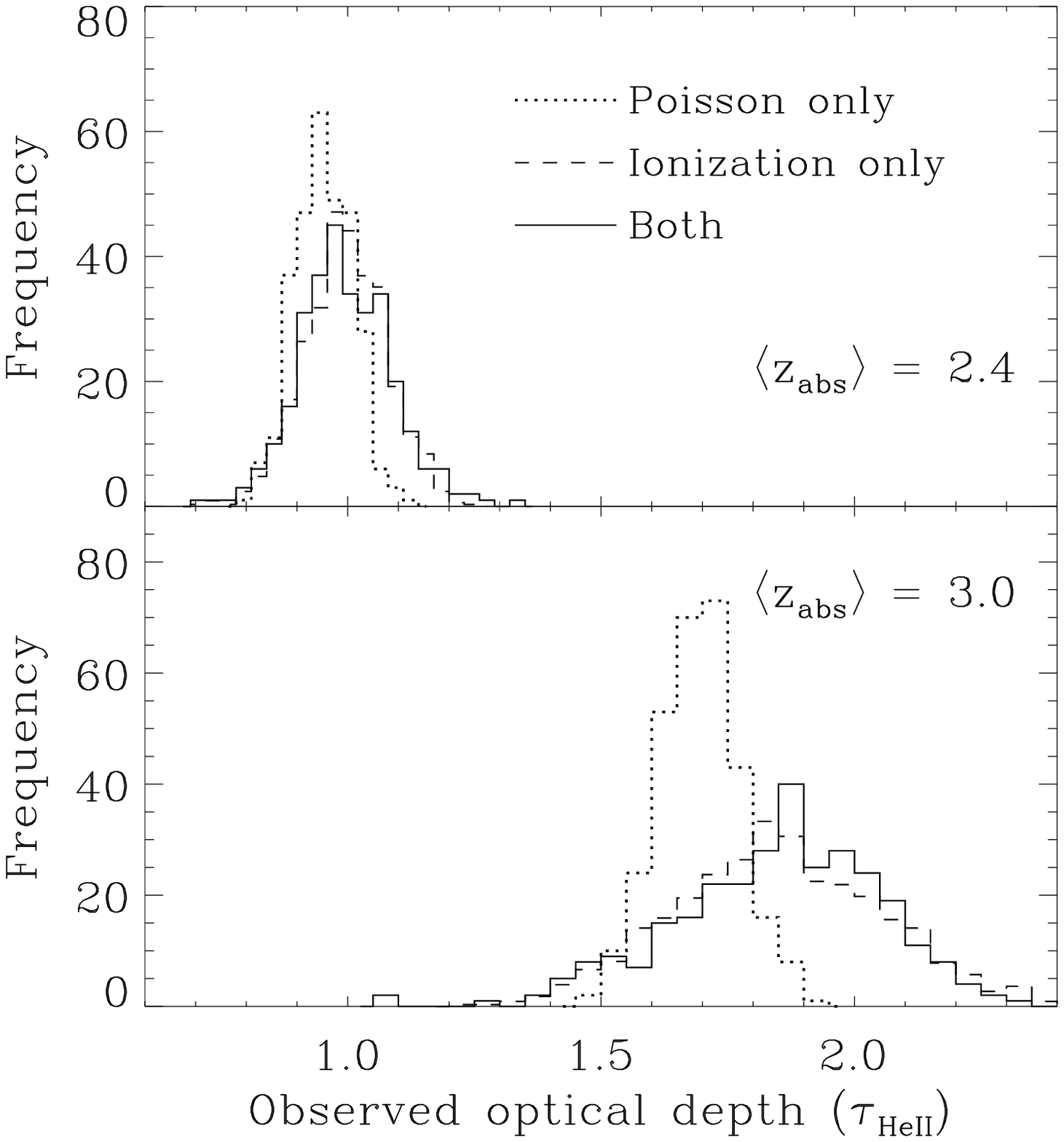}
\caption{
\label{fig:tau-histogram}
Histogram from Monte Carlo simulations of the observed effective optical 
depth $\tau\subHe = -\ln(\bar{T})$ in the He~II ``$D_A$ window'',  
from 262--294~\AA\ in the rest frame of a quasar.  Top panel: towards a 
quasar with $z_{\it em} = 2.7$ like \HS.  
Bottom panel: towards a quasar with $z_{\it em} = 3.37$, giving
a mean absorption redshift of $z_{\it abs} = 3.0$.
The dotted line incorporates effects of
Poisson fluctuations only, the dashed line the effects of ionization 
fluctuations only (averaging over the absorbers), 
while the solid line includes both types of fluctuations.
Uses absorption model A2 and bulk Doppler widths ($\xi\subHe = 1$), 
and quasar model Q1 with spectral index $\alpha_s = 2.1$ giving $\eta = 95$.
The strength of the ionization fluctuations increases rapidly with redshift,
in contrast to the intrinsic fluctuations.  }
\end{figure}

\begin{figure}
\epsscale{.75}
\plotone{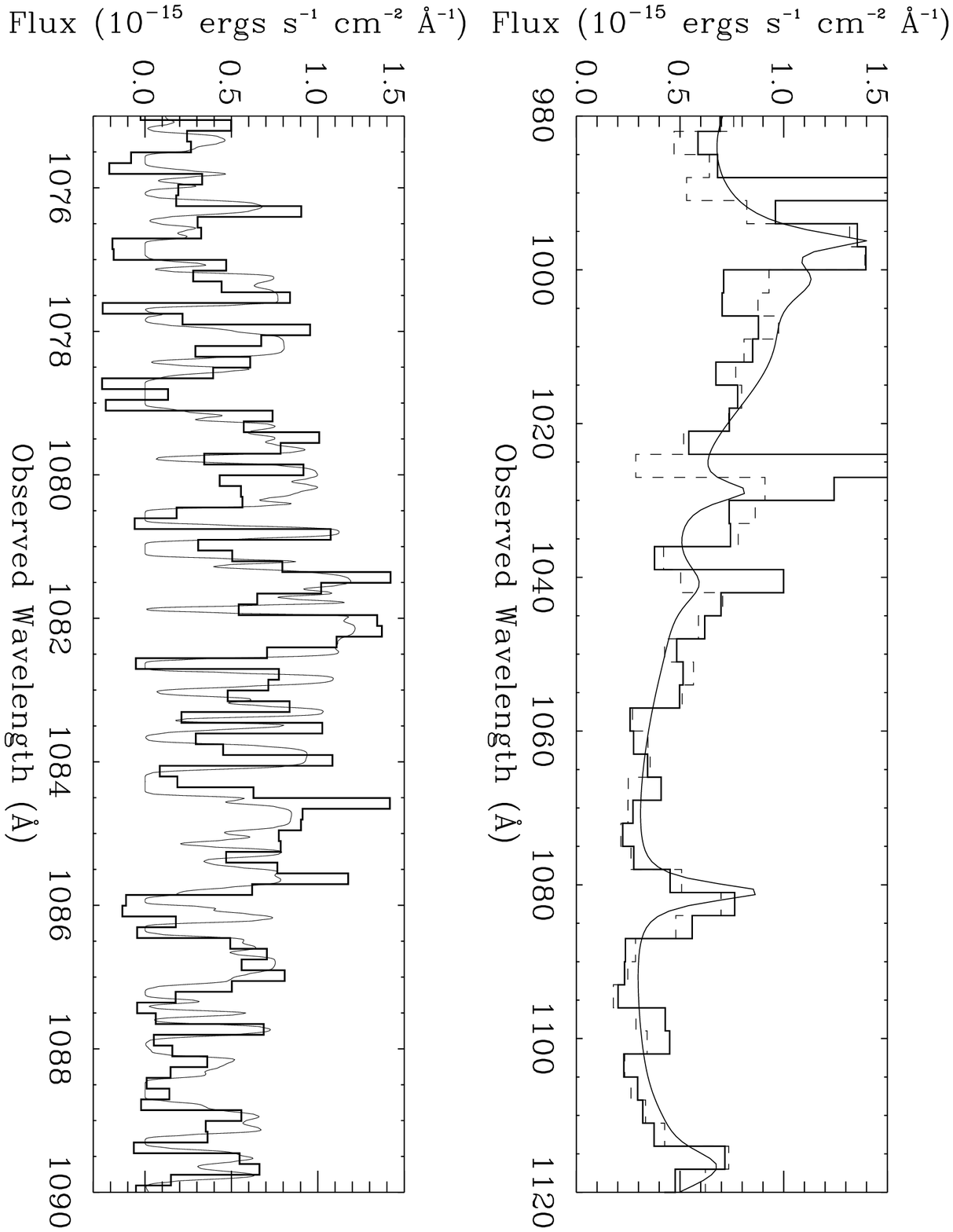}
\caption{
\label{fig:fuse-spectrum} 
Simulation of an observation of He~II \Lya\ absorption with the FUSE 
satellite, showing the spectrum of a QSO like \HS\ with $z_{\it em} = 2.7$   
and flux at the He~II Lyman limit $1.25 \times 10^{-15}$ erg s$^{-1}$ 
cm$^{-2}$ Hz$^{-1}$ \AA$^{-1}$.  Assumes absorption model A2 with thermal 
linewidths ($\xi\subHe = 1$), model Q1, $\alpha_s = 1.8$, and a substantial 
IGM with $\omigm = 0.026$; in this model $\eta = 53$.  
The exposure time is $3 \times 10^5$~s.
The upper panel shows the large-scale view, with
a solid curve indicating the mean absorption as a function of wavelength.
The dashed histogram is the
binned intrinsic spectrum, while the solid histogram incorporates
the effects of finite resolution, photon statistics, and sky and 
detector backgrounds (including an H~I $\lambda$1025.72 Ly$\beta$ line).
In this particular realization, several quasars near the line of sight
induce peaks in the ionizing background and thus in the transmission.
The lower panel shows a smaller view with 0.15 \AA\ pixels, centered
on the peak at 1082 \AA.  In this panel the light solid line is the
intrinsic spectrum and the heavy histogram is the observed spectrum.
Even with thermal widths, the structure in the spectrum is resolved by
the FUSE detector.  
}
\end{figure}

\begin{figure}
\epsscale{1.1}
\plotone{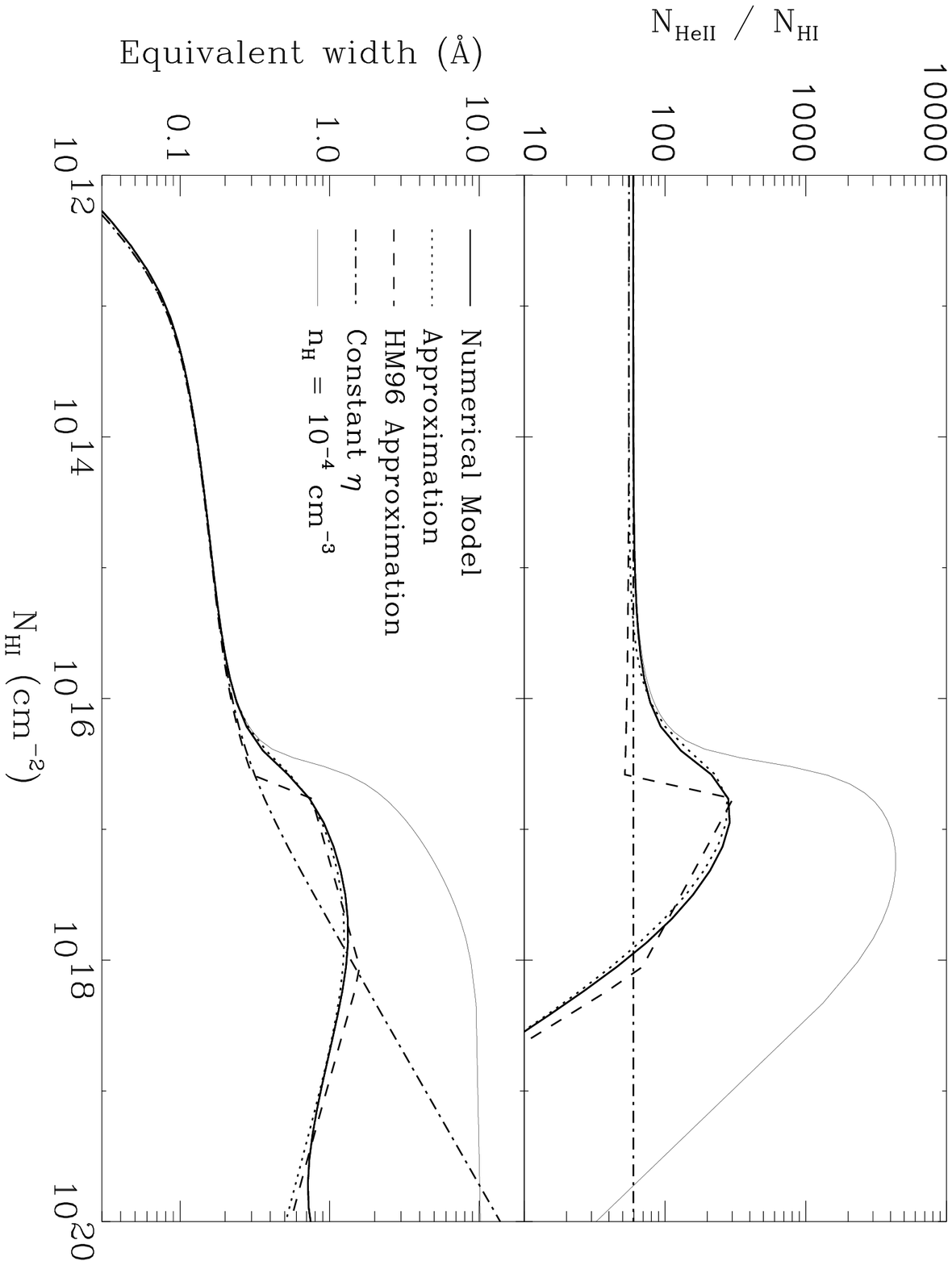}
\caption{
\label{fig:column-eta}
Column ratio and He~II curve of growth in slabs of increasing optical
depth.  Results of our numerical slab models are compared to
three approximations: a constant $\eta$, the smooth approximation of
equation~(\ref{etacol-approx}), and the step-like approximation of
HM96.  Uses our standard density model (eq. [13]), except for the top 
curve, which uses a constant density of $n_H = 10^{-4}$ cm$^{-3}$.  The 
background spectrum is taken to be a power law ($\alpha_b = 2.5$) with 
intensity at the H~I Lyman limit $J\subH = 10^{-21}$~\fluxu. 
These parameters result in an optically thin $\eta = 60$.
The columns depicted here are the observed columns,
$\Nobs = \mu_1^{-1} \Nperp$, as discussed in Appendix~\ref{app:re-emiss}.
}
\end{figure}

\end{document}